\PassOptionsToPackage{table,xcdraw}{xcolor}
\documentclass[acmsmall]{acmart}

\usepackage{makecell}
\usepackage{multirow}
\usepackage[table]{xcolor}

\AtBeginDocument{%
  }

\setcopyright{acmlicensed}
\copyrightyear{2024}
\acmYear{2024}
\acmDOI{XXXXXXX.XXXXXXX}

\acmJournal{CSUR}
\acmVolume{XX}
\acmNumber{X}
\acmMonth{4}

\begin{document}

\title{Survey on Adversarial Attack and Defense for Medical Image Analysis: Methods and Challenges}

\author{Junhao Dong}
\email{dongjh8@mail2.sysu.edu.cn}
\orcid{0000-0002-6232-9157}

\author{Junxi Chen}
\email{chenjx353@mail2.sysu.edu.cn}
\orcid{0009-0000-2512-4438}

\author{Xiaohua Xie}
\email{xiexiaoh6@mail.sysu.edu.cn}
\orcid{0000-0002-0310-4679}
\authornote{Corresponding author.}

\author{Jianhuang Lai}
\email{stsljh@mail.sysu.edu.cn}
\orcid{0000-0003-3883-2024}
\affiliation{%
	\institution{School of Computer Science and Engineering, Sun Yat-sen University, and Guangdong Province Key Laboratory of Information Security Technology}
	\city{Guangzhou}
	\country{China}
}

\author{Hao Chen}
\authornotemark[1]
\affiliation{%
	\institution{Department of Computer Science and Engineering, Department of Chemical and Biological Engineering, and Division of Life Science, $\!\!$ Hong Kong University of Science and Technology}
	\country{$\!$Hong Kong,$\!$ China}}
\email{jhc@cse.ust.hk}
\orcid{0000-0002-8400-3780}

\renewcommand{\shortauthors}{Dong et al.}

\begin{abstract}
Deep learning techniques have achieved superior performance in computer-aided medical image analysis, yet they are still vulnerable to imperceptible adversarial attacks, resulting in potential misdiagnosis in clinical practice. Oppositely, recent years have also witnessed remarkable progress in defense against these tailored adversarial examples in deep medical diagnosis systems. In this exposition, we present a comprehensive survey on recent advances in adversarial attacks and defenses for medical image analysis with a systematic taxonomy in terms of the application scenario. We also provide a unified framework for different types of adversarial attack and defense methods in the context of medical image analysis. For a fair comparison, we establish a new benchmark for adversarially robust medical diagnosis models obtained by adversarial training under various scenarios. To the best of our knowledge, this is the first survey paper that provides a thorough evaluation of adversarially robust medical diagnosis models. By analyzing qualitative and quantitative results, we conclude this survey with a detailed discussion of current challenges for adversarial attack and defense in medical image analysis systems to shed light on future research directions. Code is available on \href{https://github.com/tomvii/Adv_MIA}{\color{red}{GitHub}}.
\end{abstract}



\received{15 April 2024}
\received[revised]{12 July 2024}
\received[accepted]{19 October 2024}

\maketitle

\section{Introduction}
\label{sec:1}
Driven by the success of Deep Neural Networks (DNNs) in natural image processing tasks \cite{he2016deep, wang2020deep}, they have also been demonstrated to have expert-level performance for various medical imaging tasks, including but not limited to skin lesion diagnosis \cite{ge2017skin}, diabetic retinopathy detection \cite{gondal2017weakly}, and tumor segmentation \cite{pereira2016brain}. Among these medical applications, Artificial Intelligence (AI)-based diabetic retinopathy detection system was the first that was approved for marketing by the US Food and Drug Administration \cite{FDA2018}. In clinical practice, deep learning-driven diagnosis models can save the overall cost of manual work and even improve patient outcomes by early detection \cite{alzubaidi2021role}.

Although deep learning has emerged as a promising technique for fundamental research in multiple disciplines \cite{dong2021visually, zemskova2022deep}, it still suffers from adversarial examples \cite{SzegedyZSBEGF13}, which can induce a catastrophic disruption to DNNs. Generally, the adversary can be obtained by adding a visually imperceptible perturbation to the legitimate example, which makes it easy to bypass the manual check \cite{carlini2017adversarial}. The existence of such tailored examples becomes one of the major hindrances to practically applying DNNs in safety-critical scenarios, \textit{e.g.} medical image analysis \cite{paschali2018generalizability_A1, finlayson2019adversarial_A3, ma2021understanding_D10}. In particular, the adversarial vulnerability can be even worse for diagnosis systems, which may result in misdiagnosis, insurance fraud, and even a crisis of confidence in AI-based medical technology \cite{shah2018susceptibility_A4, finlayson2019adversarial_A3}. Moreover, the diagnosis system is complex and unlikely to update, which can be difficult to imagine how these adversarial attacks could be operationalized. Recent works have revealed the adversarial vulnerability of diagnosis models under various scenarios. Consequently, adversarial robustness has been considered a new measurement for the security of medical applications.

Considering the huge healthcare economy and pervasive medical fraud, numerous studies have been conducted to defend against these malicious examples in medical imaging. Generally, early research efforts focus on adversarial training for improving network robustness \cite{ren2019brain_D11, vatian2019impact_D13, zhou2021ssmd_D41} or adversarial detection to identify adversarial examples \cite{ma2021understanding_D10, park2020robustification_D19, watson2021attack_D7}. Besides, a small fraction of works incorporate image-level pre-processing \cite{xu2022medrdf_D2, liu2020defending_D23} or feature enhancement \cite{taghanaki2019kernelized_D20, xue2019improving_D22} into the adversarial defense framework. The family of adversarial defense methods has been demonstrated to be effective in establishing adversarial robustness for unified pattern recognition in medical diagnosis \cite{li2021defending_D1, xu2021towards_D9}. However, there exists a considerable gap in the research-oriented setting and also evaluation between various defense methods, inducing difficulties in comparisons. This thus motivates us to construct a systematical survey, developing the benchmark implementation and evaluation for the most effective adversarial defense approach, \textit{a.k.a.} adversarial training.

\begin{figure}[t]
	\centering
	\includegraphics[width=0.6\linewidth]{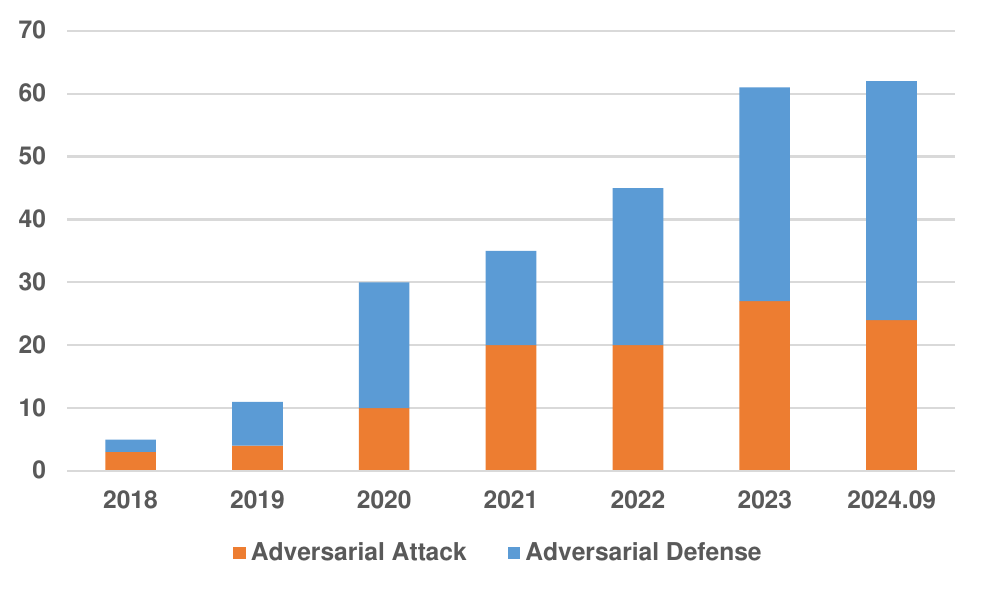}
	\vspace{-6mm}
	\caption{Number of publications per year related to adversarial attack and defense for medical image analysis, inclusive of data up to September 2024.}
	\vspace{-6mm}
	\label{fig:1}
\end{figure}

Several survey papers \cite{sipola2020model, apostolidis2021survey, shamshiri2022security, kaviani2022adversarial} have so far summarized the adversarial attack and defense for medical image analysis. However, many of them only focus on a certain medical task, e.g., Coronavirus 2019 (COVID-19) analysis \cite{shamshiri2022security}. Otherwise, these surveys do not provide a detailed taxonomy or a comprehensive evaluation of existing attack and defense methods for computer-aided diagnosis models. Furthermore, recent advances in adversarial attack and defense for medical image analysis systems have not been covered in these surveys. In this paper, we aim to address these gaps and provide a taxonomic overview of recent advances in adversarial attack and defense for medical image analysis with a discussion of their benefits and limitations. Although a considerable number of adversarial attack and defense methods emerge every year in the field of medical image analysis (see Fig. \ref{fig:1}), there still remains a lack of unified and fair measurement for various defense methods. Hence, we construct a benchmark evaluation of adversarial training methods for future development. The main contributions of this work lie in the following aspects:

\begin{itemize}
	\item We provide a comprehensive review of adversarial attack and defense methods in the field of medical image analysis, including a family of attack and defense methods with a systematic taxonomy based on the application scenario.
	\item We establish a unified framework for different types of adversarial attack and defense approaches with respect to diverse medical imaging tasks.
	\item To the best of our knowledge, this is the first survey that establishes a benchmark for adversarially robust diagnosis models under various scenarios. We further present a systematic analysis of adversarial attack and defense in the context of medical images.
	\item We identify current challenges and provide insightful guidance for future research in adversarial attack and defense for medical image analysis.
\end{itemize}

\begin{figure}[!t]
	\centering
	\includegraphics[width=0.93\linewidth]{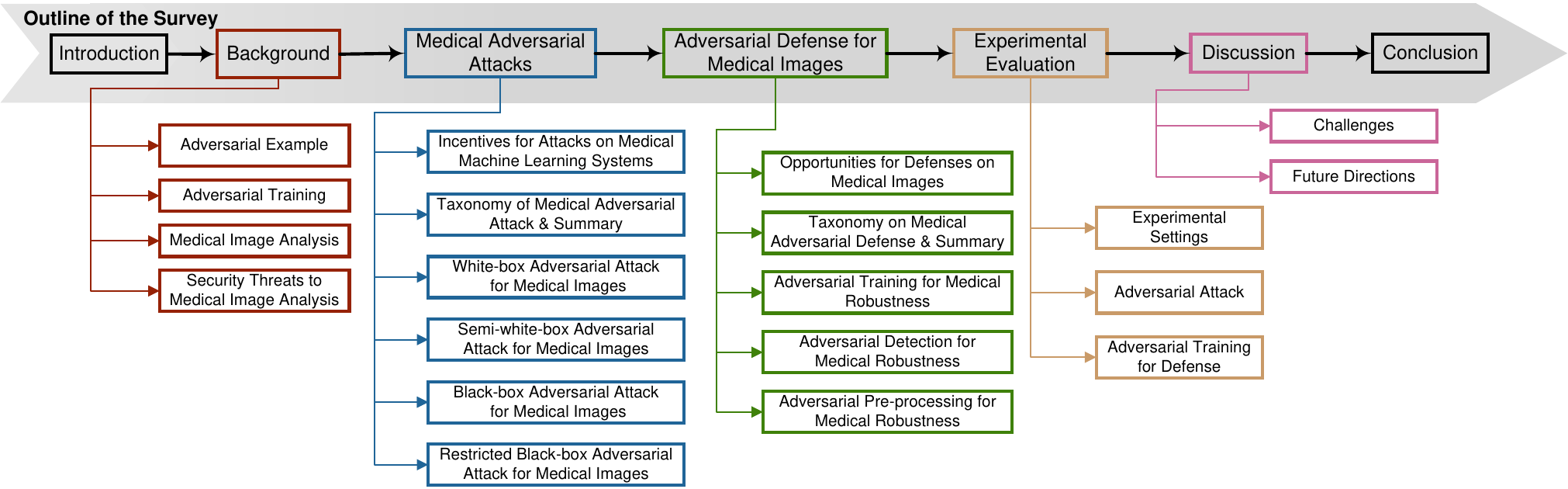}
	\vspace{-4mm}
	\caption{Outline of our Survey: Starting with the background of adversarial machine learning and medical image analysis, we comprehensively review recent advancements in medical adversarial attack and defense mechanisms. We also evaluate the robustness of medical diagnosis models against adversarial attacks and conclude with a discussion on current challenges, providing insights into potential research directions.}
	\vspace{-4mm}
	\label{fig:new-2}
\end{figure}

A pivotal motivation for focusing on adversarial learning within medical image analysis is the distinct nature of medical images compared to natural images, which necessitates specific modifications in both attack and defense mechanisms. This divergence is rooted in several unique characteristics of medical images: \textbf{1. Modality-Specific Characteristics}: Medical images, derived from varied modalities such as MRI, CT, X-ray, and ultrasound, possess unique imaging attributes including intensity distributions and channel configurations. These attributes demand tailored adversarial strategies that account for modality-specific nuances rather than general approaches used for natural images. \textbf{2. Limited and Imbalanced Datasets}: The scarcity and imbalance typical of medical datasets—stemming from privacy concerns, the rarity of certain conditions, and the high costs of data acquisition—pose distinct challenges. These challenges exacerbate the risk of overfitting and adversarial manipulation, necessitating robust and specialized defensive techniques. \textbf{3. Clinical Significance and Ethical Considerations}: The direct impact of medical image analysis on patient care prioritizes the need for accuracy and reliability. The severe consequences of adversarial attacks in this context not only heighten the stakes but also introduce profound ethical considerations, driving the need for exceptionally robust defense mechanisms. \textbf{4. Anatomical Structures and Domain-Specific Knowledge}: The consistent presence of specific anatomical structures across individuals introduces additional layers of complexity. Adversarial attacks can exploit these structures to induce subtle yet impactful misdiagnoses, while defenses must enhance their focus on clinically relevant regions to ensure diagnostic accuracy.

The remainder of this survey is structured as follows: We start with a unified framework to formally define both medical adversarial attack and defense in Section \ref{sec:2}. Next, we give a comprehensive review of adversarial attack techniques targeting medical machine learning systems with a taxonomy based on the attack scenarios (Section \ref{sec:3}). In Section \ref{sec:4}, we summarize and categorize existing adversarial defense methods in the context of medical image analysis. Then, we present benchmark evaluation results of medical attack and defense algorithms in various settings (Section \ref{sec:5}). Lastly, we discuss current challenges and future research directions in Section \ref{sec:6}. A tree diagram is also presented in Figure \ref{fig:new-2} to summarize our survey and aid reader comprehension.

\vspace{-3mm}
\section{Background}
\vspace{-0.6mm}
\label{sec:2}
Although the deep learning paradigm has made significant breakthroughs in numerous AI fields, it is still vulnerable to various security threats. Among them, adversarial attack has attracted the most attention from the community of deep learning security, as it raises a series of potential safety issues to the application of deep learning. Apart from disrupting the inference stage of DNNs, adversarial attacks can easily bypass the manual check, since the adversarial example is visually similar to its clean counterpart. This insidious security threat can further be magnified for computer-aided diagnosis models, which may result in catastrophic misdiagnosis and even a crisis of social confidence \cite{finlayson2019adversarial_A3}. In this section, we formally define adversarial attack and defense for medical image analysis and highlight the vulnerability of medical images to adversarial attacks.

We first clarify the notations and definitions used in this survey. Considering a specific dataset $(\mathbf{x}, y) \sim \mathcal{D}$ where $\mathcal{D}$ is a data distribution over pairs of given examples $\mathbf{x}$ and their corresponding labels $y$. We denote $f_{\boldsymbol{\theta}} (\cdot)$ as the deep learning-based medical analysis model with network parameter $\boldsymbol{\theta}$. Generally, adversarial examples $\mathbf{\hat{x}}$ are typically created by appending an imperceptible noise $\boldsymbol{\delta}$ to clean examples $\mathbf{x}$, which can be formally defined as below: 
\begin{equation}
	\begin{aligned}
		\mathbf{\hat{x}} := \mathbf{x} + \boldsymbol{\delta} \textrm{~~with~} f_{\boldsymbol{\theta}} (\mathbf{\hat{x}}) \neq y \textrm{~and~} d(\mathbf{x}, \mathbf{\hat{x}}) \leq \epsilon,
	\end{aligned}
	\label{eq:1}
\end{equation}
where $d(\cdot, \cdot)$ is the distance metric, and $\epsilon$ is the maximum allowed perturbation bound for imperceptibility. By definition, adversarial examples $\mathbf{\hat{x}}$ need to be close to their legitimate counterparts $\mathbf{x}$ under a certain distance metric, e.g., $\ell_{p}$ distance. In other words, the adversarial perturbation is $\ell_{p}$-norm bounded as $\left\| \boldsymbol{\delta} \right\|_{p} \leq \epsilon$. In this survey, we mainly focus on attacks under $\ell_{\infty}$-norm threat model, which can further be formulated as the following optimization problem:
\begin{equation}
	\begin{aligned}
		\max\limits_{\left\| \boldsymbol{\delta} \right\|_{\infty} \leq \epsilon}\mathcal{L}\left( f_{\boldsymbol{\theta}} \left( \mathbf{x}+\boldsymbol{\delta}\right) , y \right) ,
		\label{eq:2}
	\end{aligned}
\end{equation}
where $\mathcal{L}$ primarily depends on a certain task (such as the cross-entropy loss for classification). Normally, the above-mentioned optimization problem can be solved by quasi-Newton methods \cite{SzegedyZSBEGF13, carlini2017towards} or gradient descent-based algorithms. Specifically, we can obtain the worst-case example (or the strongest adversarial example) by iterative Projected Gradient Descent (PGD) \cite{MadryMSTV18} on the negative loss function with step size $\alpha$, as follows:
\begin{equation}
	\begin{aligned}
		\mathbf{\hat{x}}^{t+1} = \Pi_{\mathbb{B}(\mathbf{x}, \epsilon)} \left(  \mathbf{\hat{x}}^{t} + \alpha \cdot \operatorname{sign} \left( \nabla_{\mathbf{\hat{x}}^{t}}\mathcal{L}\left( \mathbf{\hat{x}}^{t}, y \right)  \right)  \right),
		\label{eq:3}
	\end{aligned}
\end{equation}
where $\mathbf{\hat{x}}^{t}$ represents the $t^\text{th}$ iteration update, and $\mathbb{B}(\mathbf{x}, \epsilon)$ denotes $\ell_{\infty}$-norm bound with radius $\epsilon$ around clean examples $\mathbf{x}$. Note that adversarial examples can be generated by numerous types of methods, including the Limited-memory BFGS method \cite{SzegedyZSBEGF13} and Fast Gradient Sign Method (FGSM) \cite{GoodfellowSS14}. Instead of the aforementioned white-box attack that can access the full knowledge of the target model, there also exist some other threat models, e.g., unrestricted attack \cite{chen2024content}, black-box attack \cite{bai2023query, ma2021finding}, which poses a more significant security issue to computer-aided diagnosis models in real-world scenario \cite{paschali2018generalizability_A1, bms2022analysis_A36}. Note that the aforementioned adversarial attack framework can be generalized to diverse tasks by merely altering the target loss function $\mathcal{L}$. We will discuss further details of all these threat models for medical image analysis in Section \ref{sec:3}.

Numerous types of defense methods have been proposed to enhance the robustness of diagnosis models against adversarial examples \cite{asgari2018vulnerability_D17, li2020robust_D24, tripathi2020fuzzy_D4}. Among them, adversarial training \cite{GoodfellowSS14, MadryMSTV18} has received the greatest attention, which can improve the robustness via augmenting adversaries as training data. The adversarially trained model is expected to correctly predict both clean and adversarial examples during the inference stage. Specifically, the standard adversarial training \cite{MadryMSTV18} can be extended based on Equation (\ref{eq:2}) as the following min-max optimization problem:
\begin{equation}
\small
	\begin{aligned}
		\min\limits_{\boldsymbol{\theta}} \mathbb{E}_{\left( \mathbf{x}, y\right)\sim \mathcal{D} }\left[ \max\limits_{\left\| \boldsymbol{\delta} \right\|_{\infty} \leq \epsilon}\mathcal{L}\left( f_{\boldsymbol{\theta}} \left( \mathbf{x}+\boldsymbol{\delta}\right) , y \right)  \right].
		\label{eq:4}
	\end{aligned}
\end{equation}
The inner maximization is to search for the worst-case adversarial examples to disrupt the target network. The outer minimization mainly focuses on optimizing empirical adversarial risk over network parameter $\theta$. Generally, adversarial training improves the intrinsic robustness of DNNs while preserving their performance on clean inputs during the inference stage, which can be improved by integrating additional regularization modules \cite{terzi2021adversarial, zhou2022improving, xu2022infoat, yu2022understanding, zhou2022modeling}.

Besides enhancing the robustness against adversaries, several defense methods focus on data pre-processing (both clean and adversarial examples) without affecting the subsequent diagnosis systems \cite{xu2022medrdf_D2, kansal2022defending_D51}. Generally, data pre-processing aims at transforming adversaries into their benign versions for the subsequent inference while making clean inputs remain unchanged, achieving superior generalization ability against unforeseen adversaries \cite{yoon2021adversarial, zhou2021removing, dai2022deep, zhou2021towards, zhou2023eliminating, lee2023robust, sun2023critical}. We can thus formulate a pre-processing-based defense as the following optimization:
\begin{equation}
	\small
	\begin{aligned}
		\min\limits_{\psi} \mathbb{E}_{\left( \mathbf{x}, y\right)\sim \mathcal{D} }\left[ \mathcal{L}\left( f_{\boldsymbol{\theta}} \left( \psi\left( \mathbf{x}+\boldsymbol{\delta}\right) \right) , y \right) + \lambda \cdot \mathcal{L}\left( f_{\boldsymbol{\theta}} \left( \psi\left( \mathbf{x}\right) \right) , y \right) \right].
		\label{eq:5}
	\end{aligned}
\end{equation}
where $\lambda$ is the weighting factor, and $\psi$ denotes the pre-processing module, which can be a parametric or non-parametric operator to mitigate the effect of adversarial perturbations. However, discriminative regions in medical images usually occupy only a few pixels. Thus, the pre-processing methods in the context of medical images still suffer from a higher risk of discriminative feature loss than that of natural images.

Different from medical adversarial defense methods that aim at inferencing adversaries correctly, some works focus on the detection of input adversarial examples \cite{watson2021attack_D7, li2020robust_D24, yang2022defense_D67}. Adversarial detection can thus be regarded as a binary classification task to distinguish legitimate and adversarial examples. The main objective of adversarial detection is formalized as the following optimization: 
\begin{equation}
	\small
	\begin{aligned}
		\min\limits_{\boldsymbol{\omega}} \mathbb{E}_{\left( \mathbf{x}, y\right)\sim \mathcal{D} }\left[ \mathcal{L}_\text{CE}\left( f_{\boldsymbol{\omega}} \left(  \mathbf{x}+\boldsymbol{\delta} \right) , 1 \right) + \beta \cdot \mathcal{L}_\text{CE}\left( f_{\boldsymbol{\omega}} \left(  \mathbf{x}\right) , 0 \right) \right].
		\label{eq:6}
	\end{aligned}
\end{equation}
where $f_{\boldsymbol{\omega}}$ is the binary classifier with parameter $\boldsymbol{\omega}$ for adversarial detection, and $\mathcal{L}_\text{CE}$ represents the cross-entropy loss. $\beta$ is the weighting factor to determine the focus of detection models. The adversarial detector is desired to correctly distinguish legitimate examples (0) and their adversarial counterparts (1). Note that adversarial detection can be either parametric (learnable) or non-parametric. Other than adversarial training and pre-processing-based defense, adversarial detection mainly focuses on detecting adversarial examples in advance but not robust inference. Note that adversarial detection is effective for all medical imaging tasks by establishing an input-level defense. Similarly, adversarial training can also be generalized to different medical imaging tasks by rebuilding its target loss function. We will further discuss the details of these adversarial defense methods and some other types of approaches in Section \ref{sec:4}.

Numerous adversarial attack and defense methods have been demonstrated to yield excellent performance for natural images \cite{aldahdooh2022adversarial}. However, there exist fundamental differences between natural vision tasks and medical imaging tasks in terms of data sizes, features, and task patterns. Therefore, it is difficult to transfer adversarial attack and defense methods for natural images directly to the medical domain. In addition, several works have shown that medical images can even suffer from more severe adversarial attacks than natural images \cite{finlayson2019adversarial_A3, ma2021understanding_D10, yao2021hierarchical_A15, rasaee2021explainable_A44}. Considering the massive healthcare industry and the significant impact of computer-aided diagnosis, it is necessary to pay close attention to the security and reliability of computer-aided diagnosis models. Furthermore, we emphasize that medical diagnosis demands high robustness and explainability, which are crucial for maintaining social trust in medical AI applications. Mistakes induced by adversarial attacks can lead to serious consequences, including patient harm and erosion of confidence in automated systems. Hence, we summarize this survey to highlight the significance of focusing on adversarial attacks and defenses for medical image analysis, a domain where the stakes are exceptionally high.

Substantial efforts have also been made to understand and explain the existence of adversarial examples \cite{GoodfellowSS14, ilyas2019adversarial, ZhangGL000S022}. Departing from the initial discovery of adversarial vulnerabilities in deep learning-based models \cite{SzegedyZSBEGF13}, Goodfellow \textit{et al.} \cite{GoodfellowSS14} proposed that the linear nature of neural networks, despite their inherent non-linearities, contributes significantly to their susceptibility to adversarial perturbations. This perspective suggests that small, carefully crafted perturbations can lead to significant changes in the output due to linear behavior in high-dimensional spaces. Building upon this, Ilyas \textit{et al.} \cite{ilyas2019adversarial} introduced the notion that adversarial examples are not merely bugs but arise from non-robust yet useful features that models exploit from the data. They further showed that adversarial perturbations, while disruptive, are a result of the model's capacity to leverage any available signal to maximize performance on the training data, suggesting that these features are not bugs but intrinsic to the dataset. Further advancing the theoretical understanding, Zhang \textit{et al.} \cite{ZhangGL000S022} made the first step in a causal perspective for understanding and mitigating adversarial examples. They show the significance of the correlation between style variables and ground-truth labels by constructing the causal graph based on adversary generation.

In the context of medical image analysis, efforts to explain adversaries have also been undertaken. Studies such as \cite{bortsova2021adversarial_A18, ma2021understanding_D10, watson2021attack_D7} have empirically explored various factors that contribute to the generation of adversarial examples against deep diagnostic models. These works delve into the specific vulnerabilities of medical imaging systems, considering factors like data distribution, model architecture, and the unique characteristics of medical images that may influence diagnosis models.

\vspace{-3mm}
\section{Medical Adversarial Attacks}
\label{sec:3}
\textbf{Medical adversarial attacks} are designed to deliberately generate adversarial examples that compromise the integrity of medical diagnosis models during the inference stage. These attacks pose significant challenges to the reliability and safety of computer-aided diagnosis systems, which are increasingly reliant on deep learning-based models. The primary incentive for conducting these attacks could be to assess the robustness of these diagnosis systems or, in a more adversarial scenario, to manipulate medical outcomes. We further establish a comprehensive taxonomy of adversarial attacks tailored to medical image analysis, guiding readers through the various attack modalities that mirror real-world threats. We categorize these into four main types based on the attacker's knowledge and access levels. Each category involves distinct methodologies and presents unique challenges in terms of detection and mitigation. By analyzing these categories modularly, we can better understand how adversarial attacks can be tailored to specific medical image processing tasks and scenarios, providing a clearer picture of both the vulnerabilities exposed by these attacks and the potential defense mechanisms in the context of medical image analysis.


\begin{figure}[t]
	\centering
	\includegraphics[width=0.6\linewidth]{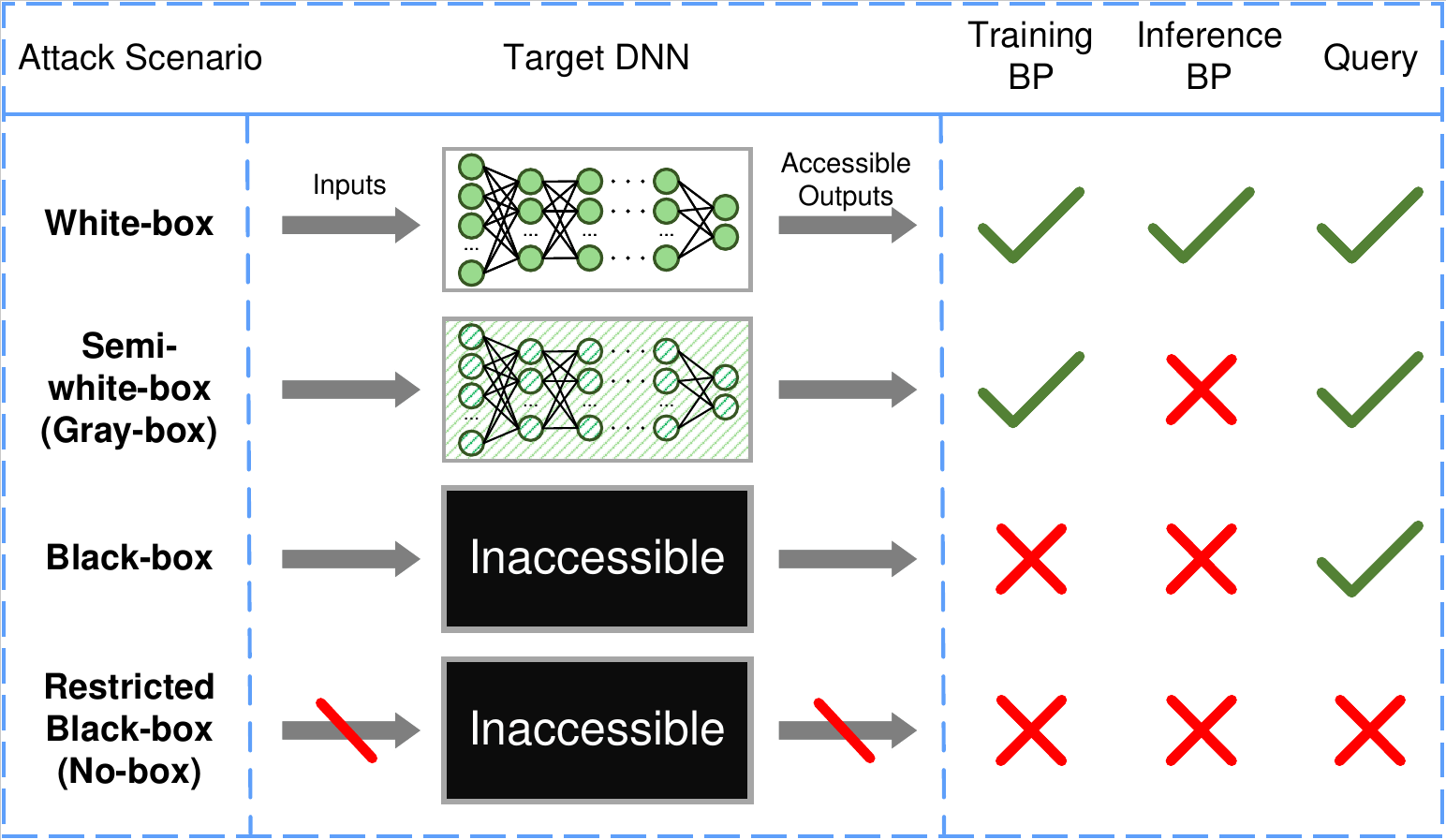}
	\vspace{-3mm}
	\caption{Taxonomy of medical adversarial attacks in terms of application scenarios. Following \cite{chen2017zoo, dong2022restricted}, we categorize adversarial attack methods into four classes according to the degrees of accessible knowledge, including Backward Propagation (BP) gradients of target DNN during the training and inference stage of the adversary generator. ``Query'' represents the accessibility to outputs of the target DNN.}
	\vspace{-4mm}
	\label{fig:2}
\end{figure}

\vspace{-3mm}
\subsection{Taxonomy of Adversarial Attacks}
In the past few years, various methods have been proposed to generate adversarial examples against computer-aided diagnosis models. According to several studies on the motivations for conducting medical adversarial attacks \cite{finlayson2019adversarial_A3, levy2022security_A27}, we consider the financial gain can be the biggest incentive to attack healthcare systems, as health insurance frauds have grown on a large scale over the years \cite{villegas2021fourteen}. The attacker might try to deliberately cause misdiagnosis for AI-based medical reimbursement systems to make an erroneous decision. In addition, the adversarial vulnerability of the computer-aided diagnosis mechanism can even be involved in terrorism and unfair competition in the future. Most importantly, a growing number of attacks against medical image diagnosis will induce a crisis of confidence in the autonomous AI diagnostic system. Existing attack methods concentrate on different vulnerabilities of DNNs and also diverse medical imaging tasks. More importantly, they are designed in terms of application scenarios (degrees of accessible knowledge), as shown in Fig. \ref{fig:2}. Consistent with the taxonomy of adversarial attacks for natural images in a general setting \cite{chen2017zoo, pitropakis2019taxonomy, dong2022restricted}, we categorize all medical attack methods into four classes (order by accessible knowledge to target model in decreasing order): white-box attack, semi-white-box (gray-box) attack, black-box attack, and restricted black-box (no-box) attack.  

In addition to the taxonomy of adversarial attacks for diverse model-accessible scenarios, we here also delineate an alternative taxonomy based on the adversary generation (optimization) strategies for adversarial attack. We now categorize adversary generation strategies into four main classes: 

\noindent\textbf{(1) Gradient-based optimization.}
In a white-box setting, this approach leverages gradient information to derive corresponding adversarial perturbations/transformations. Previous studies in medical adversarial attacks have predominantly employed (iterative) gradient ascent to maximize objective functions \cite{finlayson2019adversarial_A3, liu2019robustifying_A31, morshuis2022adversarial_A26}. Such a gradient-based optimization technique can further be extended to a universal (image-agnostic) setting, which means that a set of examples are possible to be disrupted by a universal adversarial perturbation \cite{cheng2020adversarial_A2, hirano2021universal_A7_D55, minagi2022natural_A19}.

\noindent\textbf{(2) Zeroth-order optimization.}
The gradient information is generally inaccessible in the black-box setting, while the only accessible information is the corresponding outputs of the target model. Thus, a possible way is to approximate the gradient of the target model via a classic type of derivative-free optimization method\textemdash the zeroth-order method using differences of function values. This was initially applied to the black attacks against natural images \cite{chen2017zoo}, and then extended to medical adversarial attacks in the black-box setting \cite{byra2020adversarial_A11, li2022query_A50}.

\noindent\textbf{(3) Evolutionary optimization.}
Evolutionary optimization typically utilizes bio-inspired algorithms to efficiently solve optimization problems where the underlying functions are not explicitly known or are too complex to model directly \cite{nguyen2012evolutionary}. Such an optimization strategy is thus applied to select the optimal perturbation of one pixel in the context of natural images \cite{su2019one} and medical images \cite{tsai2023adversarial_A2023_3} under the white-box scenario. Specifically in the context of black-box medical adversarial attacks, Cui \textit{et al.} \cite{cui2021deattack_A48} further designed a differential evolution attack against medical image segmentation models in the black-box setting. 

\noindent\textbf{(4) Deep generative models.}
In addition to adversary generation through optimization, a series of methods have been proposed to conduct one-stage adversary generation via deep generative models for natural images \cite{xiao2018generating, jandial2019advgan++}. By learning the medical-specific domain knowledge, such generative models can also be used to produce visually undetectable adversarial perturbations corresponding to the original medical images \cite{allyn2020adversarial_A6, wang2021adversarial_A37}. The generated adversaries can obtain a terribly high attack success rate on diverse computer-aided diagnosis tasks.

Under the white-box scenario, the attacker has full information about the target model, which can also access inference outputs of the target DNN unlimited times. Nonetheless, the semi-white-box (gray-box) setting \cite{xiao2018generating} mainly concentrates on constructing a generative model to produce adversaries against the target DNN. Generally, the gray-box attackers train the adversary generator in the white-box setting, while they do not require the accessibility of the target DNN during the inference (adversary generation) stage. The standard black-box adversarial attack can only access the DNN output (probabilities) using a limited number of queries. In contrast, the restricted black-box (no-box) attack does not need even a single-time query to the inaccessible DNN. 

Both white-box and gray-box settings are ideal for the medical adversarial attack, which rely heavily on the accessibility of prior knowledge about the target DNN. Specifically, the attackers focus on the backward gradients to update adversaries. Black-box attacks concentrate on zeroth-order optimization according to numerous times of query outputs of the target DNN. In comparison, no-box attacks highly depend on the transferability of adversaries among different models. Note that Black-box and no-box scenarios are close to computer-aided diagnosis systems in practice, which poses a more significant threat to model deployment. In the following sections, we will survey recent advances in these four types of adversarial attacks for medical image analysis.

\vspace{-3mm}
\subsection{Summary of Medical Attack Works}
\label{sec:3_Sum}
To facilitate future research activities, we present an overview of adversarial attack works for medical image analysis with a detailed taxonomy in Table \ref{tab:1}. Particularly, we include attack methods designed for natural images and medical images (in \textbf{bold}). We list adversarial attack methods tailored for natural images that are extended to these medical works, including L-BFGS \cite{SzegedyZSBEGF13}, CW attack \cite{carlini2017towards}, DeepFool \cite{moosavi2016deepfool}, Universal Adversarial Perturbation (UAP) \cite{moosavi2017universal}, Jacobian Saliency Map Attack (JSMA)\cite{papernot2016limitations}, Zeroth Order Optimization (ZOO) \cite{chen2017zoo}, One-Pixel Attack (OPA) \cite{su2019one}, etc.

\begin{table*}[t!]
	\centering
	\tiny 
	\renewcommand{\arraystretch}{0.7}
	\caption{Summary of adversarial attack works in the context of medical image analysis (time ascending). The newly proposed adversarial attack methods tailored for medical images are in \textbf{bold}.}
	\vspace{-4mm}
	\label{tab:1}
	\rowcolors{2}{gray!0}{gray!8}
	\resizebox{\linewidth}{!}{  
		
		\begin{tabular}{|c|c|c|c|c|c|}
			\hline
			Reference & Year & Task &  Attack scenario  &  Method  &  Data Modality\\ 
			\hline
			
			\cite{paschali2018generalizability_A1} & 2018 & Classification, Segmentation &  White-box, No-box  & FGSM, DeepFool, JSMA & MRI, Dermoscopy \\ 
			
			\cite{finlayson2019adversarial_A3} & 2018 & Classification &  White-box, No-box  & PGD, AdvPatch &\begin{tabular}[c]{@{}c@{}}Fundoscopy \\ X-ray, Dermoscopy \end{tabular}  \\ 
			
			\cite{shah2018susceptibility_A4} & 2018 & Classification &  No-box  & FGSM & Fundoscopy \\ 
			
			\cite{liu2019robustifying_A31} & 2019 & Segmentation &  White-box  & FGSM, I-FGSM, TI-FGSM & MRI \\ 
			
			\cite{chen2019intelligent_A13} & 2019 & Segmentation &  White-box  & \textbf{Multi-task VAE} & CT \\ 
			
			\cite{ozbulak2019impact_A12} & 2019 & Segmentation &  White-box  & \textbf{Adaptive Segmentation Mask Attack} & Fundoscopy, Dermoscopy \\
			
			\cite{kovalev2019influence_A5} & 2019 & Classification &  White-box  & PGD & X-ray, Histology \\
			
			\cite{rao2020thorough_A45_D60} & 2020 & Classification &  White-box, No-box  & \begin{tabular}[c]{@{}c@{}}FGSM, PGD, MI-FGSM, \\ DAA, DII-FGSM \end{tabular} & X-ray  \\
			
			\cite{cheng2020adversarial_A43} & 2020 & Classification &  White-box, No-box  & \textbf{Adversarial Exposure Attack} & Fundoscopy  \\
			
			\cite{rahman2020adversarial_A22} & 2020 & Classification, Object detection &  White-box, Gray-box, No-box  & \begin{tabular}[c]{@{}c@{}}FGSM, DeepFool, CW,  \\ BIM, L-BFGS, PGD, JSMA \end{tabular}  & CT, X-ray \\
			
			\cite{byra2020adversarial_A11} & 2020 & Classification &  Black-box  & ZOO & Ultrasound \\ 
			
			\cite{yao2020miss_A34} & 2020 & Landmark detection &  White-box  & \textbf{Adaptive Targeted I-FGSM} & MRI, X-ray  \\
			
			\cite{yoo2020outcomes_A32} & 2020 & Classification &  White-box, No-box  & FGSM & Fundoscopy \\ 
			
			\cite{cheng2020adversarial_A2} & 2020 & Segmentation &  White-box  & UAP & MRI \\ 
			
			\cite{allyn2020adversarial_A6} & 2020 & Classification &  White-box  & Generative model & Dermoscopy \\ 
			
			\cite{gongye2020new_A24} & 2020 & Classification &  White-box  & FGSM, PGD & X-ray \\ 
			
			\cite{hirano2021universal_A7_D55} & 2021 & Classification &  White-box  & UAP & OCT, X-ray, Dermoscopy \\ 
			
			\cite{joel2021adversarial_A33} & 2021 & Classification &  White-box  & FGSM, BIM, PGD & CT, MRI, X-ray  \\
			
			\cite{chen2021adversarial_A30} & 2021 & Segmentation &  White-box  & \textbf{IND and OOD Attacks} & MRI \\ 
			
			\cite{QiGS0Z21_A17} & 2021 & Classification, Segmentation &  White-box  & \textbf{Stabilized Medical Attack} & CT, Endoscopy, Fundoscopy \\ 
			
			\cite{bortsova2021adversarial_A9} & 2021 & Segmentation &  White-box, No-box  & PGD & X-ray \\
			
			\cite{kovalev2021biomedical_A51} & 2021 & Classification &  White-box, No-box  & CW & CT, X-ray, Microscopy  \\  
			
			\cite{pal2021vulnerability_A8} & 2021 & Classification &  White-box  & FGSM & CT, X-ray \\ 
			
			\cite{shao2021target_A16} & 2021 & Segmentation &  White-box  & \textbf{Multi-scale Attack} & Fundoscopy, Dermoscopy \\ 
			
			\cite{tian2021bias_A14} & 2021 & Classification &  White-box, No-box  & \textbf{Adversarial Bias Field Attack} & X-ray \\ 
			
			\cite{foote2021now_A39} & 2021 & Classification &  White-box  & PGD, UAP & Microscopy  \\ 
			
			\cite{yilmaz2021assessment_A10} & 2021 & Classification &  White-box  & FGSM & X-ray \\ 
			
			\cite{kulkarni2021kryptonite_A28} & 2021 & Classification &  White-box  & \textbf{Kryptonite Attack} & MRI, Dermoscopy \\ 
			
			\cite{gougeh2021adversarial_A23} & 2021 & Classification &  White-box  & FGSM, PGD, CW, ST & X-ray \\
			
			\cite{koga2021simple_A29} & 2021 & Classification &  Black-box  & \textbf{Black-box UAP} & X-ray, Dermoscopy, Fundoscopy \\ 
			
			\cite{rasaee2021explainable_A44} & 2021 & Classification &  White-box  & I-FGSM & Ultrasound  \\ 
			
			\cite{yao2021hierarchical_A15} & 2021 & Classification &  White-box  & \textbf{Hierarchical Feature Constraint} & X-ray, Fundoscopy \\ 
			
			\cite{bortsova2021adversarial_A18} & 2021 & Classification &  No-box  & FGSM, PGD & X-ray, Fundoscopy,  Microscopy \\
			
			\cite{diyasa2021grasping_A38} & 2021 & Classification &  White-box  & \begin{tabular}[c]{@{}c@{}}FGSM, BIM, CW, RFGSM \\ PGD, FAB, DeepFool, SparseFool \end{tabular} & Microscopy  \\
			
			\cite{cui2021deattack_A48} & 2021 & Segmentation &  Black-box  & \textbf{Differential Evolution Attack} & CT, Dermoscopy, Ultrasound  \\ 
			
			\cite{zhou2021machine_A46} & 2021 & Classification &  Gray-box  & \textbf{GAN-based Adversary Generator} & X-ray  \\
			
			\cite{wang2021adversarial_A37} & 2022 & Classification &  Black-box  & \textbf{AmdGAN} & \begin{tabular}[c]{@{}c@{}}CT, OCT, X-ray, Fundoscopy,  \\ Dermoscopy, Ultrasound, Microscopy \end{tabular}    \\
			
			\cite{levy2022security_A27} & 2022 & Classification, Segmentation &  White-box  & \begin{tabular}[c]{@{}c@{}} \textbf{Modified FGSM with} \\ \textbf{with tricks to break defences} \end{tabular} & \begin{tabular}[c]{@{}c@{}}CT, MRI, X-ray,  \\ Dermoscopy, Fundoscopy \end{tabular}  \\ 
			
			\cite{minagi2022natural_A19} & 2022 & Classification &  White-box  & UAP & X-ray, Fundoscopy, Dermoscopy \\
			
			\cite{patel2022predictive_A49} & 2022 & Classification &  White-box, No-box  & \textbf{Attention-based I-FGSM} & CT  \\ 
			
			\cite{kwon2022advu_A35} & 2022 & Segmentation &  White-box  & FGSM & Microscopy  \\
			
			\cite{de2022evaluation_A40} & 2022 & Classification &  White-box  & FGSM & X-ray  \\ 
			
			\cite{apostolidis2022digital_A21} & 2022 & Classification &  No-box  & \textbf{Digital Watermarking} & CT, MRI, X-ray \\
			
			\cite{wei2022analysis_A42} & 2022 & Classification &  White-box, No-box  & \begin{tabular}[c]{@{}c@{}}FGSM, BIM, PGD \\ \textbf{No-sign Operation} \end{tabular} & X-ray  \\
			
			\cite{selvakkumar2022addressing_A25} & 2022 & Classification &  White-box  & FGSM & Fundoscopy \\
			
			\cite{ahmed2022failure_A47} & 2022 & Classification &  White-box  & FGSM, PGD, CW & CT, Dermoscopy, Microscopy  \\
			
			\cite{li2022query_A50} & 2022 & Segmentation &  Black-box  & \textbf{Improved Adaptive Square Attack} & X-ray  \\ 
			
			\cite{bms2022analysis_A36} & 2022 & Classification &  Black-box  & FGSM, BIM, PGD, MI-FGSM & CT, Fundoscopy  \\ 
			
			\cite{morshuis2022adversarial_A26} & 2022 & Reconstruction &  White-box  & \begin{tabular}[c]{@{}c@{}}\textbf{Adversarial k-space Noise},  \\ \textbf{Adversarial Rotation} \end{tabular} & MRI \\ 
			
			\cite{bharath2022analysis_A41} & 2022 & Classification &  White-box  & FGSM, L-BFGS & Fundoscopy  \\ 
			
			\cite{wang2022feature_A20} & 2022 & Classification &  Gray-box  & \textbf{Feature Space-Restricted Attention Attack} & X-ray, Fundoscopy, Dermoscopy \\ 
			
			\cite{yao2023adversarial_A2023_1} & 2023 & Classification & White-box & \textbf{Hierarchical Feature Constraint} & X-ray, Fundoscopy \\
			
			\cite{ding2023vith_A2023_2} & 2023 & Classification & White-box, No-box & \textbf{Vision Transformer Hashing} & X-ray, Dermoscopy \\
			
			\cite{tsai2023adversarial_A2023_3} & 2023 & Classification & White-box & OPA, \textbf{Differential Evolution Attack} & OCT, X-ray, Dermoscopy \\
			\cite{chen2023frequency_A2023_4} & 2023 & Classification & White-box, No-box & \textbf{Frequency Constraint-based Attack} & X-ray, Ultrasound \\
			\cite{li2023threat_A2023_5} & 2023 & Classification & White-box & FGSM & CT \\
			\cite{lee2024adversarial_A2024_1} & 2024 & Classification & White-box & \textbf{Dynamic Adaptive Instance Normalization} & CT \\
			\cite{chen2024rae_A2024_2} & 2024 & \makecell{Classification, Segmentation, \\ Object detection} & No-box & \textbf{Visible Watermark Perturbation} & \makecell{MRI, X-ray, Fundoscopy, \\ Dermoscopy, Microscopy} \\
			\hline
		\end{tabular}
	}
	\vspace{-6mm}
\end{table*}

\vspace{-3mm}
\subsection{White-box Attacks}
Most works of adversarial attack for medical image analysis focus on the white-box scenario. In particular, these works primarily consider the adversarial vulnerability of computer-aided diagnosis models in various medical imaging tasks with full knowledge of medical DNNs. Specifically, the attacker can treat the target diagnosis DNN as a locally deployed model during adversary generation.

To the best of our knowledge, Paschali \textit{et al.} \cite{paschali2018generalizability_A1} were the first to systematically evaluate the white-box attack against a variety of medical imaging models of several medical tasks, including skin lesion classification and whole brain segmentation. Concretely speaking, they transfer adversarial attack methods for natural images \cite{GoodfellowSS14, moosavi2016deepfool, papernot2016limitations} directly to medical images, which poses an underlying threat to modern diagnosis DNN models. In the meantime, the attack success rate can be further reduced when attackers are in the face of deeper diagnosis models.

Generally, white-box adversarial attacks are conducted by solving a certain optimization problem. Nevertheless, Allyn \textit{et al.} \cite{allyn2020adversarial_A6} leveraged a generative model to produce visually undetectable adversarial perturbations corresponding to original input images. The generated adversarial examples can obtain a terribly high attack success (fooling) rate to dermoscopic image recognition systems, presenting a significant risk of misdiagnosis in real-world scenarios.

In addition to generating the adversarial perturbation corresponding to a certain medical image, several works \cite{hirano2021universal_A7_D55, foote2021now_A39, minagi2022natural_A19} consider obtaining a universal perturbation that is adaptive to a large proportion of medical images from a certain dataset, causing a security hole in the computer-aided clinical diagnosis. Hirano \textit{et al.} \cite{hirano2021universal_A7_D55} conducted a universal adversarial attack for medical image classification under both targeted and non-targeted scenarios. Note that the main goal of the targeted attack is to enable adversaries to be misclassified as a specified target class. Oppositely, non-targeted attacks just focus on making DNN models output wrong results. They discovered that non-targeted adversarial attacks can achieve better transferability than targeted attacks. 

Numerous researchers spare no effort to design custom adversarial attack methods \cite{tian2021bias_A14, yao2021hierarchical_A15, QiGS0Z21_A17, kulkarni2021kryptonite_A28, yao2023adversarial_A2023_1, lee2024adversarial_A2024_1} to better adapt to medical imaging tasks. By theoretically investigating the vulnerability of medical image representations, Yao \cite{yao2021hierarchical_A15} designed a novel hierarchical feature constraint as auxiliary guidance to hide the adversarial representations in the clean feature domain, which can further be a plug-and-play module for existing attack methods to reduce the risk of being detected. Considering various modalities of medical images, Qi \textit{et al.} \cite{QiGS0Z21_A17} proposed a new medical adversary generation method by optimizing a well-defined objective function that is composed of a deviation loss term and a stabilization loss term. Specifically, the deviation aims at enlarging the prediction gap between adversarial outputs and their corresponding ground truth. The stabilization term can be regarded as a regularization to constrain adversarial perturbations to low variance, which avoids the local optima induced by instance-wise noise during optimization. 

Apart from white-box adversarial attacks for medical classification tasks, there also exist various researchers that explore the vulnerabilities of other medical imaging tasks. The majority of these works concentrate on medical segmentation \cite{chen2019intelligent_A13, ozbulak2019impact_A12, cheng2020adversarial_A2, bortsova2021adversarial_A9, shao2021target_A16, chen2021adversarial_A30, kwon2022advu_A35}. Ozbulak \textit{et al.} \cite{ozbulak2019impact_A12} started the first attempt to expose the adversarial vulnerability of medical segmentation tasks by designing the Adaptive Segmentation Mask Attack (ASMA) method. The proposed ASMA incorporates the adaptive segmentation mask and the dynamic perturbation multiplier to generate targeted adversaries. Beyond the superior disruption against various DNNs, the authors also demonstrated the generalizability of the proposed method with different distance metrics for adversarial perturbations. Based on feature-level ASMA attack \cite{ozbulak2019impact_A12}, Shao \textit{et al.} \cite{shao2021target_A16} further incorporated multi-scale gradients to generate adversarial perturbations to biomedical image segmentation models. The above-mentioned medical segmentation attacks mainly depend on in-distribution adversarial examples. In contrast, the proposed method in \cite{chen2021adversarial_A30} considers out-of-distribution adversarial attacks against the lumbar disk shape reconstruction problem. Specifically, PGD \cite{MadryMSTV18} is applied to optimize the out-of-distribution perturbation to bypass defense methods.

In addition, Yao \textit{et al.} \cite{yao2020miss_A34} proposed an Adaptive Targeted Iterative FGSM (ATI-FGSM) with a comprehensive evaluation to study the vulnerability of multiple landmark detection systems. ATI-FGSM aims at moving a cohort of landmarks by dynamically assigning a weight for the loss term of a specific landmark in each iteration, facilitating the convergence of adversary generation. In addition, Rahman et al. \cite{rahman2020adversarial_A22} examined the adversarial robustness of nine deep diagnosis applications for COVID-19 detection, which demonstrates their extreme susceptibility to adversarial attacks.

Apart from adversarial attacks against medical pattern recognition systems, a recent work \cite{morshuis2022adversarial_A26} investigated the adversarial vulnerability of Magnetic Resonance (MR) image reconstruction from k-space data. In addition to appending adversarial perturbation directly to k-space measurements, the authors also consider a visually slight rotation during the acquisition to obtain the adversarial effect, which can also be optimized by the iterative PGD approach.

\vspace{-4mm}
\subsection{Semi-white-box Attacks}
\vspace{-1mm}
Semi-white-box (Gray-box) attacks have been widely explored for natural images \cite{xiao2018generating, wang2019gan, jandial2019advgan++}. However, there remain rare academic works to explore this attack scenario for medical image analysis \cite{rahman2020adversarial_A22, zhou2021machine_A46, wang2022feature_A20}. Generally, the semi-white-box adversarial attack involves two stages: 1) the attacker trains a generative model to produce adversarial examples against the target DNN model. During the training process, the attacker has full access to the target model, including gradients of backward propagation. 2) During the application stage, the adversary generator can directly obtain adversarial examples against the target model with the input of legitimate images, which do not requires any information about the target model as in the totally black-box scenario.

Generative Adversarial Networks (GANs) have also been shown to have superior performance in generating highly plausible adversarial images \cite{xiao2018generating, wang2019gan, jandial2019advgan++, dong2022restricted}. To investigate the safety issues for AI-based computer-aided diagnosis of breast cancer on digital mammograms, Zhou \textit{et al.} \cite{zhou2021machine_A46} developed two GAN-based models at two different resolutions, which can synthesize highly plausible adversarial images to further induce a wrong diagnosis of breast cancer. The authors also pointed out another potential problem that GAN-based adversarial attack methods can also be trained using external medical data, as mammography is the widely available imaging modality in clinical practice. Recently, the  Feature-Space-Restricted Attention Attack \cite{wang2022feature_A20} is proposed to efficiently generate adversarial examples for various medical modalities with less visual perturbation. Specifically, the authors added the feature-level restriction to make sure that adversarial examples are close to the classification boundary in the feature space. Furthermore, an attention mechanism constraint was also proposed to regularize the image-level perturbations to focus on the lesion area, which incorporates class-wise attention information into the adversarial perturbation generation.

\subsection{Black-box Attacks}

Existing white-box adversarial attack mainly requires multiple backward gradients of target models. In other words, the attacker treats the target DNN as the locally deployed model to generate corresponding adversarial examples. However, the white-box setting can be unreliable in the real-world scenario, as it needs complete knowledge of the DNN model to attack. In comparison, the general black-box scenario can be a more suitable setting to simulate practical adversarial attacks. Numerous works have been proposed to explore black-box attacks for natural images \cite{chen2017zoo, ilyas2019prior, andriushchenko2020square, yatsura2021meta}. On account of the inaccessibility to backward gradients (or first-order information) of the target DNN, black-box adversarial attacks primarily resort to the zeroth order optimization for the gradient estimation, which requires querying the output probabilities of the target DNN with numerous times. There also exist several studies that focus on the black-box adversarial attack in the context of medical image analysis \cite{byra2020adversarial_A11, wang2021adversarial_A37, bms2022analysis_A36, cui2021deattack_A48, li2022query_A50}.

To explore the practical security threat against medical image analysis tasks, Byra \textit{et al.} \cite{byra2020adversarial_A11} devised a black-box adversarial attack approach that is specific to ultrasound image reconstruction. In comparison to existing attack methods that manipulate pixels of medical images, the proposed adversarial perturbation mainly concentrates on the reconstruction parameters. Moreover, the parameter-level attack is conducted for each radio-frequency data frame separately based on zeroth-order optimization \cite{chen2017zoo}. On account of the difficulty in accessing the target diagnosis model during the black-box scenario, Wang \textit{et al.} \cite{wang2021adversarial_A37} employed the knowledge distillation method to learn a GAN-based model to generate adversarial examples. Specifically, the distillation model aims at simulating the inference outputs of target black-box DNN with multiple queries. Subsequently, the proposed generator can efficiently produce adversarial examples based on any clean inputs against the targeted diagnosis model.

Considering the scarcity of adversarial attacks against medical image segmentation, Cui \textit{et al.} \cite{cui2021deattack_A48} proposed a Differential Evolution Attack (DEAttack) against medical image segmentation models in the black-box setting with only tampering a few pixels of medical images. Moreover, the DEAttack remains a much higher efficiency in creating adversarial examples than directly applying the differential evolution algorithm via incorporating pre-selected sensitive regions and colors from original medical images. The authors also demonstrated that the DEAttack can be conducted in the black-box setting by only compensating the modification for 1\% of pixels of medical images. Furthermore, Li \textit{et al.} \cite{li2022query_A50} improved the square attack by designing a more accurate gradient estimation for better convergence, which efficiently alters the predictions of foreground pixels with a few queries to the medical segmentation model. A learnable variance of the adaptive distribution is also enabled to limit sampling regions to a small area for improved efficiency.

\vspace{-3mm}
\subsection{Restricted Black-box (No-box) Attacks}
The above-mentioned methods either rely on having the whole knowledge of the targeted diagnosis model or require multiple queries to the black-box model. However, the attacker might not be able to access the target diagnosis model directly in most real-world situations. Particularly, the restricted black-box (no-box) setting can effectively represent the hardest (worst) case for the practical adversarial attack, which even do not require querying the target black-box DNN. The no-box attack mainly depends on the transferability \cite{LiuCLS17, ilyas2019adversarial, demontis2019adversarial} of adversarial examples among diverse DNN models. For instance, the no-box attacker can craft adversarial images based on a locally deployed surrogate model, which can directly transfer to target medical diagnosis systems. A well-established study has demonstrated that restricted black-box adversarial attacks pose a more stealthy security threat for natural vision tasks \cite{cheng2019improving, jiang2019black, WeiLCC19, lu2020enhancing, YinWYGKDLL21}. Nevertheless, existing studies related to no-box adversarial attacks for medical image analysis are still lacking \cite{bortsova2021adversarial_A18, apostolidis2022digital_A21, ding2023vith_A2023_2, chen2023frequency_A2023_4} and mainly based on approaches for natural images \cite{shah2018susceptibility_A4, rao2020thorough_A45_D60}, which requires further research efforts.

To uncover the potential security threat of medical diagnosis systems, Shah \textit{et al.} \cite{shah2018susceptibility_A4} first investigated the black-box adversarial vulnerability of diabetic retinopathy detection models. The adversarial examples generated against a specific diagnosis DNN can also be transferred to other models, which can cause grave consequences for practical diagnostic prediction. Furthermore, hybrid lesion-based algorithms \cite{chiem2007novel} that are composed of multiple lesion detectors are demonstrated a greater ability to defend against transferable adversarial examples. To further deploy a reliable diagnosis system in practice, Bortsova \textit{et al.} \cite{bortsova2021adversarial_A18} studied several unexplored factors affecting the no-box adversarial attack against medical image analysis systems. In addition, the transferable attack success rate can be enhanced by bridging the gap in training data and also network architectures between different diagnosis models. Hence, it is essential to consider the above-mentioned factors during the development of security-critical medical image analysis systems in clinical practice.

Unfortunately, very few works have investigated custom black-box adversarial attack methods for medical image analysis. Innovatively, Apostolidis \textit{et al.} \cite{apostolidis2022digital_A21} highlighted a novel view of digital watermarking to restricted black-box attacks in the context of medical image analysis. Specifically, the Krawtchouk orthogonal moments \cite{yap2003image} are also incorporated to generate adversarial watermarks against three different medical modalities. Massive experiments demonstrated that CT scans are extremely vulnerable against no-box attacks across various models.

Other than no-box adversarial attacks against medical image classification, there also exist scarce studies that explore the no-box adversarial vulnerability of medical image segmentation models. Bortsova \textit{et al.} \cite{bortsova2021adversarial_A9} investigated both the white-box and no-box adversarial attacks against medical image segmentation and their relationships. A surrogate network is used to generate targeted and untargeted adversarial images that can be further transferred to black-box segmentation models. Note that the training datasets of the surrogate model and the black-box model have no intersection to simulate clinical practice. The attacker can effectively misguide the target model to produce specific segmentation results, \textit{e.g.}, a heart symbol. Particularly, the transferability of restricted black-box attack relies on a high-level adversarial noise, which is also in alignment with \cite{paschali2018generalizability_A1}.

\vspace{-3mm}
\section{Medical Adversarial Defenses}
\label{sec:4}
\textbf{Medical adversarial defense} focuses on developing strategies to protect diagnosis models from adversaries, which are malicious inputs designed to mislead models into making incorrect diagnoses. Given the critical role of reliability and trust in healthcare applications, robust defense mechanisms are essential to safeguard against threats to the healthcare industry posed by adversaries. Thus, we highlight the significance of establishing robust defenses within computer-aided diagnosis systems to ensure patient safety and maintain trust in automated medical assessments. Furthermore, we introduce a systematic taxonomy of medical adversarial defense methods, categorized based on the operational stage of the defense mechanism. This categorization aids in understanding where each defense method applies within the pipeline\textemdash from input handling to final decision making\textemdash and how each method contributes to overall system robustness for trustworthy medical image analysis.

\vspace{-3mm}
\subsection{Taxonomy of Adversarial Defenses}
On account of catastrophic failures caused by adversarial examples, various methods have been proposed to build trustworthy AI systems for natural images \cite{carlini2017adversarial, MadryMSTV18, XieWZRY18, dong2022improving, dong2023enemy}. In the meantime, establishing robust computer-aided diagnosis models for clinical applications also contributes to the delivery of reliable healthcare services to millions of people. It is thus significant to investigate adversarially robust models for medical image analysis. Considering its application prospect, we can imagine that healthcare or automatic diagnosis systems claim they can provide robust and reliable services to billions of people. Below, we systematically categorize adversarial defense methods for medical image analysis into five classes: adversarial training, adversarial detection, image-level pre-processing, feature enhancement, and knowledge distillation.

Adversarial training and its variants have been demonstrated to be the most effective methods to defend against adversarial examples. The primary goal of adversarial training is to improve the robustness of a certain model against adversarial attacks, which can be easily achieved by augmenting adversarial examples into training examples \cite{MadryMSTV18, zhang2019theoretically, wang2019improving}. In comparison, adversarial detection aims at distinguishing adversarial examples in advance to prevent the deep learning-based system from further catastrophes. Note that we have formally defined both adversarial training and detection in Section \ref{sec:2}. Image-level pre-processing aims to remove the visually imperceptible perturbation from adversarial examples while not affecting the subsequent inference of clean examples. Furthermore, feature enhancement methods focus on feature-level processing to mitigate the adversarial effect, which can be conducted by additional modules or novel frameworks. There also exists another type of defense method that distills a lightweight and adversarially robust model from a large-scale model for deployment in real-time healthcare applications.

Note that adversarial detection methods are incapable of conducting a correct diagnosis for adversaries, which mainly rely on distinguishing adversaries in advance to reject these malicious inputs. In comparison, the other four types of methods aim to establish a robust medical system, which treats all inputs equally for subsequent diagnosis. In the following, we provide a comprehensive review related to these five types of adversarial defense methods for medical imaging.

\subsection{Summary of Medical Defense Works}
Overall, we provide a summary of adversarial defense works in the context of medical image analysis in Table \ref{tab:2}. In addition to a significant number of adversarial attack methods mentioned in Section \ref{sec:3_Sum}, several attack approaches also serve as robustness evaluation metrics for adversarial defense in the context of medical images. We thus provide a list of used evaluation metrics for reference: Dense Adversary Generation (DAG) \cite{xie2017adversarial}, Generative Adversarial Perturbations (GAP) \cite{poursaeed2018generative}, Distributionally Adversarial Attack (DAA) \cite{zheng2019distributionally}, Diverse Inputs I-FGSM (DII-FGSM) \cite{xie2019improving}, HopSkipJumpAttack \cite{chen2020hopskipjumpattack}, AutoPGD \cite{croce2020reliable}, SPSA \cite{uesato2018adversarial}, Stabilized Medical Attack (SMA) \cite{QiGS0Z21_A17}.

\begin{table*}[t!]
	\centering
	\tiny 
	\renewcommand{\arraystretch}{0.7}
	\caption{Summary of adversarial defense methods for medical image analysis (time ascending). The newly proposed adversarial attack approaches tailored for robustness evaluation of medical images are in \textbf{bold}.}
	\vspace{-4mm}
	\label{tab:2}
	\rowcolors{2}{gray!0}{gray!8}
	\resizebox{\linewidth}{!}{  
		
		\begin{tabular}{|c|c|c|c|c|c|}
			\hline
			Reference & Year & Task &  Defense type  &  Evaluation metrics  &  Data Modality\\ 
			\hline
			\cite{asgari2018vulnerability_D17} & 2018 & Classification & Feature Enhancement & \begin{tabular}[c]{@{}c@{}} FGSM, PGD, BIM,\\L-BFGS, DeepFool \end{tabular} & X-ray \\
			
			\cite{huang2018some_D21} & 2018 & Reconstruction & Adversarial Training & FGSM, BIM & CT \\
			
			\cite{ren2019brain_D11} & 2019 & Segmentation & Adversarial Training & FGSM & MRI \\
			
			\cite{xue2019improving_D22} & 2019 & Classification & Feature Enhancement & FGSM, I-FGSM, CW & X-ray, Dermoscopy \\
			
			\cite{taghanaki2019kernelized_D20} & 2019 & \begin{tabular}[c]{@{}c@{}} Classification, Segmentation, \\ Object Detection \end{tabular} & Feature Enhancement & \begin{tabular}[c]{@{}c@{}} FGSM, CW, PGD, BIM,\\GN, SPSA, MI-FGSM\end{tabular} & X-ray, Dermoscopy \\
			
			\cite{vatian2019impact_D13} & 2019 & Classification & Adversarial Training & GN & CT, MRI \\
			
			\cite{he2019non_D16} & 2019 & Segmentation & Feature Enhancement & I-FGSM & X-ray, Dermoscopy \\
			
			\cite{li2019volumetric_D46} & 2019 & Segmentation & Adversarial Training & FGSM, I-FGSM & CT \\
			
			\cite{kotia2020risk_D49} & 2019 & Classification & Adversarial Training & FGSM & MRI \\
			
			\cite{stimpel2019multi_D62} & 2019 & Low-level vision & Feature Enhancement & \textbf{Optimization-based Attack} & X-ray, MRI \\
			
			\cite{park2020robustification_D19} & 2020 & Segmentation & Adversarial Detection & DAG & MRI \\
			
			\cite{li2020anatomical_D40} & 2020 & Regression & Feature Enhancement & FGSM, I-FGSM & MRI \\
			
			\cite{rao2020thorough_A45_D60} & 2020 & Classification & Adversarial Training, Pre-processing & \begin{tabular}[c]{@{}c@{}} FGSM, PGD, DAA, \\ MI-FGSM, DII-FGSM \end{tabular} & X-ray \\
			
			\cite{paul2020mitigating_D8}  & 2020 & Classification & Adversarial Training & OPA, FGSM & CT \\
			
			\cite{wu2020classification_D15} & 2020 & Classification & Adversarial Training & PGD & Fundoscopy \\
			
			\cite{anand2020self_D48} & 2020 & Classification, Segmentation & Adversarial Training & PGD, FGSM & X-ray, MRI \\
			
			\cite{ma2020increasing_D5}  & 2020 & Classification, Segmentation & Adversarial Training & PGD, I-FGSM & CT, MRI \\
			
			\cite{cheng2020addressing_D18} & 2020 & Reconstruction & Adversarial Training & \textbf{False-Negative Adversarial Feature} & MRI \\
			
			\cite{raj2020improving_D61} & 2020 & Reconstruction & Adversarial Training & \textbf{GAN-based Attack} & CT, X-ray\\
			
			\cite{huq2020analysis_D47} & 2020 & Classification & Adversarial Training & PGD, FGSM & Dermoscopy \\
			
			\cite{tripathi2020fuzzy_D4}  & 2020 & Classification & Pre-processing & \begin{tabular}[c]{@{}c@{}} PGD, BIM, CW, \\ FGSM, DeepFool \end{tabular} & CT, X-ray \\
			
			\cite{liu2020no_D12} & 2020 & Classification & Adversarial Training & PGD & CT \\
			
			\cite{chen2020realistic_D14} & 2020 & Segmentation & Adversarial Training & \textbf{Adversarial Bias Attack} & MRI \\
			
			\cite{liu2020defending_D23} & 2020 & Segmentation & Pre-processing & ASMA & Fundoscopy, Dermoscopy \\
			
			\cite{shen2020robustness_D36} & 2020 & Classification & Adversarial Training & \textbf{Random and Optimized Attacks} & CT \\
			
			\cite{li2020robust_D24} & 2020 & Classification & Adversarial Detection & FGSM, PGD, BIM, MI-FGSM & X-ray \\
			
			\cite{ma2021understanding_D10} & 2021 & Classification & Adversarial Detection & FGSM, BIM, PGD, CW & X-ray, Fundoscopy, Dermoscopy \\
			
			\cite{watson2021attack_D7}  & 2021 & Classification & Adversarial Detection & PGD, CW & X-ray \\
			
			\cite{carannante2021trustworthy_D54} & 2021 & Segmentation & Feature Enhancement & PGD, FGSM & CT, MRI \\
			
			\cite{hirano2021universal_A7_D55} & 2021 & Classification & Adversarial Training & UAP & OCT, X-ray, Dermoscopy \\
			
			\cite{li2021defending_D1} & 2021 & Classification & Adversarial Training, Adversarial Detection & FGSM, PGD, CW & OCT \\
			
			\cite{xu2021towards_D9}  & 2021 & Classification & Adversarial Training & PGD, GAP & X-ray, Fundoscopy, Dermoscopy \\
			
			\cite{lal2021adversarial_D3}  & 2021 & Classification & Adversarial Training, Feature Enhancement & \begin{tabular}[c]{@{}c@{}} FGSM, DeepFool, \\ \textbf{Speckle Noise Attack} \end{tabular} & X-ray, Fundoscopy \\
			
			\cite{pervin2021adversarial_D25} & 2021 & Segmentation & Adversarial Training & FGSM & CT \\
			
			\cite{uwimana2021out_D26} & 2021 & Classification                                 & Adversarial Detection & FGSM, BIM, CW, DeepFool & Microscopy \\
			
			\cite{han2021advancing_D35} & 2021 & Classification & Feature Enhancement & PGD & CT, MRI, X-ray \\
			
			\cite{zhou2021ssmd_D41} & 2021 & Object Detection & Adversarial Training & FGSM & CT, Microscopy \\
			
			\cite{liu2021robustifying_D6}  & 2021 & Segmentation & Distillation & FGSM, I-FGSM, TI-FGSM & MRI \\
			
			\cite{daza2021towards_D27} & 2021 & Segmentation & Feature Enhancement, Adversarial Training & PGD, AA & CT, MRI \\
			
			\cite{chen2021enhancing_D33} & 2021 & Classification & Feature Enhancement & PGD, FGSM & X-ray \\
			
			\cite{yao2021medical_D45} & 2021 & Classification & Adversarial Detection, Pre-processing & \begin{tabular}[c]{@{}c@{}} FGSM, BIM, PGD, CW, PGD,\\MI-FGSM, TI-FGSM, DI-FGSM \end{tabular} & X-ray, Fundoscopy \\
			
			\cite{gupta2022vulnerability_D56} & 2022 & Classification & Adversarial Training & FGSM & MRI \\
			
			\cite{kansal2022defending_D51} & 2022 & Classification & Pre-processing & FGSM, PGD & X-ray \\
			
			\cite{jaiswal2022ros_D66} & 2022 & Classification & Distillation & FGSM, PGD & X-ray, Dermoscopy \\
			
			\cite{yang2022defense_D67} & 2022 & Classification & Adversarial Detection & HopSkipJumpAttack & MRI, X-ray, Microscopy \\
			
			\cite{xu2022medrdf_D2}  & 2022 & Classification & Pre-processing & I-FGSM, PGD, CW & X-ray, Dermoscopy \\
			
			\cite{joel2022using_D32} & 2022 & Classification & Adversarial Training & FGSM, PGD, BIM & CT, MRI, X-ray \\
			
			\cite{alatalo2022detecting_D63} & 2022 & Classification & Adversarial Detection & OPA & Microscopy \\
			
			\cite{hu2022adversarial_D29} & 2022 & Classification & Adversarial Training & DDN & MRI \\
			
			\cite{rodriguez2022role_D30} & 2022 & Classification & Adversarial Training & FGSM, PGD & OCT, X-ray, Dermoscopy \\
			
			\cite{ma2022adaptive_D53} & 2022 & \begin{tabular}[c]{@{}c@{}} Segmentation, Object Detection, \\ Landmark Detection \end{tabular} & Adversarial Training & PGD, I-FGSM & MRI, X-ray, Microscopy \\
			
			\cite{xie2022effective_D38} & 2022 & Classification & Adversarial Training & FGSM, PGD, BIM & CT, MRI, X-ray  \\
			
			\cite{wang2022fight_D43} & 2022 & Classification & Pre-processing & \begin{tabular}[c]{@{}c@{}} FGSM, BIM, CW, \\ PGD, AA, DI-FGSM \end{tabular} & Dermoscopy \\
			
			\cite{ghaffari2022adversarial_D28} & 2022 & Classification & Feature Enhancement & FGSM, PGD, FAB, SquareAttack & Microscopy \\
			
			\cite{maliamanis2022resilient_D39} & 2022 & Classification, Segmentation & Adversarial Training & \begin{tabular}[c]{@{}c@{}} FGSM, PGD, SquareAttack, \\ \textbf{Moment-based Adversarial Attack} \end{tabular} & X-ray, Microscopy \\
			
			\cite{almalik2022self_D42} & 2022 & Classification & Adversarial Detection, Feature Enhancement & FGSM, PGD, BIM, AutoPGD & X-ray, Fundoscopy \\
			
			\cite{SunXVG22_D44} & 2022 & Classification & Adversarial Training, Feature Enhancement & FGSM, PGD, CW & Ultrasound \\
			
			\cite{pandey2022adversarially_D52} & 2022 & Segmentation & Feature Enhancement & FGSM, PGD, SMA & CT \\
			
			\cite{bharath2022analysis_D58} & 2022 & Classification & Adversarial Training, Distillation & L-BFGS, FGSM & Fundoscopy \\
			
			\cite{roh2022impact_D59} & 2022 & Classification & Adversarial Training & FGSM & Microscopy \\
			
			\cite{daanouni2022nsl_D64} & 2022 & Classification                                 & Feature Enhancement & FGSM & Fundoscopy \\
			
			\cite{le2022efficient_D65} & 2022 & Segmentation & Pre-processing & DAG, I-FGSM & MRI, X-ray, Fundoscopy \\
			
			\cite{chen2022enhancing_D31} & 2022 & Segmentation & Adversarial Training & PGD & MRI \\
			
			\cite{shi2022robust_D57} & 2022 & Classification & Feature Enhancement & FGSM, PGD & X-ray, Fundoscopy \\
			
			\cite{dai2023improving_D2023_1} & 2023 & Classification & Feature Enhancement & BIM, PGD, Deepfool, AA & X-ray, Fundoscopy, Dermoscopy \\
			\cite{wang2023reversing_D2023_2} & 2023 & Classification & Pre-processing & \makecell{FGSM, BIM, PGD, \\ CW, DI-FGSM, AA} & Dermoscopy \\
			\cite{xiang2023toward_D2023_3} & 2023 & Classification & Adversarial Training, Feature Enhancement & FGSM, PGD, MIM, AA & CT \\
			\cite{wang2023adversarial_D2023_5} & 2023 & Classification & Adversarial Detection & FGSM, PGD, SPSA & Ultrasound \\

			\cite{li2024dynamic_D2024_4} & 2024 & Classification & Adversarial Training & I-FGSM, PGD & Dermoscopy \\
			\cite{zafar2024robust_D2024_6} & 2024 & Classification & Adversarial Training, Pre-processing & FGSM, PGD, BIM & X-ray \\

			\hline
		\end{tabular}
	}
	\vspace{-6mm}
\end{table*}

\subsection{Adversarial Training}

The majority of medical adversarial defense methods concentrate on adversarial training to establish robust diagnosis systems. Among them, a large proportion of works extend existing adversarial training methods for natural images to medical classification tasks \cite{paul2020mitigating_D8, xu2021towards_D9, vatian2019impact_D13, wu2020classification_D15, hu2022adversarial_D29, joel2022using_D32, xiang2023toward_D2023_3, li2024dynamic_D2024_4, zafar2024robust_D2024_6}. Vatian \textit{et al.} \cite{vatian2019impact_D13} investigated adversarial examples for medical imaging and tried several approaches to defend against these malicious examples. Specifically, both FGSM \cite{GoodfellowSS14} and JSMA \cite{papernot2016limitations} are incorporated to generate adversarial examples during the adversarial training stage. To further achieve better robustness, the authors also utilized the Gaussian noise to augment adversarially trained data and replaced the original Rectified Linear Units (ReLU) with Bounded ReLU. In addition to providing a robustness evaluation of several computer-aided diagnosis models, Xu \textit{et al.} adapt PGD-based adversarial training \cite{MadryMSTV18} and Misclassification Aware adveRsarial Training (MART) \cite{wang2019improving} to robustify computer-aided diagnosis models. A new medical dataset named \textit{Robust-Benchmark} is also proposed to evaluate the robustness of common perturbations. The above-mentioned works mainly focus on enhancing the robustness of a single model. An ensemble adversarial training method is also employed to establish a robust malignancy prediction system based on lung nodules \cite{paul2020mitigating_D8}. Specifically, various models with diverse architectures are adversarially trained with different initialized weights in advance. The ensemble prediction results can be further obtained by averaging the output probabilities from adversarially trained models.

In addition to transferring natural defense methods to diagnosis models, there also exist several adversarial defense methods that are tailored for medical image analysis \cite{lal2021adversarial_D3, liu2020no_D12, SunXVG22_D44, maliamanis2022resilient_D39}. Liu \textit{et al.} \cite{liu2020no_D12} investigate three types of adversarial augmentation examples that can be added to the training dataset for robustness improvement. Projected Gradient Descent (PGD) \cite{MadryMSTV18} is utilized to iteratively search the worst latent code to synthesize adversarial nodules that the target diagnosis model fails to detect. Furthermore, image-level adversarial perturbations are also generated to augment the training data. Extensive experiments demonstrate the effectiveness of the proposed method in enhancing the detection performance on real CT data and also robustness to unforeseen noise perturbations. Moreover, the Multi-Instance Robust Self-Training method with Drop-Max layer (MIRST-DM) \cite{SunXVG22_D44} is proposed to learn a smooth decision boundary for robust classification of breast cancer. In particular, MIRST-DM employs a sequence of adversarial images produced during the iterative adversary generation stage to accelerate the model convergence and also boost the robustness. The drop-max layer is proposed to discard the maximum value and forward the second-largest value to the next layer, which can depress unstable features to enhance the adversarial robustness. Note that speckle noise is one of the most common noises in retinal fundus images, which may hinder subsequent detection.

Although an array of adversarial training methods have been explored in the context of medical image analysis, the majority of them concentrate on supervised learning. Nonetheless, the manual annotation cost for medical images can sometimes be extremely expensive. To explore a low-cost training paradigm to enhance adversarial robustness for computer-aided diagnosis models, Li \textit{et al.} \cite{li2021defending_D1} proposed a novel medical defense method based on Semi-Supervised Adversarial Training (SSAT). Typically, SSAT produces the pseudo labels for unlabeled images in advance and then minimizes the empirical risk to enhance the network robustness. Furthermore, a systematic investigation of the adversarial robustness in the context of biomedical image analysis \cite{anand2020self_D48} has indicated that self-supervised learning can exhibit better adversarial robustness and also natural performance compared with transfer learning schemes on small medical datasets. 

Except for adversarial training for biomedical image classification, several academic works explore improving the intrinsic network robustness for medical segmentation \cite{ren2019brain_D11, li2019volumetric_D46, chen2022enhancing_D31, chen2020realistic_D14}. Generally, adversarial training can result in a clean accuracy drop \cite{zhang2019theoretically, XieY20, RadeM22}, which is unacceptable for various medical diagnosis applications. To mitigate this issue, Ma \textit{et al.} \cite{ma2020increasing_D5} proposed the Increasing Margin Adversarial (IMA) training method to adjust the upper bound of each adversarial perturbation during the training stage so that the intensity of the perturbation can be adaptively reduced. Similar to friendly adversarial training \cite{zhang2020attacks}, this work is inspired by the hypothesis that excessively strong adversarial examples might mislead the training process to cause a clean performance drop. The generated proper adversarial examples can effectively reduce the occurrence of overfitting decision boundary that impairs the standard performance of biomedical image applications. Chen \textit{et al.} \cite{chen2020realistic_D14} proposed a novel adversarial augmentation method for enhancing intrinsic network robustness via simulating underlying artifacts in clinical MR imaging. Specifically, the authors utilize the PGD method to search the specific control points to produce an adversarial bias field that can disrupt the intensity of original images by multiplication. Augmenting these physical adversarial examples into the training set can further promote robust feature learning for MR image segmentation, which is also effective in both low-data and cross-population settings.

Aside from adversarial training for medical pattern recognition, several researchers spare no effort to improving the robustness for medical image reconstruction \cite{cheng2020addressing_D18, raj2020improving_D61}. Especially for limited angle tomography, Huang \textit{et al.} \cite{huang2018some_D21} investigated the vulnerability of U-Net \cite{ronneberger2015u} against the Poisson noise during the artifact reduction. The reconstruction robustness can be further enhanced by adding projection-domain Poisson perturbation to training data. False-Negative Adversarial Feature (FNAF) \cite{cheng2020addressing_D18} is designed to simulate the worst-case perceptible small features in the clinical diagnosis setting, which disappear after MRI reconstruction. In the meantime, FNAF can also be embedded into the robust training process to improve MRI reconstruction for small and infrequent structures. Furthermore, Raj \textit{et al.} \cite{raj2020improving_D61} proposed a robust learning strategy with theoretical analysis to improve the robustness of various image-level reconstruction tasks, including face reconstruction and CT reconstruction. Unlike previous adversary generation methods that resort to iterative optimization, an auxiliary generative network is utilized to produce adversaries during the training stage. The image reconstruction model thus minimizes the target loss of adversary generation. 

\subsection{Adversarial Detection}
Different from establishing robustness during the training stage of computer-aided diagnosis models, adversarial detection aims at distinguishing adversarial examples from input examples during the application stage. In order to prevent subsequent misdiagnosis caused by adversarial examples, various adversarial detection methods have been proposed in the context of biomedical image analysis \cite{li2020robust_D24, alatalo2022detecting_D63, almalik2022self_D42}. In particular, medical adversarial detection can also be regarded as anomaly detection, which can be solved by incorporating explainability techniques \cite{watson2021attack_D7}. Based on the observation that adversarial attacks induce a shift in the distribution of SHAP values for medical imaging and electronic health record data, the authors thus proposed both full- and semi-supervised detection trained on SHAP values. Ma \textit{et al.} \cite{ma2021understanding_D10} provided a comprehensive analysis of medical adversarial examples and their relations to natural imaging. Moreover, two subspace distance-based methods in terms of kernel density and local intrinsic dimensionality, respectively. The robustness of Vision Transformers (ViT) for medical image classification is first investigated in \cite{almalik2022self_D42}. Specifically, the Kullback-Leibler (KL) divergence between various Multi-Layer Perceptrons (MLPs) can be applied to distinguish medical adversaries against ViT.

Although adversarially trained models can improve the in-distribution robustness of DNNs, they still suffer from a significant robustness drop against Out-Of-Distribution (OOD) adversarial examples. To further mitigate the disruption from OOD adversarial examples, Li \textit{et al.} \cite{li2021defending_D1} proposed an unsupervised adversarial detection method based on the latent features from the penultimate layer of computer-aided diagnosis models. Specifically, a Gaussian mixture model is employed to estimate the probability density for these latent features of clean examples. During the inference stage, the unsupervised detector rejects OOD adversarial examples based on the deviation of the extracted latent features from their corresponding probability density.

Most existing adversarial detection methods are designed for the medical classification scenario. There still remains limited research devoted to detecting adversarial examples for medical segmentation models. Park \textit{et al.} \cite{park2020robustification_D19} resorted to the frequency domain to better distinguish adversaries based on the reconstruction error. In addition, the reformer network is also incorporated to purify medical images to the manifold of legitimate examples for the subsequent segmentation stage.

\vspace{-4mm}
\subsection{Image-level Pre-processing}
\vspace{-1mm}
In general, an adversarial image consists of a clean image and its corresponding adversarial perturbation. Hence, denoising the adversarial example to get rid of the perturbation part can further be an effective strategy to facilitate the subsequent network diagnosis. Thus, there is no need to re-train or modify medical models when applying image-level pre-processing, which can be convenient and safe for biomedical image analysis. In this section, we introduce several image-level pre-processing works concentrating on protecting computer-aided diagnosis systems against adversarial examples. 

The vast majority of pre-processing-based defense methods are designed for medical classification tasks \cite{yao2021medical_D45, rao2020thorough_A45_D60, kansal2022defending_D51, wang2023reversing_D2023_2}. Hence, the image-level pre-processing is required not to destroy distinguishable parts for the subsequent diagnosis. The Medical Retrain-less Diagnostic Framework (MedRDF) \cite{xu2022medrdf_D2} is proposed to convert a pre-trained non-robust diagnosis model into its robust counterpart during inference. Specifically, MedRDF creates various copies of medical images that are perturbed by isotropic noises. These copies are then predicted via majority voting after denoising by a customized denoiser. A robust metric is also proposed to provide the confidence score of MedRDF, which further assists doctors in clinical practice. Beyond pixel-level denoising, Kansal \textit{et al.} \cite{kansal2022defending_D51} extended a high-level representation guided denoiser to protect medical applications against adversaries. The guidance of such high-level information further facilitates the image-level elimination of the adversarial effect to the final diagnosis result rather than visual disruptions. 

Current research on adversarial pre-processing for robust biomedical segmentation systems is still lacking and time-consuming. To this end, Liu \textit{et al.} \cite{liu2020defending_D23} proposed a low-cost image compression-based method to eliminate image-level adversarial perturbations against biomedical segmentation. In particular, a fine-grained frequency refinement approach is utilized to redesign the quantization table in JPEG compression-based pre-processing. Moreover, the optimized quantization step constraints are carefully set to prioritize defense efficiency and also compensate for the accuracy reduction caused by quantization errors. Furthermore, Le \textit{et al.} \cite{le2022efficient_D65} introduced a learnable adversarial denoising method by utilizing the U-Net \cite{ronneberger2015u} as the defender model. The defender first pre-processes medical images before feeding them into the subsequent biomedical segmentation models that are required to be effective for both clean examples and their adversarial counterparts.

\subsection{Feature Enhancement}
Adversarial examples have been demonstrated to be attributed to non-robust features (extracted from certain patterns in the data distribution) \cite{ilyas2019adversarial}, resulting in the mismatching of robustness between human and machine vision. Therefore, it is more than necessary to enhance the feature representation for robust inference. For a formal definition, we regard the modification of architectures or mapping functions as the feature enhancement in this paper. A myriad of feature enhancement methods have been designed to boost the robustness of medical classification models \cite{shi2022robust_D57, han2021advancing_D35, chen2021enhancing_D33, daanouni2022nsl_D64}. Taghanaki \textit{et al.} \cite{taghanaki2019kernelized_D20} modified the medical classification networks by replacing max-pooling layers with average-pooling layers. This modification significantly boosts the robustness against adversarial examples for different network architectures. The plausible reason for the robustness enhancement is that average-pooling can capture more global-level contextual information than the max-pooling, which increases the difficulty of adversarial attacks. Moreover, AutoEncoder (AE) can also be embedded into computer-aided diagnosis models for the feature-level denoising \cite{xue2019improving_D22}, which is independent of the image-level pre-processing procedure. In the meantime, the guidance of feature invariance is incorporated to reduce the model sensitivity against adversaries. Han \textit{et al.} \cite{han2021advancing_D35} introduced a dual-batch normalization to adversarial training, which improves the robustness of diagnostic models without the degradation of clean accuracy.

Apart from the mainstream medical classification task, feature enhancement has also been applied in several medical imaging tasks, including segmentation \cite{carannante2021trustworthy_D54}, object detection \cite{taghanaki2019kernelized_D20}, and low-level vision \cite{stimpel2019multi_D62}. Non-Local Context Encoder (NLCE) \cite{he2019non_D16} is proposed to improve the robustness of biomedical image segmentation models as a plug-and-play module. Similar to the observation from \cite{taghanaki2019kernelized_D20}, the NLCE module aims at capturing the global-level spatial dependencies and also contexts to strengthen features, which can be easily applied to various DNN-based medical image segmentation models. Stimpel \textit{et al.} \cite{stimpel2019multi_D62} utilize the guided filter with a learned guidance map for medical image super resolution and denoising. The guided filter exhibits a solid ability to limit the effectiveness of unforeseen adversarial examples on their generated outputs.

\vspace{-3mm}
\subsection{Knowledge Distillation}
In the machine learning community, knowledge distillation can be an effective technique to transfer the learned knowledge from a complex (teacher) model to a lightweight (student) model. Accordingly, self-distillation refers specifically to the situation when the teacher and student models share the same network architecture. Moreover, adversarial knowledge distillation has also been widely explored for the natural imaging domain \cite{goldblum2020adversarially, zi2021revisiting, ZhuY0ZL0ZXY22}, which transfers the adversarial robustness from the large teacher model to a lightweight student model. 

There also exist a few works that concentrate on enhancing adversarial robustness by knowledge distillation in the context of medical image analysis. The Robust Stochastic Knowledge Distillation (RoS-KD) framework \cite{jaiswal2022ros_D66} is designed to distill robust knowledge from multiple teacher models to a specific student model. Note that the distillation is conducted on noisy labeled data to efficiently simulate practical adversarial examples. The smooth parameter update mechanism is also proposed using weight averaging on multiple checkpoints. Furthermore, Liu \textit{et al.} \cite{liu2021robustifying_D6} adapted the defensive distillation to brain tumor segmentation for MRI data. The authors also showed that the defensive distillation achieves a better robust performance against FGSM attacks than adversarial training.

\begin{figure}[t]
	\centering
	\includegraphics[width=0.6\linewidth]{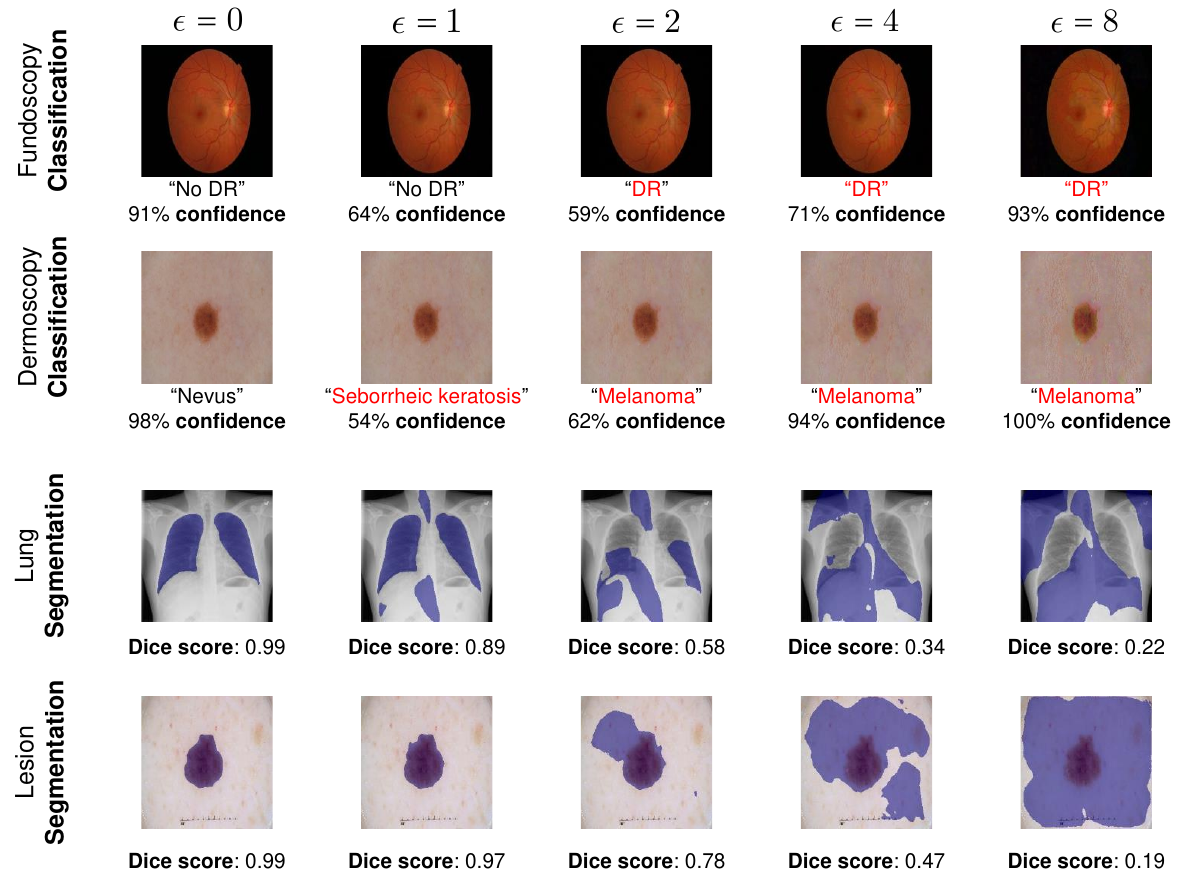}
	\vspace{-4mm}
	\caption{Visualization of medical adversarial examples with predictions under diverse perturbation size $\epsilon$. The generated segmentation masks are superimposed on the original images for visualization.}
	\vspace{-4mm}
	\label{fig:3}
\end{figure}

\vspace{-3mm}
\section{Experimental Evaluation}
\label{sec:5}
In this section, we provide a systematic evaluation of both adversarial attack and defense for computer-aided diagnosis models. We first introduce our experimental settings, including datasets and measurements. Moreover, we measure the effect of diverse adversarial attack methods on several medical imaging tasks in various scenarios. We also establish a benchmark of adversarial training for biomedical image analysis systems to facilitate future research. 

\begin{table}[!t]
	\centering
	\caption{White-box robustness results under different attack configurations using ResNet-18 for medical classification. We present accuracy (\%) and percentage decrease for binary and multi-class classification tasks.}
	\vspace{-4mm}
	\renewcommand{\arraystretch}{0.6}
		\resizebox{0.6\linewidth}{!}{  
			\begin{tabular}{cccccc}
				\toprule
				\multirow{2}{*}{Adversarial Type}&\multirow{2}{*}{$\epsilon$}&\multicolumn{2}{c}{Messidor (Fundoscopy)}&\multicolumn{2}{c}{ISIC (Dermoscopy)} \\
				\cmidrule(lr){3-4} \cmidrule(lr){5-6}
				&&Binary&Multi-class&Binary&Multi-class\\
				\midrule
				None & 0 & 71.3 & 50.0 & 64.9 & 60.0\\
				\midrule
				\multirow{4}{*}{\makecell{FGSM \cite{GoodfellowSS14}}} & 1/255 & 47.1$_\textit{(-33.9\%)}$ & 22.1$_\textit{(-55.8\%)}$ & 41.5$_\textit{(-36.1\%)}$ & 12.5$_\textit{(-79.2\%)}$ \\
				& 2/255 & 38.8$_\textit{(-45.6\%)}$ & 13.8$_\textit{(-72.4\%)}$ & 37.6$_\textit{(-42.1\%)}$ & 4.8$_\textit{(-92.0\%)}$ \\
				& 4/255 & 27.9$_\textit{(-60.9\%)}$ & 9.2$_\textit{(-81.6\%)}$ & 35.7$_\textit{(-45.0\%)}$ & 4.5$_\textit{(-92.5\%)}$ \\
				& 8/255 & 24.2$_\textit{(-66.1\%)}$ & 20.4$_\textit{(-59.2\%)}$ & 34.5$_\textit{(-46.8\%)}$ & 8.0$_\textit{(-86.7\%)}$ \\
				\midrule
				\multirow{4}{*}{\makecell{PGD \cite{MadryMSTV18}}} & 1/255 & 21.3$_\textit{(-70.1\%)}$ & 12.5$_\textit{(-75.0\%)}$ & 38.4$_\textit{(-40.8\%)}$ & 8.9$_\textit{(-85.1\%)}$ \\
				& 2/255 & 8.8$_\textit{(-87.7\%)}$ & 4.6$_\textit{(-90.8\%)}$ & 22.8$_\textit{(-64.9\%)}$ & 2.6$_\textit{(-95.7\%)}$ \\
				& 4/255 & 1.7$_\textit{(-97.6\%)}$ & 0.4$_\textit{(-99.2\%)}$ & 12.5$_\textit{(-80.7\%)}$ & 0.8$_\textit{(-98.7\%)}$ \\
				& 8/255 & 0.0$_\textit{(-100.0\%)}$ & 0.0$_\textit{(-100.0\%)}$ & 12.8$_\textit{(-80.3\%)}$ & 0.4$_\textit{(-99.3\%)}$ \\
				\midrule
				\multirow{4}{*}{\makecell{CW \cite{carlini2017towards}}} & 1/255 & 15.8$_\textit{(-77.8\%)}$ & 12.5$_\textit{(-75.0\%)}$ & 37.5$_\textit{(-42.2\%)}$ & 9.1$_\textit{(-84.8\%)}$ \\
				& 2/255 & 7.1$_\textit{(-90.0\%)}$ & 6.7$_\textit{(-86.6\%)}$ & 22.5$_\textit{(-65.3\%)}$ & 2.5$_\textit{(-95.8\%)}$ \\
				& 4/255 & 1.3$_\textit{(-98.2\%)}$ & 2.1$_\textit{(-95.8\%)}$ & 9.3$_\textit{(-85.7\%)}$ & 0.9$_\textit{(-98.5\%)}$ \\
				& 8/255 & 0.0$_\textit{(-100.0\%)}$ & 0.0$_\textit{(-100.0\%)}$ & 2.1$_\textit{(-96.8\%)}$ & 0.4$_\textit{(-99.3\%)}$ \\
				\midrule
				\multirow{4}{*}{\makecell{SMA \cite{QiGS0Z21_A17}}} & 1/255 & 14.6$_\textit{(-79.5\%)}$ & 12.3$_\textit{(-75.4\%)}$ & 37.2$_\textit{(-42.7\%)}$ & 8.7$_\textit{(-85.5\%)}$ \\
				& 2/255 & 5.7$_\textit{(-92.0\%)}$ & 5.2$_\textit{(-89.6\%)}$ & 22.6$_\textit{(-65.2\%)}$ & 2.5$_\textit{(-95.8\%)}$ \\
				& 4/255 & 0.0$_\textit{(-100.0\%)}$ & 0.5$_\textit{(-99.0\%)}$ & 9.9$_\textit{(-84.7\%)}$ & 1.0$_\textit{(-98.3\%)}$ \\
				& 8/255 & 0.0$_\textit{(-100.0\%)}$ & 0.0$_\textit{(-100.0\%)}$ & 1.7$_\textit{(-97.4\%)}$ & 0.3$_\textit{(-99.5\%)}$ \\
				\midrule
				\multirow{4}{*}{\makecell{AA \cite{croce2020reliable}}} & 1/255 & 12.8$_\textit{(-82.1\%)}$ & 6.7$_\textit{(-86.6\%)}$ & 36.9$_\textit{(-43.1\%)}$ & 8.5$_\textit{(-85.8\%)}$ \\
				& 2/255& 3.8$_\textit{(-94.6\%)}$ & 2.5$_\textit{(-95.0\%)}$ & 22.1$_\textit{(-66.0\%)}$ & 2.4$_\textit{(-96.0\%)}$ \\
				& 4/255& 0.0$_\textit{(-100.0\%)}$ & 0.0$_\textit{(-100.0\%)}$ & 8.4$_\textit{(-87.1\%)}$ & 0.8$_\textit{(-98.7\%)}$ \\
				& 8/255& 0.0$_\textit{(-100.0\%)}$ & 0.0$_\textit{(-100.0\%)}$ & 1.6$_\textit{(-97.5\%)}$ & 0.1$_\textit{(-99.8\%)}$ \\
				\bottomrule
				
			\end{tabular}
		}
	\vspace{-4mm}
	\label{tab:3}
\end{table}

\subsection{Experimental Settings}
\subsubsection{Datasets}
In this paper, we mainly use four standard benchmark datasets to explore adversarial attack and defense for medical image analysis: 1) \textbf{Messidor}\footnote{Kindly provided by the Messidor program partners (see https://www.adcis.net/en/third-party/messidor/).} dataset \cite{decenciere2014feedback}, containing 1,200 eye fundus color numerical images for detecting diabetic retinopathy of four classes according to retinopathy grade; 2) International Skin Imaging Collaboration (\textbf{ISIC 2017}) dataset \cite{codella2018skin} of 2,750 dermoscopic images of three classes for skin lesion classification and segmentation. 3) \textbf{ChestX-ray 14} dataset \cite{wang2017chestx}, consisting of 112,120 frontal-view X-ray images of 14 thorax diseases. 4) \textbf{COVID-19} database \cite{chowdhury2020can}, comprising 21,165 chest X-ray images with lung masks for segmentation. 

For medical classification, we consider both binary and multi-class classification tasks. Following \cite{xu2021towards_D9}, we randomly select 960 fundus images from Messidor dataset as the training set. All the fundus images are processed with several data augmentation operations, including random rotating and flipping. We also conduct pre-processing on skin lesion images from ISIC 2017 dataset with resizing and center-cropping. Due to the computational budget, we evenly sample 10,000 X-ray images with random flipping and normalization from ChestX-ray 14 dataset and then randomly choose 8,000 examples from them as the training set. Considering the computational cost, we sample 2,750 X-ray images and corresponding lung masks from the COVID-19 database for medical segmentation. Following the dataset division from ISIC 2017 dataset, 2,000 chest X-ray images are utilized for training, and the rest are used for evaluation.

\begin{table}[!t]
	\centering
	\begin{minipage}{0.51\textwidth}
	\centering
	\vspace{-2mm}
	\caption{White-box robustness (\%) against PGD attacks in diverse setups with U-Net for medical segmentation.}
	\vspace{-4mm}
	\renewcommand{\arraystretch}{0.6}
		\resizebox{1\linewidth}{!}{  
			\begin{tabular}{cccccc}
				\toprule
				\multirow{2}{*}{Adversarial Loss}&\multirow{2}{*}{$\epsilon$}&\multicolumn{2}{c}{ISIC (Dermoscopy)}&\multicolumn{2}{c}{COVID-19 (X-ray)} \\
				\cmidrule(lr){3-4} \cmidrule(lr){5-6}
				&&mIOU&Dice&mIOU&Dice\\
				\midrule
				None & 0 & 0.818 & 0.883 & 0.977 & 0.988\\
				\midrule
				\multirow{4}{*}{BCE} & 1/255 & 0.580 & 0.680 & 0.747 & 0.840\\
				& 2/255 & 0.405 & 0.517 & 0.559 & 0.690\\
				& 4/255 & 0.248 & 0.354 & 0.355 & 0.492\\
				& 8/255 & 0.167 & 0.255 & 0.230 & 0.350\\
				\midrule
				\multirow{4}{*}{IOU} & 1/255 & 0.542 & 0.672 & 0.695 & 0.829\\
				& 2/255 & 0.328 & 0.487 & 0.454 & 0.648\\
				& 4/255 & 0.145 & 0.305 & 0.239 & 0.418\\
				& 8/255 & 0.073 & 0.193 & 0.136 & 0.283\\
				\midrule
				\multirow{4}{*}{Dice} & 1/255 & 0.561 & 0.663 & 0.704 & 0.807\\
				& 2/255 & 0.340 & 0.450 & 0.473 & 0.610\\
				& 4/255 & 0.158 & 0.243 & 0.265 & 0.391\\
				& 8/255 & 0.082 & 0.140 & 0.152 & 0.246\\
				\bottomrule
				
			\end{tabular}
		}
	\vspace{-6mm}
	\label{tab:4}
	\end{minipage}
	\hfill
	\begin{minipage}{0.44\textwidth}
	\centering
	\vspace{-2mm}
	\caption{Black-box and no-box adversarial robustness (\%) under different settings with ResNet-18 for medical classification.}
	\vspace{-4mm}
	\renewcommand{\arraystretch}{0.3}
		\resizebox{1\linewidth}{!}{  
			\begin{tabular}{cccccc}
				\toprule
				\multirow{2}{*}{Adversarial Type}&\multirow{2}{*}{$\epsilon$}&\multicolumn{2}{c}{Messidor (Fundoscopy)}&\multicolumn{2}{c}{ISIC (Dermoscopy)} \\
				\cmidrule(lr){3-4} \cmidrule(lr){5-6}
				&&Binary&Multi-class&Binary&Multi-class\\
				\midrule
				None & 0 & 71.3 & 50.0 & 64.9 & 60.0\\
				\midrule
				\multirow{4}{*}{\makecell{Square Attack \cite{andriushchenko2020square} \\ (\textbf{Black-box})}} & 1/255 & 13.3 & 9.6 & 45.1 & 19.3\\
				& 2/255 & 5.8 & 3.8 & 30.3 & 6.0\\
				& 4/255 & 0.0 & 0.4 & 15.6 & 1.6\\
				& 8/255 & 0.0 & 0.0 & 2.9 & 0.5\\
				\midrule
				\multirow{4}{*}{\makecell{PGD \cite{MadryMSTV18} \\ (\textbf{No-box})}} & 1/255 & 47.5 & 41.3 & 59.1 & 58.5\\
				& 2/255 & 36.7 & 37.9 & 49.7 & 55.6\\
				& 4/255 & 27.9 & 27.9 & 35.3 & 50.4\\
				& 8/255 & 10.4 & 18.7 & 21.2 & 38.5\\
				\midrule
				\multirow{4}{*}{\makecell{CW \cite{carlini2017towards} \\ (\textbf{No-box})}} & 1/255 & 42.1 & 37.1 & 59.1 & 58.0\\
				& 2/255 & 35.4 & 37.1 & 48.9 & 52.9\\
				& 4/255 & 35.4 & 40.4 & 34.8 & 41.7\\
				& 8/255 & 19.6 & 12.9 & 20.5 & 19.3\\
				\midrule
				\multirow{4}{*}{\makecell{SMA \cite{QiGS0Z21_A17} \\ (\textbf{No-box})}} & 1/255 & 43.0 & 38.6 & 60.5 & 56.8\\
				& 2/255 & 37.5 & 35.2 & 51.8 & 49.9\\
				& 4/255 & 35.5 & 29.3 & 40.6 & 36.1\\
				& 8/255 & 17.9 & 12.6 & 21.9 & 17.2\\
				\bottomrule
				
			\end{tabular}
		}
	\vspace{-6mm}
	\label{tab:5}
	\end{minipage}
\end{table}

\vspace{-2mm}
\subsubsection{Evaluation Metrics}
In the context of our study, adversarial attacks aim to degrade the performance of neural networks by causing misclassifications or incorrect segmentations. For medical image classification tasks, we use classification accuracy and the Area Under the Receiver Operating Characteristic Curve (AUC) as the primary evaluation metrics, as they quantify the model's ability to correctly classify images and its discriminative power across various thresholds.

For image segmentation\textemdash especially critical in biomedical applications\textemdash it is essential to quantify how accurately the predicted segmentation maps correspond to the ground truth. To achieve this, we employ two widely recognized evaluation metrics: Mean Intersection over Union (mIoU) and Dice Coefficient. mIoU measures the average overlap between the predicted segmentation and the ground truth mask across all classes. The Dice coefficient, also known as the Sørensen-Dice index, is defined as twice the area of overlap divided by the total number of pixels in both the prediction and the ground truth. Both metrics evaluate the similarity between the predicted segmentation and the ground truth, providing insights into the precision of segmentation models. Thus, adversarial attacks against segmentation aims at reducing the mIoU or Dice coefficient between the generated segmentation mask and target mask, resulting in wrong segmentation results. Oppositely, adversarial defense focuses on keeping the network outputs of adversarial and clean examples as similar as possible. In other words, adversarial defense aims to enhancing the robustness of DNNs against adversaries. In this paper, we focus on building a unified benchmark for medical adversarial training to enhance the robustness of computer-aided diagnosis models.

Following the setting from RobustBench \cite{CroceASDFCM021}, we utilize ResNet-18 \cite{he2016deep} and MobileNetV2 \cite{sandler2018mobilenetv2} as target networks to conduct adversarial attack and defense for medical classification. For biomedical segmentation, we adopt U-Net \cite{ronneberger2015u} and SegNet \cite{badrinarayanan2017segnet} to generate segmentation masks. We use Stochastic Gradient Descent (SGD) optimizer with the Nesterov momentum factor \cite{Nesterov1983AMF} of 0.9 and the cyclic learning rate schedule \cite{smith2019super} with the maximum learning rate of 0.01. In this paper, we mainly consider the most common scenario, attacks under $\ell_{\infty}$-norm threat model. We conduct all the experiments on a single NVIDIA Tesla A100. We report the accuracies on clean examples as well as adversaries obtained by five strong adversarial attack methods: FGSM \cite{GoodfellowSS14}, PGD \cite{MadryMSTV18} with 20 steps with the step size of $1/255$, CW \cite{carlini2017towards}, Square Attack \cite{andriushchenko2020square}, and Auto Attack (AA) \cite{croce2020reliable}. Note that the maximum $\ell_{\infty}$-norm perturbation is set as $\epsilon=8/255$. For adversarial training, we adopt PGD to generate adversaries as the training data for 100 epochs.

\vspace{-2mm}
\subsection{Adversarial Attack}
\vspace{-1mm}
In this section, we transfer several adversarial attack methods for natural images to the medical imaging domain under various attack scenarios for comprehensive evaluation. We also consider medical classification and segmentation in diverse biomedical imaging modalities. In addition to standard adversarial attack methods, we also explore the effect of Stabilized Medical Attack (SMA) \cite{QiGS0Z21_A17} that is tailored for medical images in diverse tasks.

\begin{table}[!t]
	\centering
	\begin{minipage}{0.54\textwidth}
	\centering
	\vspace{-0.5mm}
	\caption{No-box robustness (\%) against PGD attacks in different settings with U-Net for medical segmentation.}
	\vspace{-4mm}
	\renewcommand{\arraystretch}{0.6}
		\resizebox{1\linewidth}{!}{  
			\begin{tabular}{cccccc}
				\toprule
				\multirow{2}{*}{Adversarial Loss}&\multirow{2}{*}{$\epsilon$}&\multicolumn{2}{c}{ISIC (Dermoscopy)}&\multicolumn{2}{c}{COVID-19 (X-ray)} \\
				\cmidrule(lr){3-4} \cmidrule(lr){5-6}
				&&mIOU&Dice&mIOU&Dice\\
				\midrule
				None & 0 & 0.818 & 0.883 & 0.977 & 0.988\\
				\midrule
				\multirow{4}{*}{BCE} & 1/255 & 0.801 & 0.863 & 0.961 & 0.979\\
				& 2/255 & 0.783 & 0.851 & 0.933 & 0.963\\
				& 4/255 & 0.755 & 0.823 & 0.870 & 0.922\\
				& 8/255 & 0.698 & 0.780 & 0.774 & 0.856\\
				\midrule
				\multirow{4}{*}{Dice} & 1/255 & 0.807 & 0.875 & 0.966 & 0.982\\
				& 2/255 & 0.797 & 0.866 & 0.954 & 0.976\\
				& 4/255 & 0.775 & 0.848 & 0.926 & 0.959\\
				& 8/255 & 0.722 & 0.804 & 0.883 & 0.931\\
				
				\bottomrule
				
			\end{tabular}
		}
	\vspace{-4mm}
	\label{tab:6}
	\end{minipage}
	\hfill
	\begin{minipage}{0.4\textwidth}
	\centering
	\footnotesize
	\caption{Average time cost comparison of adversary generation methods for diverse medical imaging tasks.}
	\vspace{-4mm}
	\renewcommand{\arraystretch}{0.6}
	\resizebox{\linewidth}{!}{  
			\begin{tabular}{ccc}
				\toprule
				\multicolumn{3}{l}{\textbf{Medical Classification:}} \\
				\midrule
				Adversarial Type & Messidor (Fundoscopy) & ISIC (Dermoscopy) \\
				
				\midrule
				FGSM & 0.6 s & 1.9 s\\
				PGD & 22.6 s & 40.3 s\\
				CW & 33.7 s & 61.0 s\\
				SMA & 35.8 s & 69.5 s \\
				AA & 738.6 s & 1464.0 s\\
				Square Attack & 548.1 s & 1769.3 s\\
				\midrule
				\multicolumn{3}{l}{\textbf{Medical Segmentation:}} \\
				\midrule
				Adversarial Type & ISIC (Dermoscopy) & COVID-19 (X-ray) \\
				\midrule
				PGD-BCE & 78.6 s & 78.7 s\\
				PGD-Dice & 84.5 s & 84.8 s\\
				
				\bottomrule
				
			\end{tabular}
 	}
	\vspace{-4mm}
	\label{tab:7}
	\end{minipage}
\end{table}

\begin{table}[!t]
	\centering
	\caption{White-box \textbf{B}inary (\textbf{B}) and \textbf{M}ulti-class (\textbf{M}) accuracy (\%) of adversarially trained medical classification models for fundoscopy and dermoscopy in different settings. }
	\vspace{-4mm}
	\begin{minipage}{0.49\linewidth}
	\renewcommand{\arraystretch}{0.6}
		\resizebox{\linewidth}{!}{  
			\begin{tabular}{cccccccccccccc}
				\toprule
				\multicolumn{14}{l}{\textbf{Fundoscopy Classification on Messidor \cite{decenciere2014feedback}: }} \\
				\midrule
				\multirow{2}{*}{Adv. Type} & \multirow{2}{*}{$\epsilon$} & \multicolumn{2}{c}{NAT} & \multicolumn{2}{c}{PGD-AT} & \multicolumn{2}{c}{TRADES} & \multicolumn{2}{c}{MART} & \multicolumn{2}{c}{MPAdvT} & \multicolumn{2}{c}{HAT} \\
				\cmidrule(lr){3-4} \cmidrule(lr){5-6} \cmidrule(lr){7-8} \cmidrule(lr){9-10} \cmidrule(lr){11-12} \cmidrule(lr){13-14}
				&&\textbf{B}&\textbf{M}&\textbf{B}&\textbf{M}&\textbf{B}&\textbf{M}&\textbf{B}&\textbf{M}&\textbf{B}&\textbf{M}&\textbf{B}&\textbf{M}\\
				\midrule
				None & 0 & 71.3 & 50.0 & 58.8 & 46.7 & 62.1 & 44.6 & 55.0 & 42.5 & 59.5 & 47.3 & 62.5 & 46.7 \\
				\midrule
				\multirow{4}{*}{FGSM} & 1/255 & 47.1 & 22.1 & 57.5 & 45.8 & 58.3 & 41.7 & 54.6 & 42.5 & 58.7 & 46.9 & 56.7 & 42.5 \\
				& 2/255 & 38.8 & 13.8 & 55.8 & 44.2 & 55.8 & 37.5 & 54.6 & 42.1 & 56.3 & 45.8 & 53.8 & 39.2 \\
				& 4/255 & 27.9 & 9.2 & 52.9 & 39.2 & 48.8 & 30.4 & 52.5 & 40.8 & 53.0 & 41.6 & 46.3 & 30.8 \\
				& 8/255 & 24.2 & 20.4 & 45.4 & 35.8 & 29.6 & 19.2 & 49.6 & 39.2 & 46.3 & 37.7 & 28.8 & 13.3 \\
				\midrule
				\multirow{4}{*}{PGD} & 1/255 & 21.3 & 12.5 & 57.5 & 45.8 & 58.3 & 41.7 & 54.6 & 42.5 & 58.7 & 46.9 & 56.7 & 42.5 \\
				& 2/255 & 8.8 & 4.6 & 55.8 & 42.9 & 55.8 & 37.5 & 54.6 & 42.1 & 56.3 & 45.8 & 53.8 & 38.3 \\
				& 4/255 & 1.7 & 0.4 & 52.6 & 36.7 & 46.7 & 29.2 & 52.5 & 40.4 & 52.8 & 39.7 & 46.3 & 27.9 \\
				& 8/255 & 0.0 & 0.0 & 43.3 & 30.0 & 25.8 & 15.4 & 49.2 & 37.1 & 44.5 & 34.3 & 21.3 & 11.3 \\
				\midrule
				\multirow{4}{*}{CW} & 1/255 & 15.8 & 12.5 & 57.5 & 45.4 & 58.3 & 40.8 & 54.6 & 42.5 & 58.7 & 46.3 & 56.7 & 42.5 \\
				& 2/255 & 7.1 & 6.7 & 55.4 & 41.3 & 55.8 & 35.8 & 54.6 & 41.7 & 55.9 & 43.7 & 53.8 & 36.7 \\
				& 4/255 & 1.3 & 2.1 & 52.5 & 35.4 & 46.7 & 27.1 & 52.5 & 40.4 & 52.7 & 37.5 & 46.3 & 24.6 \\
				& 8/255 & 0.0 & 0.0 & 42.5 & 27.9 & 26.3 & 13.3 & 49.2 & 36.3 & 43.6 & 30.8 & 22.5 & 7.9 \\
				\midrule
				\multirow{4}{*}{SMA} & 1/255 & 14.6 & 12.3 & 57.5 & 45.4 & 58.3 & 40.8 & 54.6 & 42.5 & 58.7 & 46.3 & 56.7 & 42.5 \\
				& 2/255 & 5.7 & 5.2 & 55.2 & 41.3 & 55.8 & 35.9 & 54.6 & 41.7 & 55.6 & 43.4 & 53.8 & 36.9 \\
				& 4/255 & 0.0 & 0.5 & 52.5 & 35.1 & 46.6 & 26.7 & 52.5 & 40.3 & 52.7 & 37.0 & 46.0 & 23.8 \\
				& 8/255 & 0.0 & 0.0 & 42.3 & 28.6 & 26.0 & 12.9 & 49.0 & 36.4 & 43.2 & 30.3 & 21.6 & 4.8 \\
				\midrule
				\multirow{4}{*}{AA} & 1/255 & 12.8 & 6.7 & 57.5 & 45.8 & 58.3 & 40.8 & 54.6 & 42.5 & 58.7 & 46.1 & 56.7 & 42.5 \\
				& 2/255 & 3.8 & 2.5 & 55.2 & 41.3 & 55.8 & 36.3 & 54.6 & 41.7 & 55.6 & 43.0 & 53.8 & 37.5 \\
				& 4/255 & 0.0 & 0.0 & 52.3 & 35.0 & 46.3 & 26.3 & 52.5 & 40.4 & 52.5 & 36.2 & 46.3 & 23.3 \\
				& 8/255 & 0.0 & 0.0 & 42.1 & 28.3 & 25.4 & 12.5 & 49.2 & 36.7 & 42.7 & 29.8 & 20.0 & 1.7 \\
				\bottomrule
			\end{tabular}
		}
	\end{minipage}
	\begin{minipage}{0.49\linewidth}
	\renewcommand{\arraystretch}{0.6}
		\resizebox{\linewidth}{!}{  
				\begin{tabular}{cccccccccccccc}
				\toprule
				\multicolumn{14}{l}{\textbf{Dermoscopy Classification on ISIC \cite{codella2018skin}:}} \\
				\midrule
				\multirow{2}{*}{Adv. Type} & \multirow{2}{*}{$\epsilon$} & \multicolumn{2}{c}{NAT} & \multicolumn{2}{c}{PGD-AT} & \multicolumn{2}{c}{TRADES} & \multicolumn{2}{c}{MART} & \multicolumn{2}{c}{MPAdvT} & \multicolumn{2}{c}{HAT} \\
				\cmidrule(lr){3-4} \cmidrule(lr){5-6} \cmidrule(lr){7-8} \cmidrule(lr){9-10} \cmidrule(lr){11-12} \cmidrule(lr){13-14}
				
				&&\textbf{B}&\textbf{M}&\textbf{B}&\textbf{M}&\textbf{B}&\textbf{M}&\textbf{B}&\textbf{M}&\textbf{B}&\textbf{M}&\textbf{B}&\textbf{M}\\
				\midrule
				None & 0 & 64.9 & 60.0 & 61.2 & 52.0 & 62.3 & 52.9 & 61.1 & 49.6 & 61.2 & 52.8 & 68.6 & 57.1 \\
				\midrule
				\multirow{4}{*}{FGSM} & 1/255 & 41.5 & 12.5 & 58.0 & 48.0 & 58.1 & 49.5 & 58.9 & 45.8 & 58.3 & 49.4 & 63.3 & 51.1 \\
				& 2/255 & 37.6 & 4.8 & 55.6 & 44.4 & 53.5 & 44.0 & 56.7 & 42.4 & 55.9 & 45.7 & 58.8 & 46.8 \\
				& 4/255 & 35.7 & 4.5 & 49.2 & 38.0 & 44.9 & 35.3 & 51.6 & 33.7 & 49.8 & 38.6 & 48.5 & 35.3 \\
				& 8/255 & 34.5 & 8.0 & 39.6 & 25.5 & 33.1 & 22.3 & 40.9 & 21.7 & 40.3 & 26.2 & 37.5 & 20.7 \\
				\midrule
				\multirow{4}{*}{PGD} & 1/255 & 38.4 & 8.9 & 57.9 & 48.0 & 58.0 & 49.2 & 58.9 & 45.6 & 58.3 & 49.4 & 63.2 & 51.1 \\
				& 2/255 & 22.8 & 2.6 & 55.6 & 44.4 & 52.9 & 43.7 & 56.5 & 41.9 & 55.7& 45.7 & 57.6 & 45.9 \\
				& 4/255 & 12.5 & 0.8 & 48.4 & 36.8 & 42.0 & 32.8 & 49.9 & 31.2 & 49.2 & 37.3 & 44.4 & 33.6 \\
				& 8/255 & 12.8 & 0.4 & 35.1 & 22.4 & 21.5 & 14.4 & 34.5 & 14.9 & 38.0 & 24.3 & 25.7 & 14.4 \\
				\midrule
				\multirow{4}{*}{CW} & 1/255 & 37.5 & 9.1 & 57.9 & 47.9 & 58.0 & 49.1 & 58.9 & 45.1 & 58.3 & 49.2 & 63.2 & 50.5 \\
				& 2/255 & 22.5 & 2.5 & 55.6 & 43.6 & 52.9 & 43.2 & 56.5 & 41.3 & 55.7 & 44.9 & 57.6 & 45.3 \\
				& 4/255 & 9.3 & 0.9 & 48.4 & 36.0 & 41.9 & 31.9 & 49.9 & 30.9 & 49.1 & 36.6 & 44.4 & 32.8 \\
				& 8/255 & 2.1 & 0.4 & 36.3 & 22.7 & 21.9 & 14.3 & 34.7 & 15.2 & 37.4 & 23.0 & 26.5 & 13.9 \\
				\midrule
				\multirow{4}{*}{SMA} & 1/255 & 37.2 & 8.7 & 57.9 & 47.9 & 58.0 & 49.1 & 58.9 & 45.7 & 58.4 & 49.2 & 63.4 & 50.8 \\
				& 2/255 & 22.6 & 2.5 & 55.6 & 43.5 & 52.9 & 43.3 & 56.4 & 41.0 & 55.7 & 44.5 & 57.7 & 45.3 \\
				& 4/255 & 9.9 & 1.0 & 48.4 & 35.8 & 41.8 & 32.0 & 50.1 & 30.4 & 49.1 & 36.3 & 44.6 & 12.9 \\
				& 8/255 & 1.7 & 0.3 & 35.0 & 21.6 & 22.3 & 16.8 & 33.7 & 14.9 & 36.2 & 23.1 & 25.5 & 14.3 \\
				\midrule
				\multirow{4}{*}{AA} & 1/255 & 36.9 & 8.5 & 57.9 & 47.9 & 58.0 & 49.1 & 58.9 & 45.5 & 58.2 & 49.1 & 63.2 & 50.7 \\
				& 2/255 & 22.1 & 2.4 & 55.6 & 43.7 & 52.7 & 43.1 & 56.4 & 41.1 & 55.7 & 43.8 & 57.6 & 45.3 \\
				& 4/255 & 8.4 & 0.8 & 48.4 & 35.5 & 41.5 & 31.6 & 49.7 & 30.3 & 48.9 & 36.2 & 44.1 & 32.8 \\
				& 8/255 & 1.6 & 0.1 & 34.8 & 20.9 & 18.9 & 12.7 & 32.3 & 14.0 & 35.7 & 22.5 & 22.3 & 12.5 \\

				\bottomrule
				
			\end{tabular}
		}
		\end{minipage}
	\vspace{-4mm}
	\label{tab:8}
\end{table}

\begin{table}[!t]
\centering
\caption{White-box Multi-class accuracy (\%) of adversarially trained medical classification models for fundoscopy and dermoscopy. Adversarial perturbations are restricted within the $\ell_\infty$-norm radius $\epsilon=8/255$.}
\vspace{-4mm}
\renewcommand{\arraystretch}{0.6}
\resizebox{\linewidth}{!}{
\begin{tabular}{ccccccccccccccccc}
\toprule
&\multicolumn{8}{c}{\textbf{Fundoscopy Classification on Messidor \cite{decenciere2014feedback}}} &\multicolumn{8}{c}{\textbf{Dermoscopy Classification on ISIC \cite{codella2018skin}}} \\
\cmidrule(lr){2-9} \cmidrule(lr){10-17}
\multirow{2}{*}{Method} & \multicolumn{4}{c}{Wide-ResNet-28-10} & \multicolumn{4}{c}{MobileNetV2} & \multicolumn{4}{c}{Wide-ResNet-28-10} & \multicolumn{4}{c}{MobileNetV2} \\ 
\cmidrule(lr){2-5} \cmidrule(lr){6-9} \cmidrule(lr){10-13} \cmidrule(lr){14-17}
& Clean & PGD & CW & AA & Clean & PGD & CW & AA & Clean & PGD & CW & AA & Clean & PGD & CW & AA \\
\midrule
NAT & 63.7 & 0.0 & 0.0 & 0.0 & 48.5 & 0.0 & 0.0 & 0.0 & 69.6 & 0.0 & 0.0 & 0.0 & 57.3 & 0.0 & 0.0 & 0.0 \\

PGD-AT \cite{MadryMSTV18} & 57.9 & 45.7 & 42.3 & 41.5 & 44.9 & 28.3 & 25.4 & 24.2 & 62.4 & 35.5 & 32.7 & 31.8 & 50.6 & 26.7 & 23.9 & 23.2 \\

TRADES \cite{zhang2019theoretically} & 58.5 & 46.0 & 44.5 & 42.8 & 43.6 & 27.5 & 26.3 & 25.0 & 63.1 & 35.9 & 33.2 & 32.3 & 51.3 & 27.1 & 24.5 & 23.7 \\

MART \cite{wang2019improving} & 56.6 & 44.7 & 41.4 & 40.3 & 40.8 & 30.4 & 27.5 & 26.8 & 61.8 & 37.6 & 34.9 & 33.8 & 49.2 & 28.5 & 26.0 & 25.3 \\

MPAdvT \cite{xu2021towards_D9} & 59.2 & 47.1 & 45.9 & 43.7 & 45.6 & 31.3 & 27.9 & 27.0 & 63.7 & 37.3 & 35.1 & 34.0 & 52.8 & 28.3 & 26.2 & 25.5 \\

HAT \cite{RadeM22} & 61.2 & 47.3 & 45.8 & 43.9 & 46.1 & 31.8 & 28.2 & 27.5 & 65.2 & 38.4 & 36.3 & 34.7 & 53.7 & 30.2 & 27.8 & 27.3 \\

\bottomrule
\end{tabular}
}
\vspace{-4mm}
\label{supp-tab:1}
\end{table}

\subsubsection{White-box Attack}
To begin with, we present several cases of the adversarial attack against diverse medical diagnosis tasks under different attack strengths in Fig. \ref{fig:3}. We can easily observe that computer-aided diagnosis models are extremely vulnerable to adversarial examples, even with a small adversarial perturbation size. The medical classification models are misguided by adversaries to make high-confidence misdiagnoses, which can interfere with clinical judgment. Moreover, incorrect segmentation results induced by adversaries can lead to false treatment suggestions.

To comprehensively measure the adversarial effect on medical classification models, we report the accuracy against several white-box attacks in both binary and multi-class classification settings, as shown in Table \ref{tab:3}. Note that we adopt the cross-entropy loss function for adversary generation. It can be seen that there exists a significant plunge of accuracy as increasing the adversarial perturbation size. In addition, we can easily observe that multi-class classification models suffer from a more severe accuracy drop than binary classification models. However, existing medical adversarial defense methods mainly focus on binary diagnosis, which is relatively easy to construct robust models. In this paper, we claim that constructing robustness for multi-class medical classification is more challenging and can further be generalized to various clinical scenarios.

Beyond adversarial attacks against medical classification, we also evaluate the performance of medical segmentation against PGD-based adversaries, as shown in Table \ref{tab:4}. We adopt Binary Cross-Entropy (BCE) loss, IOU loss and Dice loss as the adversarial loss for adversary generation. Intuitively, optimizing IOU loss during adversary generation significantly degrades mIOU, while using Dice loss as the adversarial loss notably reduces the Dice value.

\begin{table}[!t]
	\centering
	\caption{White-box robustness against PGD attacks in diverse settings using U-Net for medical segmentation.}
	\vspace{-4mm}
	\begin{minipage}{0.49\linewidth}
	\renewcommand{\arraystretch}{0.6}
		\resizebox{1\linewidth}{!}{  
			\begin{tabular}{ccccccc}
				\toprule
				\multirow{2}{*}{Task} & \multirow{2}{*}{Adversarial Loss} & \multirow{2}{*}{$\epsilon$} & \multicolumn{2}{c}{NAT} & \multicolumn{2}{c}{ADV} \\
				\cmidrule(lr){4-5} \cmidrule(lr){6-7}
				& & & mIOU & Dice & mIOU & Dice \\
				\midrule
				\multirow{14}{*}{\makecell{ISIC \\ (Dermoscopy)}} & None & 0 & 0.818 & 0.883 & 0.803 & 0.867 \\
				\cmidrule(lr){2-7}
				& \multirow{4}{*}{BCE} & 1/255 & 0.580 & 0.680 & 0.788 &0.855 \\
				& & 2/255 & 0.405 & 0.517 & 0.773 & 0.842 \\
				& & 4/255 & 0.248 & 0.354 & 0.733 & 0.810 \\
				& & 8/255 & 0.167 & 0.255 & 0.601 & 0.700 \\
				\cmidrule(lr){2-7}
				& \multirow{4}{*}{IOU} & 1/255 & 0.542 & 0.672 & 0.783 & 0.854 \\
				& & 2/255 & 0.328 & 0.487 & 0.762 & 0.840 \\
				& & 4/255 & 0.145 & 0.305 & 0.716 & 0.807 \\
				& & 8/255 & 0.073 & 0.193 & 0.573 & 0.698 \\
				\cmidrule(lr){2-7}
				& \multirow{4}{*}{Dice} & 1/255 & 0.561 & 0.663 & 0.786 & 0.853 \\
				& & 2/255 & 0.340 & 0.450 & 0.767 & 0.838 \\
				& & 4/255 & 0.158 & 0.243 & 0.723 & 0.803 \\
				& & 8/255 & 0.082 & 0.140 & 0.594 & 0.694 \\
				\bottomrule
			\end{tabular}
		}
	\end{minipage}
	\begin{minipage}{0.47\linewidth}
		\renewcommand{\arraystretch}{0.6}
		\resizebox{1\linewidth}{!}{ 
			\begin{tabular}{ccccccc}
				\toprule
				\multirow{2}{*}{Task} & \multirow{2}{*}{Adversarial Loss} & \multirow{2}{*}{$\epsilon$} & \multicolumn{2}{c}{NAT} & \multicolumn{2}{c}{ADV} \\
				\cmidrule(lr){4-5} \cmidrule(lr){6-7}
				& & & mIOU & Dice & mIOU & Dice \\
				\midrule
				\multirow{14}{*}{\makecell{COVID-19 \\ (X-ray)}} & None & 0 & 0.977 & 0.988 & 0.964 & 0.981 \\
				\cmidrule(lr){2-7}
				& \multirow{4}{*}{BCE} & 1/255 & 0.747 & 0.840 & 0.949 & 0.973 \\
				& & 2/255 & 0.559 & 0.690 & 0.931 & 0.963 \\
				& & 4/255 & 0.355 & 0.492 & 0.890 & 0.940 \\
				& & 8/255 & 0.230 & 0.350 & 0.812 & 0.887 \\
				\cmidrule(lr){2-7}
				& \multirow{4}{*}{IOU} & 1/255 & 0.695 & 0.829 & 0.943 & 0.987 \\
				& & 2/255 & 0.454 & 0.648 & 0.912 & 0.969 \\
				& & 4/255 & 0.239 & 0.418 & 0.865 & 0.938 \\
				& & 8/255 & 0.136 & 0.283 & 0.753 & 0.874 \\
				\cmidrule(lr){2-7}
				& \multirow{4}{*}{Dice} & 1/255 & 0.704 & 0.807 & 0.964 & 0.981 \\
				& & 2/255 & 0.473 & 0.610 & 0.923 & 0.958 \\
				& & 4/255 & 0.265 & 0.391 & 0.880 & 0.932 \\
				& & 8/255 & 0.152 & 0.246 & 0.774 & 0.860 \\
				
				\bottomrule
				
			\end{tabular}
		}
	\end{minipage}
	\vspace{-4mm}
	\label{tab:9}
\end{table}

\subsubsection{Black-box Attack}
Apart from white-box adversarial attacks against medical image analysis, we also evaluate the performance of computer-aided diagnosis models against black-box adversarial attacks, which can be a more practical attack scenario. We report the accuracy of medical classifiers against both Black-box and Restricted black-box (No-box) attacks in Table \ref{tab:5}. Note that the no-box attacks are conducted by transferring adversaries generated against a MobileNetV2 \cite{sandler2018mobilenetv2} model to the target ResNet-18 \cite{he2016deep} classification model. Due to the accessibility to outputs of the target model, the black-box adversarial attack can achieve a better attack success rate than no-box attacks. In the meantime, we can also observe that conducting no-box attacks for multi-class medical classification models is much harder than for binary classifiers. The plausible reason is that multi-class classifiers have more complex classification decision boundary, which varies with each architecture.

We also evaluate the effectiveness of restricted black-box adversarial attacks against medical segmentation models, as shown in Table \ref{tab:6}. The no-box attacks are obtained by transferring PGD-based adversarial examples generated against the SegNet \cite{badrinarayanan2017segnet} model to directly attack the target U-Net \cite{ronneberger2015u} segmentation model. Despite adopting the Dice loss as the adversarial loss achieves a superior performance in the white-box scenario, attacking the BCE loss can obtain more transferable adversarial examples for medical segmentation in the no-box setting.

\begin{table}[!t]
	\centering
	\footnotesize
	\caption{White-box robustness under different attack configurations using ResNet-18 for multi-label medical classification. We report the AUC score (\%) of thorax disease classification models against PGD attack.}
	\vspace{-4mm}
	\renewcommand{\arraystretch}{0.6}
		\resizebox{0.8\linewidth}{!}{  
			\begin{tabular}{cccccccc}
				\toprule
				\multirow{2}{*}{Task} & \multirow{2}{*}{Adversarial Type} & \multirow{2}{*}{$\epsilon$} & & \multicolumn{2}{c}{Training $\epsilon=8/255$} & \multicolumn{2}{c}{Training $\epsilon=4/255$} \\
				\cmidrule(lr){5-6} \cmidrule(lr){7-8}
				&&& NAT & PGD-AT & MPAdvT & PGD-AT & MPAdvT \\
				\midrule
				\multirow{5}{*}{\makecell{ChestX-ray 14 \cite{wang2017chestx} \\ (Multi-label)}} & None & 0 & 71.0 & 71.4 & 70.2 & 73.6 & 72.3 \\
				\cmidrule(lr){2-8}
				& \multirow{4}{*}{PGD} & 1/255 & 23.3 & 67.1 & 66.7 & 65.9 & 64.8 \\
				& & 2/255 & 15.7 & 62.6 & 62.4 & 60.2 & 59.6 \\
				& & 4/255 & 14.0 & 53.1 & 52.8 & 51.7 & 51.0 \\
				& & 8/255 & 11.7 & 35.5 & 34.9 & 30.4 & 29.6 \\
				
				\bottomrule
				
			\end{tabular}
		}
	\vspace{-4mm}
	\label{tab:10}
\end{table}

\begin{table}[!t]
	\centering
	\renewcommand{\arraystretch}{0.6}
	\caption{Robust medical CLIP with AUC evaluation on clean and PGD-20 adversaries in zero-shot setting.}
	\vspace{-0.4cm}
	\resizebox{0.5\linewidth}{!}{
		\begin{tabular}{ccccccc}
			\toprule
			\multirow{2}{*}{Method} & \multicolumn{2}{c}{ChestXray14} & \multicolumn{2}{c}{CheXpert} & \multicolumn{2}{c}{PadChest} \\
			\cmidrule(l){2-3} \cmidrule(l){4-5} \cmidrule(l){6-7}
			& Clean & PGD & Clean & PGD & Clean & PGD \\
			\midrule
			TeCoA \cite{MaoGYWV23} & 0.674 & 0.526 & 0.857 & 0.685 & 0.602 & 0.483 \\
			PMG-FT \cite{wang2024pre} & 0.692 & 0.538 & 0.850 & 0.688 & 0.619 & 0.495 \\
			FARE \cite{schlarmann2024robust} & 0.702 & 0.541 & 0.866 & 0.694 & 0.627 & 0.505 \\
			
			\bottomrule
		\end{tabular}
 	}
	\label{supp-tab:3}
	\vspace{-0.4cm}
\end{table}

\subsubsection{Time Cost Analysis for Adversarial Attacks}
Furthermore, we present an analysis of the computational cost related to adversary generation for medical imaging. Specifically, we measure the time cost for generating adversaries by various methods for diverse medical imaging tasks, as shown in Table \ref{tab:7}. Note that we adopt ResNet-18 for medical classification and U-Net for medical segmentation, respectively. It can be seen that both Auto attack and Square attack require relatively considerable computing resources to produce strong adversaries. The primary time cost gap of two PGD attacks for medical segmentation comes from the efficiency of different loss computations.

\vspace{-3mm}
\subsection{Adversarial Training for Defense}
In this section, we construct a unified benchmark with the most effective defense method, adversarial training to establish robustness for medical diagnosis systems for future research. Specifically, we extend several existing adversarial training methods for the natural imaging domain to medical imaging analysis. We also measure the adversarial robustness of adversarially trained diagnosis models under various attack scenarios for systematic evaluation. In addition to standard adversarial attack methods, we investigate the effect of Multi-Perturbations Adversarial Training (MPAdvT) \cite{xu2021towards_D9}, which is specially designed for deep diagnostic models in diverse tasks.

\begin{table}[!t]
	\centering
	\caption{Black-box \textbf{B}inary (\textbf{B}) and \textbf{M}ulti-class (\textbf{M}) accuracy (\%) of ResNet-18-based adversarially trained medical classification models for fundoscopy and dermoscopy.}
	\vspace{-4mm}
	\begin{minipage}{0.49\linewidth}
	\renewcommand{\arraystretch}{0.6}
		\resizebox{\linewidth}{!}{  
			\begin{tabular}{cccccccccccccc}
				\toprule
				\multicolumn{14}{l}{\textbf{Fundoscopy Classification on Messidor \cite{decenciere2014feedback}: }} \\
				\midrule
				\multirow{2}{*}{Adv. Type} & \multirow{2}{*}{$\epsilon$} & \multicolumn{2}{c}{NAT} & \multicolumn{2}{c}{PGD-AT} & \multicolumn{2}{c}{TRADES} & \multicolumn{2}{c}{MART} & \multicolumn{2}{c}{MPAdvT} & \multicolumn{2}{c}{HAT} \\
				\cmidrule(lr){3-4} \cmidrule(lr){5-6} \cmidrule(lr){7-8} \cmidrule(lr){9-10} \cmidrule(lr){11-12} \cmidrule(lr){13-14}
				&&\textbf{B}&\textbf{M}&\textbf{B}&\textbf{M}&\textbf{B}&\textbf{M}&\textbf{B}&\textbf{M}&\textbf{B}&\textbf{M}&\textbf{B}&\textbf{M}\\
				\midrule
				None & 0 & 71.3 & 50.0 & 58.8 & 46.7 & 62.1 & 44.6 & 55.0 & 42.5 & 59.5 & 47.3 & 62.5 & 46.7 \\
				\midrule
				\multirow{4}{*}{\makecell{Square Attack \cite{andriushchenko2020square} \\ (\textbf{Black-box}) }} & 1/255 & 13.3 & 9.6 & 57.5 & 45.8 & 58.3 & 42.1 & 54.6 & 42.5 & 57.9 & 46.0 & 56.7 & 42.5 \\
				& 2/255 & 5.8 & 3.8 & 55.8 & 42.9 & 55.8 & 37.9 & 54.6 & 42.5 & 56.2 & 43.8 & 56.7 & 38.3 \\
				& 4/255 & 0.0 & 0.4 & 52.9 & 37.9 & 48.8 & 29.6 & 52.5 & 41.7 & 53.5 & 38.6 & 46.7 & 24.6 \\
				& 8/255 & 0.0 & 0.0 & 43.8 & 29.2 & 28.8 & 16.3 & 50.0 & 40.4 & 44.3 & 30.7 & 21.3 & 6.3 \\
				\midrule
				\multirow{4}{*}{\makecell{PGD \cite{MadryMSTV18} \\ (\textbf{No-box})}} & 1/255 & 47.5 & 41.3 & 58.8 & 46.7 & 61.3 & 43.8 & 55.0 & 42.5 & 59.5 & 47.3 & 60.0 & 46.7 \\
				& 2/255 & 36.7 & 37.9 & 59.2 & 46.7 & 58.3 & 42.9 & 55.4 & 42.5 & 59.3 & 47.0 & 59.2 & 45.4 \\
				& 4/255 & 27.9 & 27.9 & 56.7 & 46.3 & 56.7 & 40.4 & 55.4 & 42.9 & 57.4 & 46.7 & 57.9 & 45.8 \\
				& 8/255 & 10.4 & 18.7 & 55.8 & 44.6 & 50.0 & 36.3 & 55.8 & 42.5 & 56.1 & 44.9 & 50.0 & 40.8 \\
				\midrule
				\multirow{4}{*}{\makecell{CW \cite{carlini2017towards} \\ (\textbf{No-box})}} & 1/255 & 42.1 & 37.1 & 58.8 & 46.7 & 61.3 & 43.3 & 55.0 & 42.5 & 59.5 & 47.3 & 60.0 & 46.3 \\
				& 2/255 & 35.4 & 37.1 & 59.2 & 46.7 & 58.3 & 42.5 & 55.4 & 42.5 & 59.3 & 46.9 & 59.2 & 45.4 \\
				& 4/255 & 35.4 & 40.4 & 56.7 & 45.8 & 56.7 & 40.4 & 55.4 & 42.9 & 57.2 & 46.5 & 57.9 & 43.8 \\
				& 8/255 & 19.6 & 12.9 & 56.3 & 44.2 & 50.0 & 36.3 & 55.8 & 42.5 & 56.1 & 44.2 & 52.1 & 40.8 \\
				
				\midrule
				\multirow{4}{*}{\makecell{SMA \cite{QiGS0Z21_A17} \\ (\textbf{No-box})}} & 1/255 & 43.0 & 38.6 & 58.8 & 46.6 & 61.0 & 43.7 & 55.0 & 42.5 & 59.5 & 47.3 & 59.4 & 46.0 \\
				& 2/255 & 37.5 & 35.2 & 58.5 & 46.1 & 58.2 & 41.9 & 54.8 & 42.5 & 59.2 & 46.6 & 57.9 & 44.4 \\
				& 4/255 & 35.5 & 29.3 & 56.0 & 45.8 & 56.8 & 39.8 & 54.3 & 42.0 & 57.1 & 46.3 & 56.2 & 43.0 \\
				& 8/255 & 17.9 & 12.6 & 55.7 & 44.1 & 49.6 & 35.5 & 53.7 & 41.5 & 56.0 & 43.7 & 50.9 & 40.1 \\
				
				\bottomrule
			\end{tabular}
		}
	\end{minipage}
	\begin{minipage}{0.49\linewidth}
		\renewcommand{\arraystretch}{0.6}
		\resizebox{\linewidth}{!}{  
			\begin{tabular}{cccccccccccccc}

				\toprule
				
				\multicolumn{14}{l}{\textbf{Dermoscopy Classification on ISIC \cite{codella2018skin}:}} \\
				\midrule
				\multirow{2}{*}{Adversarial Type} & \multirow{2}{*}{$\epsilon$} & \multicolumn{2}{c}{NAT} & \multicolumn{2}{c}{PGD-AT} & \multicolumn{2}{c}{TRADES} & \multicolumn{2}{c}{MART} & \multicolumn{2}{c}{MPAdvT} & \multicolumn{2}{c}{HAT} \\
				\cmidrule(lr){3-4} \cmidrule(lr){5-6} \cmidrule(lr){7-8} \cmidrule(lr){9-10} \cmidrule(lr){11-12} \cmidrule(lr){13-14}
				&&\textbf{B}&\textbf{M}&\textbf{B}&\textbf{M}&\textbf{B}&\textbf{M}&\textbf{B}&\textbf{M}&\textbf{B}&\textbf{M}&\textbf{B}&\textbf{M}\\
				\midrule
				None & 0 & 64.9 & 60.0 & 61.2 & 52.0 & 62.3 & 52.9 & 61.1 & 49.6 & 61.2 & 52.8 & 68.6 & 57.1 \\
				\midrule
				\multirow{4}{*}{\makecell{Square Attack \cite{andriushchenko2020square} \\ (\textbf{Black-box}) }} & 1/255 & 45.1 & 19.3 & 58.4 & 48.0 & 58.8 & 50.3 & 59.1 & 45.9 & 59.0 & 48.7 & 63.9 & 51.6 \\
				& 2/255 & 30.3 & 6.0 & 56.7 & 44.4 & 54.7 & 45.6 & 57.2 & 42.7 & 57.2 & 45.5 & 59.7 & 47.5 \\
				& 4/255 & 15.6 & 1.6 & 50.7 & 37.9 & 45.5 & 36.7 & 52.8 & 34.3 & 52.4 & 39.1 & 46.7 & 36.3 \\
				& 8/255 & 2.9 & 0.5 & 40.0 & 23.5 & 27.6 & 18.8 & 41.9 & 19.7 & 43.5 & 24.0 & 28.9 & 19.2 \\
				\midrule
				\multirow{4}{*}{\makecell{PGD \cite{MadryMSTV18} \\ (\textbf{No-box})}} & 1/255 & 59.1 & 58.5 & 60.8 & 51.9 & 60.1 & 52.5 & 60.0 & 49.6 & 61.2 & 52.7 & 66.1 & 56.0 \\
				& 2/255 & 49.7 & 55.6 & 60.0 & 51.5 & 59.1 & 52.0 & 59.3 & 48.9 & 60.8 & 51.7 & 64.0 & 55.1 \\
				& 4/255 & 35.3 & 50.4 & 59.1 & 50.5 & 54.0 & 51.3 & 57.3 & 47.6 & 59.6 & 51.0 & 58.5 & 54.0 \\
				& 8/255 & 21.2 & 38.5 & 58.1 & 47.5 & 47.3 & 47.6 & 55.2 & 45.6 & 58.4 & 47.8 & 52.1 & 50.7 \\
				\midrule
				\multirow{4}{*}{\makecell{CW \cite{carlini2017towards} \\ (\textbf{No-box})}} & 1/255 & 59.1 & 58.0 & 60.8 & 51.9 & 60.1 & 52.5 & 60.0 & 49.6 & 61.2 & 52.6 & 66.1 & 56.1 \\
				& 2/255 & 48.9 & 52.9 & 60.0 & 52.0 & 59.1 & 52.0 & 59.5 & 49.1 & 60.7 & 51.3 & 64.0 & 54.9 \\
				& 4/255 & 34.8 & 41.7 & 59.3 & 50.9 & 54.7 & 51.2 & 57.2 & 48.3 & 59.5 & 51.2 & 58.1 & 53.7 \\
				& 8/255 & 20.5 & 19.3 & 58.5 & 48.0 & 48.3 & 48.5 & 55.6 & 46.7 & 58.2 & 48.0 & 52.8 & 50.7 \\
				\midrule
				\multirow{4}{*}{\makecell{SMA \cite{QiGS0Z21_A17} \\ (\textbf{No-box})}} & 1/255 & 60.5 & 56.8 & 60.7 & 51.3 & 59.9 & 52.6 & 60.3 & 49.4 & 61.2 & 52.6 & 65.8 & 56.0 \\
				& 2/255 & 51.8 & 49.9 & 59.7 & 50.6 & 58.9 & 52.2 & 59.4 & 48.7 & 60.8 & 51.3 & 63.9 & 54.6 \\
				& 4/255 & 40.6 & 36.1 & 58.9 & 49.5 & 54.4 & 51.1 & 56.9 & 48.0 & 59.7 & 51.0 & 58.0 & 53.4 \\
				& 8/255 & 21.9 & 17.2 & 57.3 & 47.4 & 50.3 & 48.2 & 55.2 & 46.2 & 58.5 & 48.5 & 51.7 & 50.5 \\

				\bottomrule
				
			\end{tabular}
		}
	\end{minipage}
	\vspace{-0.6cm}
	\label{tab:11}
\end{table}

\subsubsection{Defense against White-box Attacks}
Cutting-edge adversarial training methods mainly focus on augmenting adversaries as training data to obtain a robust decision boundary for both clean and adversarial examples. In this paper, we primarily extend the adversarial robustness to the biomedical imaging domain by transferring four widely-used adversarial training methods: PGD-AT \cite{MadryMSTV18}, TRADES \cite{zhang2019theoretically}, MART \cite{wang2019improving}, and HAT \cite{RadeM22}. The adversarial robustness results for both binary and multi-class medical classification are reported in Table \ref{tab:8}. Note that the robustness results for NAtural Training (NAT) are also provided for reference. We can observe that adversarially trained models can well preserve the robustness under different attack configurations. In the meantime, the naturally trained medical models are vulnerable to high-strength adversarial attacks, especially for the multi-class classification scenario. Note that both PGD-AT and MART can relatively achieve better robustness than TRADES and HAT for medical classification. The plausible reason for the robustness gap is that these two types of methods focus on different inner adversary generation styles. Although adversarial training can significantly enhance the adversarial robustness, the clean performance still suffers from a slight drop compared with naturally trained models.

Furthermore, we expand our experimental analysis to include a broader range of model architectures, in addition to the initial evaluation on ResNet-18 \cite{he2016deep}. We now present results from experiments involving adversarial training on Wide-ResNet-28-10 (WRN-28) \cite{ZagoruykoK16} and MobileNetV2 (MNV2) \cite{sandler2018mobilenetv2}, applied to medical classification models for fundoscopy and dermoscopy across diverse settings. The evaluation results, detailed in Table \ref{supp-tab:1}, reveal that adversarial robustness correlates strongly with network capacity. Notably, the robustness of Wide-ResNet, with its deeper and wider architecture, surpasses that of the more lightweight MobileNet in both natural performance and robustness. While our current evaluations are primarily based on computational metrics, it is significant to note that the ground-truth labels in the datasets we employed are derived from clinical assessments by medical experts, following their respective database construction criteria \cite{decenciere2014feedback, codella2018skin, wang2017chestx, chowdhury2020can}. This further underscores the clinical relevance of our computational findings of both adversarial attacks and defenses in the context of medical image analysis.

Besides establishing adversarial robustness for single-label classification models, we also conduct adversarial training for medical segmentation tasks. We present the performance for adversarially trained medical segmentation models under different attack configurations in Table \ref{tab:9}. Note that we adopt the most widely-used adversarial training method, PGD-AT, to enhance the intrinsic network robustness of medical segmentation models. It can be seen that the adversarially trained segmentation models can achieve better robustness against different degrees of adversarial attacks than naturally trained models. Intuitively, we observe that optimizing the IOU loss results in a more significant degradation of the mIOU, whereas employing the Dice loss as the objective function leads to a greater decrease in the Dice value.

Generally, adversarial training can induce a mild performance drop for clean examples. However, we observe an intriguing phenomenon for multi-label classification in the context of medical images, as shown in Table \ref{tab:10}. We report AUC of multi-label thorax disease classification models against adversarial examples of different attack strengths. In particular, we observe that reducing the perturbation radius during adversarial training implicitly enhances natural performance, although it may reduce adversarial robustness. This observation suggests that utilizing a lower adversarial perturbation contributes to improved natural accuracy. We attribute this enhancement to the similar background and biomedical structures inherent in medical imaging data, where small perturbations can effectively augment original images, leading to better generalization ability. Similarly, Hu \textit{et al.} \cite{hu2024protecting} introduced targeted adversarial samples with rectal artifact-pattern noise during training to enhance the natural performance of prostate cancer classification using MRI data.

\subsubsection{Extension with Medical CLIP for Zero-Shot Robustness}
We here expand our analysis to include a comprehensive evaluation of the generalization capabilities of adversarial defense methods across diverse medical imaging datasets (distributions) and modalities. Specifically, we investigated the zero-shot adversarial robustness of computer-aided medical imaging diagnosis using the standard vision-language model CLIP, particularly within the context of chest X-ray imaging (radiology) paired with multi-label datasets. In this challenging zero-shot adversarial robustness setup \cite{MaoGYWV23}, attackers have unrestricted access to ground truth data from new datasets at the inference stage, while defenders, with no prior exposure to these datasets, are required to maintain robustness against all unforeseen adversarial images derived from them.

Following the methodology in \cite{tiu2022expert}, we utilized a Vision Transformer (ViT-B/16)-based CLIP model pre-trained specifically on radiology datasets. We conducted adversarial fine-tuning on the MIMIC dataset\textemdash a comprehensive collection of chest radiographs paired with detailed radiology text reports\textemdash to refine the CLIP model using different adversarial training approaches \cite{MaoGYWV23, wang2024pre, schlarmann2024robust}. Subsequently, we evaluated both the natural and robust accuracy of these adversarially fine-tuned CLIP models across three multi-label radiology datasets: ChestXray14 \cite{wang2017chestx}, CheXpert \cite{irvin2019chexpert}, and PadChest \cite{bustos2020padchest} in the zero-shot setting, as presented in Table \ref{supp-tab:3}. Notably, the PadChest dataset poses a significant challenge due to its long-tail distribution of 192 diseases, including rare conditions. Our findings indicate that the zero-shot robustness of medical vision-language models generalizes effectively to diverse biomedical databases. This suggests that models trained on specific modalities can maintain their robustness when applied to different datasets without additional fine-tuning, thus demonstrating the potential for broader applicability of adversarial defense methods across various medical imaging datasets (distributions).

\begin{table}[!t]
	\centering
	\caption{No-box robustness (\%) of naturally trained and adversarially trained medical segmentation models for dermoscopy and X-ray against PGD attacks using U-Net.}
	\vspace{-3mm}
	\begin{minipage}{0.49\linewidth}
	\renewcommand{\arraystretch}{0.6}
		\resizebox{1\linewidth}{!}{  
			\begin{tabular}{ccccccc}
				\toprule
				\multirow{2}{*}{Task} & \multirow{2}{*}{Adversarial Loss} & \multirow{2}{*}{$\epsilon$} & \multicolumn{2}{c}{NAT} & \multicolumn{2}{c}{ADV} \\
				\cmidrule(lr){4-5} \cmidrule(lr){6-7}
				& & & mIOU & Dice & mIOU & Dice \\
				\midrule
				\multirow{14}{*}{\makecell{ISIC \cite{codella2018skin} \\ (Dermoscopy)}} & None & 0 & 0.818 & 0.883 & 0.803 & 0.867 \\
				\cmidrule(lr){2-7}
				& \multirow{4}{*}{BCE} & 1/255 & 0.801 & 0.863 & 0.800 & 0.865 \\
				& & 2/255 & 0.783 & 0.851 & 0.799 & 0.864 \\
				& & 4/255 & 0.755 & 0.823 & 0.797 & 0.862 \\
				& & 8/255 & 0.698 & 0.780 & 0.794 & 0.860 \\
				\cmidrule(lr){2-7}
				& \multirow{4}{*}{IOU} & 1/255 & 0.805 & 0.881 & 0.802 & 0.863 \\
				& & 2/255 & 0.791 & 0.872 & 0.800 & 0.860 \\
				& & 4/255 & 0.760 & 0.859 & 0.799 & 0.854 \\
				& & 8/255 & 0.704 & 0.823 & 0.797 & 0.847 \\
				\cmidrule(lr){2-7}
				& \multirow{4}{*}{Dice} & 1/255 & 0.807 & 0.875 & 0.801 & 0.865 \\
				& & 2/255 & 0.797 & 0.866 & 0.800 & 0.864 \\
				& & 4/255 & 0.775 & 0.848 & 0.798 & 0.863 \\
				& & 8/255 & 0.722 & 0.804 & 0.796 & 0.861 \\
				\bottomrule
			\end{tabular}
		}
	\end{minipage}
	\begin{minipage}{0.49\linewidth}
		\renewcommand{\arraystretch}{0.6}
		\resizebox{1\linewidth}{!}{  
			\begin{tabular}{ccccccc}
				\toprule
				\multirow{2}{*}{Task} & \multirow{2}{*}{Adversarial Loss} & \multirow{2}{*}{$\epsilon$} & \multicolumn{2}{c}{NAT} & \multicolumn{2}{c}{ADV} \\
				\cmidrule(lr){4-5} \cmidrule(lr){6-7}
				& & & mIOU & Dice & mIOU & Dice \\
				\midrule
				\multirow{14}{*}{\makecell{COVID-19 \cite{chowdhury2020can} \\ (X-ray)}} & None & 0 & 0.977 & 0.988 & 0.964 & 0.981 \\
				\cmidrule(lr){2-7}
				& \multirow{4}{*}{BCE} & 1/255 & 0.961 & 0.979 & 0.963 & 0.981 \\
				& & 2/255 & 0.933 & 0.963 & 0.963 & 0.981 \\
				& & 4/255 & 0.870 & 0.922 & 0.963 & 0.980 \\
				& & 8/255 & 0.774 & 0.856 & 0.963 & 0.980 \\
				\cmidrule(lr){2-7}
				& \multirow{4}{*}{IOU} & 1/255 & 0.964 & 0.984 & 0.963 & 0.981 \\
				& & 2/255 & 0.949 & 0.979 & 0.963 & 0.980 \\
				& & 4/255 & 0.885 & 0.966 & 0.962 & 0.978 \\
				& & 8/255 & 0.831 & 0.940 & 0.960 & 0.975 \\
				\cmidrule(lr){2-7}
				& \multirow{4}{*}{Dice} & 1/255 & 0.966 & 0.982 & 0.963 & 0.981 \\
				& & 2/255 & 0.954 & 0.976 & 0.963 & 0.981 \\
				& & 4/255 & 0.926 & 0.959 & 0.963 & 0.981 \\
				& & 8/255 & 0.883 & 0.931 & 0.963 & 0.980 \\
				
				\bottomrule
				
			\end{tabular}
		}
	\end{minipage}
	\vspace{-13pt}
	\label{tab:12}
\end{table}

\begin{table}[!t]
	\centering
	\footnotesize
	\caption{Time cost comparison of medical adv. training methods. We report the training time per epoch.}
	\vspace{-4mm}
	\renewcommand{\arraystretch}{0.5}
	 \resizebox{0.4\linewidth}{!}{
			\begin{tabular}{ccc}
				\toprule
				\multicolumn{3}{l}{\textbf{Medical Classification:}} \\
				\midrule
				Method & Messidor (Fundoscopy) & ISIC (Dermoscopy) \\
				
				\midrule
				NAT & 16.2 s & 17.3 s\\
				PGD-AT & 53.5 s & 65.1 s\\
				TRADES & 56.0 s & 67.7 s\\
				MART & 55.0 s & 65.9 s\\
				MPAdvT & 54.7 s & 65.5 s \\
				HAT & 62.0 s & 74.9 s\\
				\midrule
				\multicolumn{3}{l}{\textbf{Medical Segmentation:}} \\
				\midrule
				Adversarial Type & ISIC (Dermoscopy) & COVID-19 (X-ray) \\
				\midrule
				NAT & 13.8 s & 14.4 s\\
				PGD-AT & 148.5 s & 155.0 s\\
				
				\bottomrule
				
			\end{tabular}
		 }
	\vspace{-6mm}
	\label{tab:13}
\end{table}

\subsubsection{Defense against Black-box Attacks}
Besides the evaluation of white-box attacks against robust diagnosis models, we also emphasize the importance of defense against black-box attacks on account of real-world scenarios. Here, we conduct black-box and no-box adversarial attacks for evaluations to simulate medical defense in clinical practice. We commence with the evaluation of robust diagnosis models in both binary and multi-class classification settings (see Table \ref{tab:11}). The adversarially trained diagnosis models effectively remain robust against adversaries of different attack strengths. We also observe a similar trend of black-box accuracy to white-box one as the attack strength increases. Hence, adversarial training still remains in effectiveness to build robustness for white-box and black-box attacks across different settings for biomedical image analysis.

Likewise, we measure the performance of robust medical segmentation models against no-box adversarial attacks for comprehensive evaluation. The no-box performance of different settings is reported in Table \ref{tab:12}. It shows that adversarial training achieves outstanding performance in enhancing robustness in lung segmentation. We observe that using the BCE loss as the adversarial loss in the no-box scenario achieves a better attack success rate than adopting the Dice loss.

\subsubsection{Time Cost Analysis for Adversarial Training}
For systematic evaluation, we also provide the time cost of several adversarial training methods for medical imaging tasks in Table \ref{tab:13}. For a fair comparison, we conduct all the training experiments on a single NVIDIA Tesla A100 GPU with the same batch size. We report the average time cost over 10 runs of several adversarial training approaches. We can easily observe that adversarial training generally takes several times as long as natural training. The time cost gap mainly comes from the iterative gradient computation during the inner adversary generation, which is further enlarged for medical segmentation.

\begin{figure}[t]
	\centering
	\includegraphics[width=0.58\linewidth]{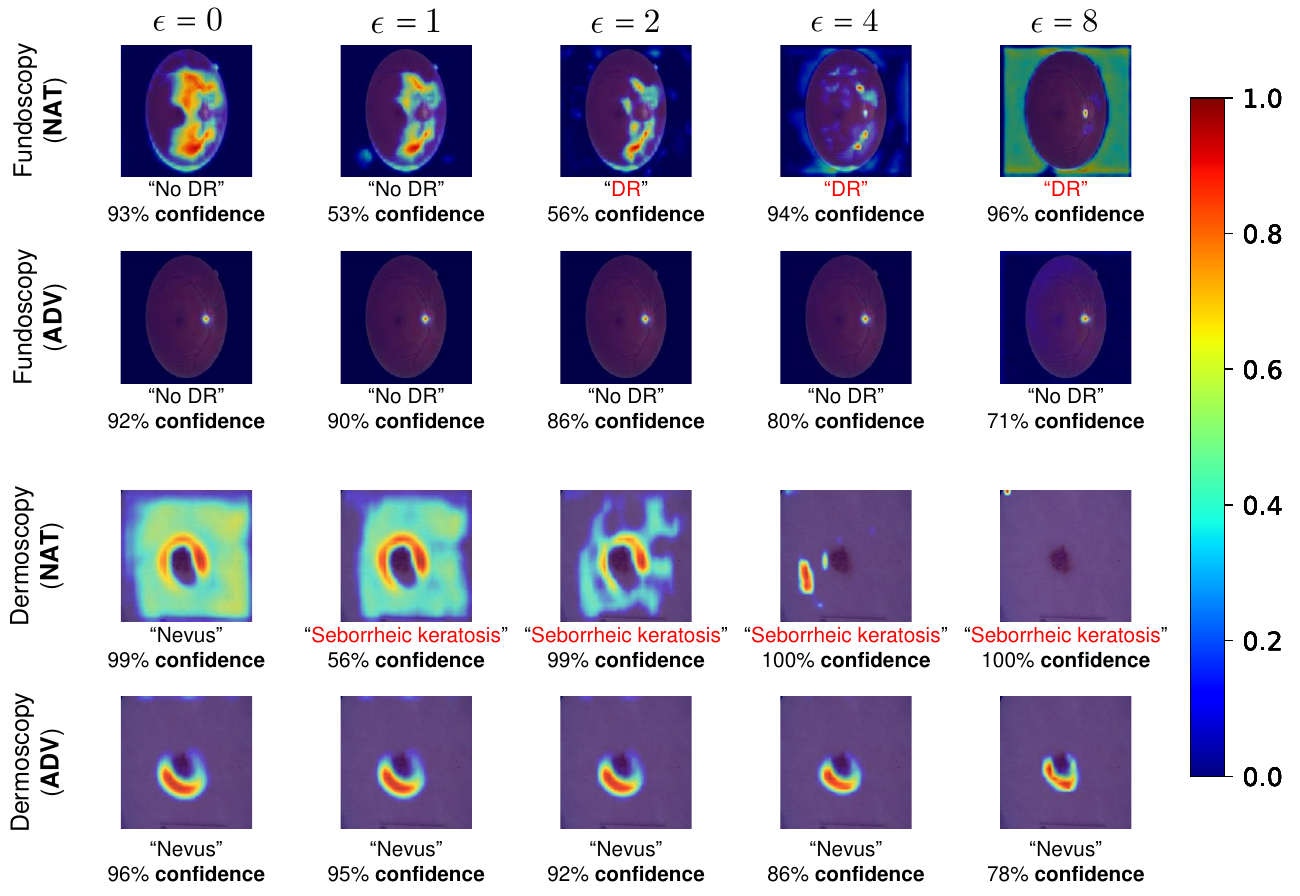}
	\vspace{-10pt}
	\caption{Heat-map visualization of medical adversarial examples under diverse attack strengths $\epsilon$ corresponding to NATurally (\textbf{NAT}) and ADVersarially (\textbf{ADV}) trained classification models.}
	\vspace{-10pt}
	\label{fig:4}
	
\end{figure}

\subsubsection{Visualization}
In general, adversaries can lead to severe mistakes during the inference stage of target medical imaging models. To better represent the negative impact of adversarial examples, we utilize Gradient-weighted Class Activation Mapping (Grad-CAM) \cite{selvaraju2017grad} to create heat maps that represent the discriminative region for medical classification (see Figure \ref{fig:4}). It can be seen that naturally trained models are extremely susceptible to adversaries, especially for high-strength attacks. The class-discriminative region for the naturally trained model varied greatly with the increase of attack strength ($\epsilon$). In comparison, the adversarially trained model can achieve robust predictions and also keep the class-discriminative region the same as the attack intensity increases.

\vspace{-3mm}
\section{Challenges and Future Directions}
\label{sec:6}
Despite the exceptional success of adversarial machine learning for medical image analysis, there still remain several challenges that are worthy of exploration. We also summarize some ongoing or future research directions below:

\noindent
\textbf{Evaluation benchmarks.}
Existing medical adversarial attack and defense methods are primarily based on customized evaluation metrics and settings, which might lead to difficulties in efficacy and efficiency assessments. Furthermore, both adversarial attack and defense are required to be realistic for clinical practice. Hence, for the sake of fair comparison and practical application, the community has to acknowledge a unified and systematical evaluation benchmark. A feasible solution is to establish an open-ended standard benchmark for robustness evaluation in the context of biomedical image analysis. Accordingly, the adversarial robustness benchmark in the medical field can follow the classical robustness benchmark for natural imaging, \textit{i.e.}, RobustBench \cite{CroceASDFCM021}. In the meantime, adversarial attacks need to be adaptive for standardized evaluation, which means that attacks are tailored for a given defense method \cite{tramer2020adaptive}. In this paper, we establish a unified benchmark for adversarial training in the context of biomedical image analysis to facilitate future research.

\noindent
\textbf{Trade-off.}
The trade-off between performance on legitimate and adversarial examples has been widely explored in the natural imaging field \cite{zhang2019theoretically, RadeM22, dong2023enemy}. The enhancement of robust accuracy inevitably induces a decrease in performance on clean examples. We also discovered such a phenomenon for adversarially trained medical diagnosis models in this survey. However, biomedical applications are usually required to be sensitive to the precision of diagnosis, especially in clinical practice. The ideal medical models should achieve excellent performance on clean examples and also remain robust against potential attacks. Therefore, finding a balance between clean and robust performance for medical diagnosis models can be an important research topic in the future. Moreover, we also observe that adversarially trained medical models can sometimes obtain the same or better performance on legitimate medical examples than regularly trained models. While an explicit index to balance accuracy and robustness remains an open research question, our results highlight the potential for tailored adversarial training strategies with a small perturbation radius in medical imaging tasks that can optimize both objectives without suffering from the trade-off issue. We believe it will be possible to further improve the adversarial robustness of medical models without compromising the clean accuracy.

\noindent
\textbf{Computational efficiency.}
Current medical adversarial defense methods mostly require an expensive computational cost, particularly for adversarial training. In general, the time cost of adversarial training can be dozens of times that of natural training due to the expensive cost of adversary generation, resulting in impediments to practical medical applications. Several fast adversarial training variants spare no effort to efficiently construct robustness in the natural imaging domain \cite{shafahi2019adversarial, andriushchenko2020understanding}, whereas there is rare research focusing on medical imaging tasks. Hence, it is necessary to put effort into efficient adversarial defense for medical image analysis. In addition, reducing computational costs can facilitate the defense within reach of organizations with modest computing resources to serve the community better.

\noindent
\textbf{Defense tailored for medical image analysis.}
Existing methods primarily transfer adversarial defense approaches for natural images directly to the medical imaging domain. Nevertheless, there remain very few defense methods tailored for medical images to enhance the adversarial robustness of computer-aided diagnosis systems. Unlike natural examples with large distribution variances, biomedical examples usually share a specific biological structure and a unified distribution, which can be further utilized as priors to construct robustness for medical image analysis systems. In the meantime, we also discover that the inner adversary generation strategy during adversarial training has a more significant impact on the robustness establishment for computer-aided diagnosis models than for natural models. Therefore, we argue that adversarial defense for the medical imaging domain has the potential to be improved by incorporating prior knowledge in biomedical images. 

\noindent
\textbf{Potential ethical issues.}
As we mentioned earlier, the performance of adversarially trained models on clean examples slightly decreases with the drastic enhancement of adversarial robustness. In other words, establishing adversarial robustness will inevitably induce a performance drop on legitimate examples. This is common in the field of natural images but raises a clinical ethical dilemma for medical imaging analysis: is a defense against potential adversarial threats more important than an accurate diagnosis on clean medical examples? In other words, how do we weigh the clean and robust accuracy of medical image analysis systems? The trade-off phenomenon and also inevitable clean accuracy drop are still unresolved issues for defense on clean images. Fortunately, we discover that adversarial training can enhance the robustness without losing natural performance for several medical imaging tasks and modalities. This also demonstrates that it is possible to simultaneously enhance adversarial robustness and also the natural performance of medical image analysis systems in the near future. Furthermore, threats of adversarial examples might also pose a crisis of confidence related to computer-aided diagnosis systems. Hence, it is essential to establish reliable and interpretable computer-aided diagnosis systems to build trust with potential users under real-world scenarios.

\vspace{-3mm}
\section{Conclusion}
In this work, we present a detailed survey of adversarial attack and defense methods for medical image analysis, which is driven by a systematic taxonomy in terms of the application scenario. This survey also incorporates a unified framework with a comprehensive analysis of different types of attack and defense methods in the context of medical images. In addition, we establish a benchmark for adversarially trained diagnosis models under various scenarios to facilitate future research. Finally, we point out representative challenges and promising future research directions in this domain. We hope this survey can further attract new efforts towards better interpretability and application of adversarial machine learning in the medical field.


\vspace{-2mm}
\begin{acks}
\vspace{-1mm}
This work was supported by the National Natural Science Foundation of China (No. 12326618, 62072482, and 62202403), Hong Kong Innovation and Technology Fund (Project No. ITS/028/21FP and MHP/002/22), and Shenzhen Science and Technology Innovation Committee Fund (Project No. SGDX20210823103201011), and the Project of Guangdong Provincial Key Laboratory of Information Security Technology (Grant No. 2023B1212060026).
\end{acks}

{
\bibliographystyle{ACM-Reference-Format}
\bibliography{egbib}


\begin{thebibliography}{226}


\ifx \showCODEN    \undefined \def \showCODEN     #1{\unskip}     \fi
\ifx \showDOI      \undefined \def \showDOI       #1{#1}\fi
\ifx \showISBNx    \undefined \def \showISBNx     #1{\unskip}     \fi
\ifx \showISBNxiii \undefined \def \showISBNxiii  #1{\unskip}     \fi
\ifx \showISSN     \undefined \def \showISSN      #1{\unskip}     \fi
\ifx \showLCCN     \undefined \def \showLCCN      #1{\unskip}     \fi
\ifx \shownote     \undefined \def \shownote      #1{#1}          \fi
\ifx \showarticletitle \undefined \def \showarticletitle #1{#1}   \fi
\ifx \showURL      \undefined \def \showURL       {\relax}        \fi
\providecommand\bibfield[2]{#2}
\providecommand\bibinfo[2]{#2}
\providecommand\natexlab[1]{#1}
\providecommand\showeprint[2][]{arXiv:#2}

\bibitem[Ahmed et~al\mbox{.}(2022)]%
        {ahmed2022failure_A47}
\bibfield{author}{\bibinfo{person}{Sabeen Ahmed}, \bibinfo{person}{Dimah Dera},
  \bibinfo{person}{Saud~Ul Hassan}, \bibinfo{person}{Nidhal Bouaynaya}, {and}
  \bibinfo{person}{Ghulam Rasool}.} \bibinfo{year}{2022}\natexlab{}.
\newblock \showarticletitle{Failure Detection in Deep Neural Networks for
  Medical Imaging}.
\newblock \bibinfo{journal}{\emph{Frontiers in Medical Technology}}
  \bibinfo{volume}{4} (\bibinfo{year}{2022}).
\newblock


\bibitem[Alatalo et~al\mbox{.}(2022)]%
        {alatalo2022detecting_D63}
\bibfield{author}{\bibinfo{person}{Janne Alatalo}, \bibinfo{person}{Tuomo
  Sipola}, {and} \bibinfo{person}{Tero Kokkonen}.}
  \bibinfo{year}{2022}\natexlab{}.
\newblock \showarticletitle{Detecting One-Pixel Attacks Using Variational
  Autoencoders}. In \bibinfo{booktitle}{\emph{World Conference on Information
  Systems and Technologies}}. Springer, \bibinfo{pages}{611--623}.
\newblock


\bibitem[Aldahdooh et~al\mbox{.}(2022)]%
        {aldahdooh2022adversarial}
\bibfield{author}{\bibinfo{person}{Ahmed Aldahdooh}, \bibinfo{person}{Wassim
  Hamidouche}, \bibinfo{person}{Sid~Ahmed Fezza}, {and}
  \bibinfo{person}{Olivier D{\'e}forges}.} \bibinfo{year}{2022}\natexlab{}.
\newblock \showarticletitle{Adversarial example detection for DNN models: A
  review and experimental comparison}.
\newblock \bibinfo{journal}{\emph{Artificial Intelligence Review}}
  (\bibinfo{year}{2022}).
\newblock


\bibitem[Allyn et~al\mbox{.}(2020)]%
        {allyn2020adversarial_A6}
\bibfield{author}{\bibinfo{person}{J{\'e}r{\^o}me Allyn},
  \bibinfo{person}{Nicolas Allou}, \bibinfo{person}{Charles Vidal},
  \bibinfo{person}{Am{\'e}lie Renou}, {and} \bibinfo{person}{Cyril Ferdynus}.}
  \bibinfo{year}{2020}\natexlab{}.
\newblock \showarticletitle{Adversarial attack on deep learning-based
  dermatoscopic image recognition systems: Risk of misdiagnosis due to
  undetectable image perturbations}.
\newblock \bibinfo{journal}{\emph{Medicine}} \bibinfo{volume}{99},
  \bibinfo{number}{50} (\bibinfo{year}{2020}).
\newblock


\bibitem[Almalik et~al\mbox{.}(2022)]%
        {almalik2022self_D42}
\bibfield{author}{\bibinfo{person}{Faris Almalik}, \bibinfo{person}{Mohammad
  Yaqub}, {and} \bibinfo{person}{Karthik Nandakumar}.}
  \bibinfo{year}{2022}\natexlab{}.
\newblock \showarticletitle{Self-Ensembling Vision Transformer (SEViT) for
  Robust Medical Image Classification}. In \bibinfo{booktitle}{\emph{MICCAI}}.
  \bibinfo{pages}{376--386}.
\newblock


\bibitem[Alzubaidi et~al\mbox{.}(2021)]%
        {alzubaidi2021role}
\bibfield{author}{\bibinfo{person}{Mahmood Alzubaidi},
  \bibinfo{person}{Haider~Dhia Zubaydi}, \bibinfo{person}{Ali~Abdulqader
  Bin-Salem}, \bibinfo{person}{Alaa~A Abd-Alrazaq}, \bibinfo{person}{Arfan
  Ahmed}, {and} \bibinfo{person}{Mowafa Househ}.}
  \bibinfo{year}{2021}\natexlab{}.
\newblock \showarticletitle{Role of deep learning in early detection of
  COVID-19: Scoping review}.
\newblock \bibinfo{journal}{\emph{Computer Methods and Programs in Biomedicine
  Update}}  \bibinfo{volume}{1} (\bibinfo{year}{2021}),
  \bibinfo{pages}{100025}.
\newblock


\bibitem[Anand et~al\mbox{.}(2020)]%
        {anand2020self_D48}
\bibfield{author}{\bibinfo{person}{Deepak Anand}, \bibinfo{person}{Darshan
  Tank}, \bibinfo{person}{Harshvardhan Tibrewal}, {and} \bibinfo{person}{Amit
  Sethi}.} \bibinfo{year}{2020}\natexlab{}.
\newblock \showarticletitle{Self-supervision vs. transfer learning: robust
  biomedical image analysis against adversarial attacks}. In
  \bibinfo{booktitle}{\emph{IEEE ISBI}}. \bibinfo{pages}{1159--1163}.
\newblock


\bibitem[Andriushchenko et~al\mbox{.}(2020)]%
        {andriushchenko2020square}
\bibfield{author}{\bibinfo{person}{Maksym Andriushchenko},
  \bibinfo{person}{Francesco Croce}, \bibinfo{person}{Nicolas Flammarion},
  {and} \bibinfo{person}{Matthias Hein}.} \bibinfo{year}{2020}\natexlab{}.
\newblock \showarticletitle{Square attack: a query-efficient black-box
  adversarial attack via random search}. In \bibinfo{booktitle}{\emph{European
  Conference on Computer Vision}}.
\newblock


\bibitem[Andriushchenko and Flammarion(2020)]%
        {andriushchenko2020understanding}
\bibfield{author}{\bibinfo{person}{Maksym Andriushchenko} {and}
  \bibinfo{person}{Nicolas Flammarion}.} \bibinfo{year}{2020}\natexlab{}.
\newblock \showarticletitle{Understanding and improving fast adversarial
  training}.
\newblock \bibinfo{journal}{\emph{Advances in Neural Information Processing
  Systems}}  \bibinfo{volume}{33} (\bibinfo{year}{2020}),
  \bibinfo{pages}{16048--16059}.
\newblock


\bibitem[Apostolidis and Papakostas(2021)]%
        {apostolidis2021survey}
\bibfield{author}{\bibinfo{person}{Kyriakos~D Apostolidis} {and}
  \bibinfo{person}{George~A Papakostas}.} \bibinfo{year}{2021}\natexlab{}.
\newblock \showarticletitle{A survey on adversarial deep learning robustness in
  medical image analysis}.
\newblock \bibinfo{journal}{\emph{Electronics}} \bibinfo{volume}{10},
  \bibinfo{number}{17} (\bibinfo{year}{2021}), \bibinfo{pages}{2132}.
\newblock


\bibitem[Apostolidis and Papakostas(2022)]%
        {apostolidis2022digital_A21}
\bibfield{author}{\bibinfo{person}{Kyriakos~D Apostolidis} {and}
  \bibinfo{person}{George~A Papakostas}.} \bibinfo{year}{2022}\natexlab{}.
\newblock \showarticletitle{Digital Watermarking as an Adversarial Attack on
  Medical Image Analysis with Deep Learning}.
\newblock \bibinfo{journal}{\emph{Journal of Imaging}} \bibinfo{volume}{8},
  \bibinfo{number}{6} (\bibinfo{year}{2022}), \bibinfo{pages}{155}.
\newblock


\bibitem[Asgari~Taghanaki et~al\mbox{.}(2018)]%
        {asgari2018vulnerability_D17}
\bibfield{author}{\bibinfo{person}{Saeid Asgari~Taghanaki},
  \bibinfo{person}{Arkadeep Das}, {and} \bibinfo{person}{Ghassan Hamarneh}.}
  \bibinfo{year}{2018}\natexlab{}.
\newblock \showarticletitle{Vulnerability analysis of chest X-ray image
  classification against adversarial attacks}.
\newblock In \bibinfo{booktitle}{\emph{Understanding and interpreting machine
  learning in medical image computing applications}}. \bibinfo{pages}{87--94}.
\newblock


\bibitem[Badrinarayanan et~al\mbox{.}(2017)]%
        {badrinarayanan2017segnet}
\bibfield{author}{\bibinfo{person}{Vijay Badrinarayanan}, \bibinfo{person}{Alex
  Kendall}, {and} \bibinfo{person}{Roberto Cipolla}.}
  \bibinfo{year}{2017}\natexlab{}.
\newblock \showarticletitle{Segnet: A deep convolutional encoder-decoder
  architecture for image segmentation}.
\newblock \bibinfo{journal}{\emph{IEEE TPAMI}} \bibinfo{volume}{39},
  \bibinfo{number}{12} (\bibinfo{year}{2017}), \bibinfo{pages}{2481--2495}.
\newblock


\bibitem[Bai et~al\mbox{.}(2023)]%
        {bai2023query}
\bibfield{author}{\bibinfo{person}{Yang Bai}, \bibinfo{person}{Yisen Wang},
  \bibinfo{person}{Yuyuan Zeng}, \bibinfo{person}{Yong Jiang}, {and}
  \bibinfo{person}{Shu-Tao Xia}.} \bibinfo{year}{2023}\natexlab{}.
\newblock \showarticletitle{Query efficient black-box adversarial attack on
  deep neural networks}.
\newblock \bibinfo{journal}{\emph{Pattern Recognition}}  \bibinfo{volume}{133}
  (\bibinfo{year}{2023}), \bibinfo{pages}{109037}.
\newblock


\bibitem[Bharath~Kumar et~al\mbox{.}(2022a)]%
        {bharath2022analysis_A41}
\bibfield{author}{\bibinfo{person}{DP Bharath~Kumar}, \bibinfo{person}{Nanda
  Kumar}, \bibinfo{person}{Snofy~D Dunston}, {and} \bibinfo{person}{V Rajam}.}
  \bibinfo{year}{2022}\natexlab{a}.
\newblock \showarticletitle{Analysis of the Impact of White Box Adversarial
  Attacks in ResNet While Classifying Retinal Fundus Images}. In
  \bibinfo{booktitle}{\emph{International Conference on Computational
  Intelligence in Data Science}}.
\newblock


\bibitem[Bharath~Kumar et~al\mbox{.}(2022b)]%
        {bharath2022analysis_D58}
\bibfield{author}{\bibinfo{person}{DP Bharath~Kumar}, \bibinfo{person}{Nanda
  Kumar}, \bibinfo{person}{Snofy~D Dunston}, {and}
  \bibinfo{person}{V~Mary~Anita Rajam}.} \bibinfo{year}{2022}\natexlab{b}.
\newblock \showarticletitle{Analysis of the Impact of White Box Adversarial
  Attacks in ResNet While Classifying Retinal Fundus Images}. In
  \bibinfo{booktitle}{\emph{Computational Intelligence in Data Science}}.
  \bibinfo{pages}{162--175}.
\newblock


\bibitem[BMS et~al\mbox{.}(2022)]%
        {bms2022analysis_A36}
\bibfield{author}{\bibinfo{person}{Pranava~Raman BMS}, \bibinfo{person}{V
  Anusree}, \bibinfo{person}{B Sreeratcha}, \bibinfo{person}{Preeti~Krishnaveni
  Ra}, \bibinfo{person}{Snofy~D Dunston}, {et~al\mbox{.}}}
  \bibinfo{year}{2022}\natexlab{}.
\newblock \showarticletitle{Analysis of the Effect of Black Box Adversarial
  Attacks on Medical Image Classification Models}. In
  \bibinfo{booktitle}{\emph{Intelligent Computing Instrumentation and Control
  Technologies}}. \bibinfo{pages}{528--531}.
\newblock


\bibitem[Bortsova et~al\mbox{.}(2021a)]%
        {bortsova2021adversarial_A9}
\bibfield{author}{\bibinfo{person}{Gerda Bortsova}, \bibinfo{person}{Florian
  Dubost}, \bibinfo{person}{Laurens Hogeweg}, \bibinfo{person}{Ioannis
  Katramados}, {and} \bibinfo{person}{Marleen de Bruijne}.}
  \bibinfo{year}{2021}\natexlab{a}.
\newblock \showarticletitle{Adversarial Heart Attack: Neural Networks Fooled to
  Segment Heart Symbols in Chest X-Ray Images}.
\newblock \bibinfo{journal}{\emph{arXiv preprint arXiv:2104.00139}}
  (\bibinfo{year}{2021}).
\newblock


\bibitem[Bortsova et~al\mbox{.}(2021b)]%
        {bortsova2021adversarial_A18}
\bibfield{author}{\bibinfo{person}{Gerda Bortsova}, \bibinfo{person}{Cristina
  Gonz{\'a}lez-Gonzalo}, \bibinfo{person}{Suzanne~C Wetstein},
  \bibinfo{person}{Florian Dubost}, \bibinfo{person}{Ioannis Katramados},
  \bibinfo{person}{Laurens Hogeweg}, \bibinfo{person}{Bart Liefers},
  \bibinfo{person}{Bram van Ginneken}, \bibinfo{person}{Josien~PW Pluim},
  \bibinfo{person}{Mitko Veta}, {et~al\mbox{.}}}
  \bibinfo{year}{2021}\natexlab{b}.
\newblock \showarticletitle{Adversarial attack vulnerability of medical image
  analysis systems: Unexplored factors}.
\newblock \bibinfo{journal}{\emph{Medical Image Analysis}}
  \bibinfo{volume}{73} (\bibinfo{year}{2021}), \bibinfo{pages}{102141}.
\newblock


\bibitem[Bustos et~al\mbox{.}(2020)]%
        {bustos2020padchest}
\bibfield{author}{\bibinfo{person}{Aurelia Bustos}, \bibinfo{person}{Antonio
  Pertusa}, \bibinfo{person}{Jose-Maria Salinas}, {and} \bibinfo{person}{Maria
  De~La Iglesia-Vaya}.} \bibinfo{year}{2020}\natexlab{}.
\newblock \showarticletitle{Padchest: A large chest x-ray image dataset with
  multi-label annotated reports}.
\newblock \bibinfo{journal}{\emph{Medical image analysis}}
  \bibinfo{volume}{66} (\bibinfo{year}{2020}), \bibinfo{pages}{101797}.
\newblock


\bibitem[Byra et~al\mbox{.}(2020)]%
        {byra2020adversarial_A11}
\bibfield{author}{\bibinfo{person}{Michal Byra}, \bibinfo{person}{Grzegorz
  Styczynski}, \bibinfo{person}{Cezary Szmigielski}, \bibinfo{person}{Piotr
  Kalinowski}, \bibinfo{person}{Lukasz Michalowski}, \bibinfo{person}{Rafal
  Paluszkiewicz}, \bibinfo{person}{Bogna Ziarkiewicz-Wroblewska},
  \bibinfo{person}{Krzysztof Zieniewicz}, {and} \bibinfo{person}{Andrzej
  Nowicki}.} \bibinfo{year}{2020}\natexlab{}.
\newblock \showarticletitle{Adversarial attacks on deep learning models for
  fatty liver disease classification by modification of ultrasound image
  reconstruction method}. In \bibinfo{booktitle}{\emph{IEEE International
  Ultrasonics Symposium (IUS)}}. \bibinfo{pages}{1--4}.
\newblock


\bibitem[Carannante et~al\mbox{.}(2021)]%
        {carannante2021trustworthy_D54}
\bibfield{author}{\bibinfo{person}{Giuseppina Carannante},
  \bibinfo{person}{Dimah Dera}, \bibinfo{person}{Nidhal~C Bouaynaya},
  \bibinfo{person}{Ghulam Rasool}, {and} \bibinfo{person}{Hassan~M
  Fathallah-Shaykh}.} \bibinfo{year}{2021}\natexlab{}.
\newblock \showarticletitle{Trustworthy Medical Segmentation with Uncertainty
  Estimation}.
\newblock \bibinfo{journal}{\emph{arXiv preprint arXiv:2111.05978}}
  (\bibinfo{year}{2021}).
\newblock


\bibitem[Carlini and Wagner(2017a)]%
        {carlini2017adversarial}
\bibfield{author}{\bibinfo{person}{Nicholas Carlini} {and}
  \bibinfo{person}{David Wagner}.} \bibinfo{year}{2017}\natexlab{a}.
\newblock \showarticletitle{Adversarial examples are not easily detected:
  Bypassing ten detection methods}. In \bibinfo{booktitle}{\emph{ACM workshop
  on artificial intelligence and security}}. \bibinfo{pages}{3--14}.
\newblock


\bibitem[Carlini and Wagner(2017b)]%
        {carlini2017towards}
\bibfield{author}{\bibinfo{person}{Nicholas Carlini} {and}
  \bibinfo{person}{David Wagner}.} \bibinfo{year}{2017}\natexlab{b}.
\newblock \showarticletitle{Towards evaluating the robustness of neural
  networks}. In \bibinfo{booktitle}{\emph{IEEE SP}}.
\newblock


\bibitem[Chen et~al\mbox{.}(2022)]%
        {chen2022enhancing_D31}
\bibfield{author}{\bibinfo{person}{Chen Chen}, \bibinfo{person}{Chen Qin},
  \bibinfo{person}{Cheng Ouyang}, \bibinfo{person}{Zeju Li},
  \bibinfo{person}{Shuo Wang}, \bibinfo{person}{Huaqi Qiu},
  \bibinfo{person}{Liang Chen}, \bibinfo{person}{Giacomo Tarroni},
  \bibinfo{person}{Wenjia Bai}, {and} \bibinfo{person}{Daniel Rueckert}.}
  \bibinfo{year}{2022}\natexlab{}.
\newblock \showarticletitle{Enhancing MR image segmentation with realistic
  adversarial data augmentation}.
\newblock \bibinfo{journal}{\emph{Medical Image Analysis}}
  \bibinfo{volume}{82} (\bibinfo{year}{2022}), \bibinfo{pages}{102597}.
\newblock


\bibitem[Chen et~al\mbox{.}(2020b)]%
        {chen2020realistic_D14}
\bibfield{author}{\bibinfo{person}{Chen Chen}, \bibinfo{person}{Chen Qin},
  \bibinfo{person}{Huaqi Qiu}, \bibinfo{person}{Cheng Ouyang},
  \bibinfo{person}{Shuo Wang}, \bibinfo{person}{Liang Chen},
  \bibinfo{person}{Giacomo Tarroni}, \bibinfo{person}{Wenjia Bai}, {and}
  \bibinfo{person}{Daniel Rueckert}.} \bibinfo{year}{2020}\natexlab{b}.
\newblock \showarticletitle{Realistic adversarial data augmentation for MR
  image segmentation}. In \bibinfo{booktitle}{\emph{MICCAI}}.
  \bibinfo{pages}{667--677}.
\newblock


\bibitem[Chen et~al\mbox{.}(2023)]%
        {chen2023frequency_A2023_4}
\bibfield{author}{\bibinfo{person}{Fang Chen}, \bibinfo{person}{Jian Wang},
  \bibinfo{person}{Han Liu}, \bibinfo{person}{Wentao Kong},
  \bibinfo{person}{Zhe Zhao}, \bibinfo{person}{Longfei Ma},
  \bibinfo{person}{Hongen Liao}, {and} \bibinfo{person}{Daoqiang Zhang}.}
  \bibinfo{year}{2023}\natexlab{}.
\newblock \showarticletitle{Frequency constraint-based adversarial attack on
  deep neural networks for medical image classification}.
\newblock \bibinfo{journal}{\emph{Computers in Biology and Medicine}}
  \bibinfo{volume}{164} (\bibinfo{year}{2023}), \bibinfo{pages}{107248}.
\newblock


\bibitem[Chen et~al\mbox{.}(2020a)]%
        {chen2020hopskipjumpattack}
\bibfield{author}{\bibinfo{person}{Jianbo Chen}, \bibinfo{person}{Michael~I
  Jordan}, {and} \bibinfo{person}{Martin~J Wainwright}.}
  \bibinfo{year}{2020}\natexlab{a}.
\newblock \showarticletitle{Hopskipjumpattack: A query-efficient decision-based
  attack}. In \bibinfo{booktitle}{\emph{IEEE Symposium on Security and Privacy
  (SP)}}. \bibinfo{pages}{1277--1294}.
\newblock


\bibitem[Chen et~al\mbox{.}(2021a)]%
        {chen2021adversarial_A30}
\bibfield{author}{\bibinfo{person}{Jiasong Chen}, \bibinfo{person}{Linchen
  Qian}, \bibinfo{person}{Timur Urakov}, \bibinfo{person}{Weiyong Gu}, {and}
  \bibinfo{person}{Liang Liang}.} \bibinfo{year}{2021}\natexlab{a}.
\newblock \showarticletitle{Adversarial robustness study of convolutional
  neural network for lumbar disk shape reconstruction from MR images}. In
  \bibinfo{booktitle}{\emph{Medical Imaging}}.
\newblock


\bibitem[Chen et~al\mbox{.}(2019)]%
        {chen2019intelligent_A13}
\bibfield{author}{\bibinfo{person}{Liang Chen}, \bibinfo{person}{Paul Bentley},
  \bibinfo{person}{Kensaku Mori}, \bibinfo{person}{Kazunari Misawa},
  \bibinfo{person}{Michitaka Fujiwara}, {and} \bibinfo{person}{Daniel
  Rueckert}.} \bibinfo{year}{2019}\natexlab{}.
\newblock \showarticletitle{Intelligent image synthesis to attack a
  segmentation CNN using adversarial learning}. In
  \bibinfo{booktitle}{\emph{International Workshop on Simulation and Synthesis
  in Medical Imaging}}. \bibinfo{pages}{90--99}.
\newblock


\bibitem[Chen et~al\mbox{.}(2021b)]%
        {chen2021enhancing_D33}
\bibfield{author}{\bibinfo{person}{Lun Chen}, \bibinfo{person}{Lu Zhao}, {and}
  \bibinfo{person}{Calvin Yu-Chian Chen}.} \bibinfo{year}{2021}\natexlab{b}.
\newblock \showarticletitle{Enhancing adversarial defense for medical image
  analysis systems with pruning and attention mechanism}.
\newblock \bibinfo{journal}{\emph{Medical Physics}}  \bibinfo{volume}{48}
  (\bibinfo{year}{2021}).
\newblock


\bibitem[Chen et~al\mbox{.}(2017)]%
        {chen2017zoo}
\bibfield{author}{\bibinfo{person}{Pin-Yu Chen}, \bibinfo{person}{Huan Zhang},
  \bibinfo{person}{Yash Sharma}, \bibinfo{person}{Jinfeng Yi}, {and}
  \bibinfo{person}{Cho-Jui Hsieh}.} \bibinfo{year}{2017}\natexlab{}.
\newblock \showarticletitle{Zoo: Zeroth order optimization based black-box
  attacks to deep neural networks without training substitute models}. In
  \bibinfo{booktitle}{\emph{ACM workshop on artificial intelligence and
  security}}. \bibinfo{pages}{15--26}.
\newblock


\bibitem[Chen et~al\mbox{.}(2024a)]%
        {chen2024rae_A2024_2}
\bibfield{author}{\bibinfo{person}{Zhen Chen}, \bibinfo{person}{Xiuli Chai},
  \bibinfo{person}{Zhihua Gan}, \bibinfo{person}{Binjie Wang}, {and}
  \bibinfo{person}{Yushu Zhang}.} \bibinfo{year}{2024}\natexlab{a}.
\newblock \showarticletitle{RAE-VWP: A Reversible Adversarial Example-Based
  Privacy and Copyright Protection Method of Medical Images for Internet of
  Medical Things}.
\newblock \bibinfo{journal}{\emph{IEEE Internet of Things Journal}}
  (\bibinfo{year}{2024}).
\newblock


\bibitem[Chen et~al\mbox{.}(2024b)]%
        {chen2024content}
\bibfield{author}{\bibinfo{person}{Zhaoyu Chen}, \bibinfo{person}{Bo Li},
  \bibinfo{person}{Shuang Wu}, \bibinfo{person}{Kaixun Jiang},
  \bibinfo{person}{Shouhong Ding}, {and} \bibinfo{person}{Wenqiang Zhang}.}
  \bibinfo{year}{2024}\natexlab{b}.
\newblock \showarticletitle{Content-based unrestricted adversarial attack}.
\newblock \bibinfo{journal}{\emph{Advances in Neural Information Processing
  Systems}}  \bibinfo{volume}{36} (\bibinfo{year}{2024}).
\newblock


\bibitem[Cheng and Ji(2020)]%
        {cheng2020adversarial_A2}
\bibfield{author}{\bibinfo{person}{Guohua Cheng} {and} \bibinfo{person}{Hongli
  Ji}.} \bibinfo{year}{2020}\natexlab{}.
\newblock \showarticletitle{Adversarial perturbation on MRI modalities in brain
  tumor segmentation}.
\newblock \bibinfo{journal}{\emph{IEEE Access}}  \bibinfo{volume}{8}
  (\bibinfo{year}{2020}).
\newblock


\bibitem[Cheng et~al\mbox{.}(2020a)]%
        {cheng2020addressing_D18}
\bibfield{author}{\bibinfo{person}{Kaiyang Cheng}, \bibinfo{person}{Francesco
  Caliv{\'a}}, \bibinfo{person}{Rutwik Shah}, \bibinfo{person}{Misung Han},
  \bibinfo{person}{Sharmila Majumdar}, {and} \bibinfo{person}{Valentina
  Pedoia}.} \bibinfo{year}{2020}\natexlab{a}.
\newblock \showarticletitle{Addressing the false negative problem of deep
  learning MRI reconstruction models by adversarial attacks and robust
  training}. In \bibinfo{booktitle}{\emph{Medical Imaging with Deep Learning}}.
  PMLR, \bibinfo{pages}{121--135}.
\newblock


\bibitem[Cheng et~al\mbox{.}(2019)]%
        {cheng2019improving}
\bibfield{author}{\bibinfo{person}{Shuyu Cheng}, \bibinfo{person}{Yinpeng
  Dong}, \bibinfo{person}{Tianyu Pang}, \bibinfo{person}{Hang Su}, {and}
  \bibinfo{person}{Jun Zhu}.} \bibinfo{year}{2019}\natexlab{}.
\newblock \showarticletitle{Improving black-box adversarial attacks with a
  transfer-based prior}.
\newblock \bibinfo{journal}{\emph{Advances in neural information processing
  systems}}  \bibinfo{volume}{32} (\bibinfo{year}{2019}).
\newblock


\bibitem[Cheng et~al\mbox{.}(2020b)]%
        {cheng2020adversarial_A43}
\bibfield{author}{\bibinfo{person}{Yupeng Cheng}, \bibinfo{person}{Felix
  Juefei-Xu}, \bibinfo{person}{Qing Guo}, \bibinfo{person}{Huazhu Fu},
  \bibinfo{person}{Xiaofei Xie}, \bibinfo{person}{Shang-Wei Lin},
  \bibinfo{person}{Weisi Lin}, {and} \bibinfo{person}{Yang Liu}.}
  \bibinfo{year}{2020}\natexlab{b}.
\newblock \showarticletitle{Adversarial exposure attack on diabetic retinopathy
  imagery}.
\newblock \bibinfo{journal}{\emph{arXiv preprint arXiv:2009.09231}}
  (\bibinfo{year}{2020}).
\newblock


\bibitem[Chiem et~al\mbox{.}(2007)]%
        {chiem2007novel}
\bibfield{author}{\bibinfo{person}{Andy Chiem}, \bibinfo{person}{Adel
  Al-Jumaily}, {and} \bibinfo{person}{Rami~N Khushaba}.}
  \bibinfo{year}{2007}\natexlab{}.
\newblock \showarticletitle{A novel hybrid system for skin lesion detection}.
  In \bibinfo{booktitle}{\emph{International Conference on Intelligent Sensors,
  Sensor Networks and Information}}.
\newblock


\bibitem[Chowdhury et~al\mbox{.}(2020)]%
        {chowdhury2020can}
\bibfield{author}{\bibinfo{person}{Muhammad~EH Chowdhury},
  \bibinfo{person}{Tawsifur Rahman}, \bibinfo{person}{Amith Khandakar},
  \bibinfo{person}{Rashid Mazhar}, \bibinfo{person}{Muhammad~Abdul Kadir},
  \bibinfo{person}{Zaid~Bin Mahbub}, \bibinfo{person}{Khandakar~Reajul Islam},
  \bibinfo{person}{Muhammad~Salman Khan}, \bibinfo{person}{Atif Iqbal},
  \bibinfo{person}{Nasser Al~Emadi}, {et~al\mbox{.}}}
  \bibinfo{year}{2020}\natexlab{}.
\newblock \showarticletitle{Can AI help in screening viral and COVID-19
  pneumonia?}
\newblock \bibinfo{journal}{\emph{IEEE Access}}  \bibinfo{volume}{8}
  (\bibinfo{year}{2020}), \bibinfo{pages}{132665--132676}.
\newblock


\bibitem[Codella et~al\mbox{.}(2018)]%
        {codella2018skin}
\bibfield{author}{\bibinfo{person}{Noel~CF Codella}, \bibinfo{person}{David
  Gutman}, \bibinfo{person}{M~Emre Celebi}, \bibinfo{person}{Brian Helba},
  \bibinfo{person}{Michael~A Marchetti}, \bibinfo{person}{Stephen~W Dusza},
  \bibinfo{person}{Aadi Kalloo}, \bibinfo{person}{Konstantinos Liopyris},
  \bibinfo{person}{Nabin Mishra}, \bibinfo{person}{Harald Kittler},
  {et~al\mbox{.}}} \bibinfo{year}{2018}\natexlab{}.
\newblock \showarticletitle{Skin lesion analysis toward melanoma detection: A
  challenge at the 2017 international symposium on biomedical imaging (isbi),
  hosted by the international skin imaging collaboration (isic)}. In
  \bibinfo{booktitle}{\emph{IEEE ISBI}}. \bibinfo{pages}{168--172}.
\newblock


\bibitem[Croce et~al\mbox{.}(2021)]%
        {CroceASDFCM021}
\bibfield{author}{\bibinfo{person}{Francesco Croce}, \bibinfo{person}{Maksym
  Andriushchenko}, \bibinfo{person}{Vikash Sehwag}, \bibinfo{person}{Edoardo
  Debenedetti}, \bibinfo{person}{Nicolas Flammarion}, \bibinfo{person}{Mung
  Chiang}, \bibinfo{person}{Prateek Mittal}, {and} \bibinfo{person}{Matthias
  Hein}.} \bibinfo{year}{2021}\natexlab{}.
\newblock \showarticletitle{RobustBench: a standardized adversarial robustness
  benchmark}. In \bibinfo{booktitle}{\emph{NeurIPS}}.
\newblock


\bibitem[Croce and Hein(2020)]%
        {croce2020reliable}
\bibfield{author}{\bibinfo{person}{Francesco Croce} {and}
  \bibinfo{person}{Matthias Hein}.} \bibinfo{year}{2020}\natexlab{}.
\newblock \showarticletitle{Reliable evaluation of adversarial robustness with
  an ensemble of diverse parameter-free attacks}. In
  \bibinfo{booktitle}{\emph{International Conference on Machine Learning}}.
  \bibinfo{pages}{2206--2216}.
\newblock


\bibitem[Cui et~al\mbox{.}(2021)]%
        {cui2021deattack_A48}
\bibfield{author}{\bibinfo{person}{Xiangxiang Cui}, \bibinfo{person}{Shi
  Chang}, \bibinfo{person}{Chen Li}, \bibinfo{person}{Bin Kong},
  \bibinfo{person}{Lihua Tian}, \bibinfo{person}{Hongqiang Wang},
  \bibinfo{person}{Peng Huang}, \bibinfo{person}{Meng Yang},
  \bibinfo{person}{Yenan Wu}, {and} \bibinfo{person}{Zhongyu Li}.}
  \bibinfo{year}{2021}\natexlab{}.
\newblock \showarticletitle{DEAttack: A differential evolution based attack
  method for the robustness evaluation of medical image segmentation}.
\newblock \bibinfo{journal}{\emph{Neurocomputing}}  \bibinfo{volume}{465}
  (\bibinfo{year}{2021}), \bibinfo{pages}{38--52}.
\newblock


\bibitem[Daanouni et~al\mbox{.}(2022)]%
        {daanouni2022nsl_D64}
\bibfield{author}{\bibinfo{person}{Othmane Daanouni}, \bibinfo{person}{Bouchaib
  Cherradi}, {and} \bibinfo{person}{Amal Tmiri}.}
  \bibinfo{year}{2022}\natexlab{}.
\newblock \showarticletitle{NSL-MHA-CNN: A Novel CNN Architecture for Robust
  Diabetic Retinopathy Prediction Against Adversarial Attacks}.
\newblock \bibinfo{journal}{\emph{IEEE Access}}  \bibinfo{volume}{10}
  (\bibinfo{year}{2022}).
\newblock


\bibitem[Dai et~al\mbox{.}(2022)]%
        {dai2022deep}
\bibfield{author}{\bibinfo{person}{Tao Dai}, \bibinfo{person}{Yan Feng},
  \bibinfo{person}{Bin Chen}, \bibinfo{person}{Jian Lu}, {and}
  \bibinfo{person}{Shu-Tao Xia}.} \bibinfo{year}{2022}\natexlab{}.
\newblock \showarticletitle{Deep image prior based defense against adversarial
  examples}.
\newblock \bibinfo{journal}{\emph{Pattern Recognition}}  \bibinfo{volume}{122}
  (\bibinfo{year}{2022}), \bibinfo{pages}{108249}.
\newblock


\bibitem[Dai et~al\mbox{.}(2023)]%
        {dai2023improving_D2023_1}
\bibfield{author}{\bibinfo{person}{Yinyao Dai}, \bibinfo{person}{Yaguan Qian},
  \bibinfo{person}{Fang Lu}, \bibinfo{person}{Bin Wang},
  \bibinfo{person}{Zhaoquan Gu}, \bibinfo{person}{Wei Wang},
  \bibinfo{person}{Jian Wan}, {and} \bibinfo{person}{Yanchun Zhang}.}
  \bibinfo{year}{2023}\natexlab{}.
\newblock \showarticletitle{Improving adversarial robustness of medical imaging
  systems via adding global attention noise}.
\newblock \bibinfo{journal}{\emph{Computers in Biology and Medicine}}
  \bibinfo{volume}{164} (\bibinfo{year}{2023}), \bibinfo{pages}{107251}.
\newblock


\bibitem[Daza et~al\mbox{.}(2021)]%
        {daza2021towards_D27}
\bibfield{author}{\bibinfo{person}{Laura Daza}, \bibinfo{person}{Juan~C
  P{\'e}rez}, {and} \bibinfo{person}{Pablo Arbel{\'a}ez}.}
  \bibinfo{year}{2021}\natexlab{}.
\newblock \showarticletitle{Towards robust general medical image segmentation}.
  In \bibinfo{booktitle}{\emph{MICCAI}}.
\newblock


\bibitem[de~Aguiar et~al\mbox{.}(2022)]%
        {de2022evaluation_A40}
\bibfield{author}{\bibinfo{person}{Erikson~J de Aguiar},
  \bibinfo{person}{Karem~D Marcomini}, \bibinfo{person}{Felipe~A Quirino},
  \bibinfo{person}{Marco~A Gutierrez}, \bibinfo{person}{Caetano Traina~Jr},
  {and} \bibinfo{person}{Agma~JM Traina}.} \bibinfo{year}{2022}\natexlab{}.
\newblock \showarticletitle{Evaluation of the impact of physical adversarial
  attacks on deep learning models for classifying covid cases}. In
  \bibinfo{booktitle}{\emph{Medical Imaging 2022: Computer-Aided Diagnosis}}.
\newblock


\bibitem[Decenci{\`e}re et~al\mbox{.}(2014)]%
        {decenciere2014feedback}
\bibfield{author}{\bibinfo{person}{Etienne Decenci{\`e}re},
  \bibinfo{person}{Xiwei Zhang}, \bibinfo{person}{Guy Cazuguel},
  \bibinfo{person}{Bruno Lay}, \bibinfo{person}{B{\'e}atrice Cochener},
  \bibinfo{person}{Caroline Trone}, \bibinfo{person}{Philippe Gain},
  \bibinfo{person}{Richard Ordonez}, \bibinfo{person}{Pascale Massin},
  \bibinfo{person}{Ali Erginay}, {et~al\mbox{.}}}
  \bibinfo{year}{2014}\natexlab{}.
\newblock \showarticletitle{Feedback on a publicly distributed image database:
  the Messidor database}.
\newblock \bibinfo{journal}{\emph{Image Analysis \& Stereology}}
  (\bibinfo{year}{2014}).
\newblock


\bibitem[Demontis et~al\mbox{.}(2019)]%
        {demontis2019adversarial}
\bibfield{author}{\bibinfo{person}{Ambra Demontis}, \bibinfo{person}{Marco
  Melis}, \bibinfo{person}{Maura Pintor}, \bibinfo{person}{Matthew Jagielski},
  \bibinfo{person}{Battista Biggio}, \bibinfo{person}{Alina Oprea},
  \bibinfo{person}{Cristina Nita-Rotaru}, {and} \bibinfo{person}{Fabio Roli}.}
  \bibinfo{year}{2019}\natexlab{}.
\newblock \showarticletitle{Why do adversarial attacks transfer? explaining
  transferability of evasion and poisoning attacks}. In
  \bibinfo{booktitle}{\emph{USENIX security symposium}}.
  \bibinfo{pages}{321--338}.
\newblock


\bibitem[Ding et~al\mbox{.}(2023)]%
        {ding2023vith_A2023_2}
\bibfield{author}{\bibinfo{person}{Weiping Ding}, \bibinfo{person}{Chuansheng
  Liu}, \bibinfo{person}{Jiashuang Huang}, \bibinfo{person}{Chun Cheng}, {and}
  \bibinfo{person}{Hengrong Ju}.} \bibinfo{year}{2023}\natexlab{}.
\newblock \showarticletitle{ViTH-RFG: Vision Transformer Hashing with Residual
  Fuzzy Generation for Targeted Attack in Medical Image Retrieval}.
\newblock \bibinfo{journal}{\emph{IEEE TFS}} (\bibinfo{year}{2023}).
\newblock


\bibitem[Diyasa et~al\mbox{.}(2021)]%
        {diyasa2021grasping_A38}
\bibfield{author}{\bibinfo{person}{I~Gede Susrama~Mas Diyasa},
  \bibinfo{person}{Radical~Rakhman Wahid}, {and} \bibinfo{person}{Brilian~Putra
  Amiruddin}.} \bibinfo{year}{2021}\natexlab{}.
\newblock \showarticletitle{Grasping Adversarial Attacks on Deep Convolutional
  Neural Networks for Cholangiocarcinoma Classification}. In
  \bibinfo{booktitle}{\emph{International Conference on e-Health and
  Bioengineering (EHB)}}. \bibinfo{pages}{1--4}.
\newblock


\bibitem[Dong et~al\mbox{.}(2023a)]%
        {dong2023enemy}
\bibfield{author}{\bibinfo{person}{Junhao Dong}, \bibinfo{person}{Seyed-Mohsen
  Moosavi-Dezfooli}, \bibinfo{person}{Jianhuang Lai}, {and}
  \bibinfo{person}{Xiaohua Xie}.} \bibinfo{year}{2023}\natexlab{a}.
\newblock \showarticletitle{The enemy of my enemy is my friend: Exploring
  inverse adversaries for improving adversarial training}. In
  \bibinfo{booktitle}{\emph{Proceedings of the IEEE/CVF Conference on Computer
  Vision and Pattern Recognition}}. \bibinfo{pages}{24678--24687}.
\newblock


\bibitem[Dong et~al\mbox{.}(2023b)]%
        {dong2022restricted}
\bibfield{author}{\bibinfo{person}{Junhao Dong}, \bibinfo{person}{Yuan Wang},
  \bibinfo{person}{Jianhuang Lai}, {and} \bibinfo{person}{Xiaohua Xie}.}
  \bibinfo{year}{2023}\natexlab{b}.
\newblock \showarticletitle{Restricted Black-Box Adversarial Attack Against
  DeepFake Face Swapping}.
\newblock \bibinfo{journal}{\emph{IEEE Transactions on Information Forensics
  and Security}}  \bibinfo{volume}{18} (\bibinfo{year}{2023}),
  \bibinfo{pages}{2596--2608}.
\newblock


\bibitem[Dong et~al\mbox{.}(2022)]%
        {dong2022improving}
\bibfield{author}{\bibinfo{person}{Junhao Dong}, \bibinfo{person}{Yuan Wang},
  \bibinfo{person}{Jian-Huang Lai}, {and} \bibinfo{person}{Xiaohua Xie}.}
  \bibinfo{year}{2022}\natexlab{}.
\newblock \showarticletitle{Improving adversarially robust few-shot image
  classification with generalizable representations}. In
  \bibinfo{booktitle}{\emph{Proceedings of the IEEE/CVF CVPR}}.
  \bibinfo{pages}{9025--9034}.
\newblock


\bibitem[Dong and Xie(2021)]%
        {dong2021visually}
\bibfield{author}{\bibinfo{person}{Junhao Dong} {and} \bibinfo{person}{Xiaohua
  Xie}.} \bibinfo{year}{2021}\natexlab{}.
\newblock \showarticletitle{Visually maintained image disturbance against
  deepfake face swapping}. In \bibinfo{booktitle}{\emph{2021 IEEE International
  Conference on Multimedia and Expo (ICME)}}. IEEE, \bibinfo{pages}{1--6}.
\newblock


\bibitem[FDA(2018)]%
        {FDA2018}
\bibfield{author}{\bibinfo{person}{FDA}.} \bibinfo{year}{2018}\natexlab{}.
\newblock \bibinfo{title}{FDA Permits Marketing of Artificial
  Intelligence-based Device to Detect Certain Diabetes-related Eye Problems}.
\newblock \bibinfo{howpublished}{[Online]}.
\newblock
\newblock
\shownote{https://www.fda.gov/news-events/press-announcements/fda-permits-marketing-artificial-intelligence-based-device-detect-certain-diabetes-related-eye}.


\bibitem[Finlayson et~al\mbox{.}(2019)]%
        {finlayson2019adversarial_A3}
\bibfield{author}{\bibinfo{person}{Samuel~G Finlayson}, \bibinfo{person}{John~D
  Bowers}, \bibinfo{person}{Joichi Ito}, \bibinfo{person}{Jonathan~L Zittrain},
  \bibinfo{person}{Andrew~L Beam}, {and} \bibinfo{person}{Isaac~S Kohane}.}
  \bibinfo{year}{2019}\natexlab{}.
\newblock \showarticletitle{Adversarial attacks on medical machine learning}.
\newblock \bibinfo{journal}{\emph{Science}} \bibinfo{volume}{363},
  \bibinfo{number}{6433} (\bibinfo{year}{2019}), \bibinfo{pages}{1287--1289}.
\newblock


\bibitem[Foote et~al\mbox{.}(2021)]%
        {foote2021now_A39}
\bibfield{author}{\bibinfo{person}{Alex Foote}, \bibinfo{person}{Amina Asif},
  \bibinfo{person}{Ayesha Azam}, \bibinfo{person}{Tim Marshall-Cox},
  \bibinfo{person}{Nasir Rajpoot}, {and} \bibinfo{person}{Fayyaz Minhas}.}
  \bibinfo{year}{2021}\natexlab{}.
\newblock \showarticletitle{Now You See It, Now You Dont: Adversarial
  Vulnerabilities in Computational Pathology}.
\newblock \bibinfo{journal}{\emph{arXiv preprint arXiv:2106.08153}}
  (\bibinfo{year}{2021}).
\newblock


\bibitem[Ge et~al\mbox{.}(2017)]%
        {ge2017skin}
\bibfield{author}{\bibinfo{person}{Zongyuan Ge}, \bibinfo{person}{Sergey
  Demyanov}, \bibinfo{person}{Rajib Chakravorty}, \bibinfo{person}{Adrian
  Bowling}, {and} \bibinfo{person}{Rahil Garnavi}.}
  \bibinfo{year}{2017}\natexlab{}.
\newblock \showarticletitle{Skin disease recognition using deep saliency
  features and multimodal learning of dermoscopy and clinical images}. In
  \bibinfo{booktitle}{\emph{Medical image computing and computer-assisted
  intervention}}.
\newblock


\bibitem[Ghaffari~Laleh et~al\mbox{.}(2022)]%
        {ghaffari2022adversarial_D28}
\bibfield{author}{\bibinfo{person}{Narmin Ghaffari~Laleh},
  \bibinfo{person}{Daniel Truhn}, \bibinfo{person}{Gregory~Patrick Veldhuizen},
  \bibinfo{person}{Tianyu Han}, \bibinfo{person}{Marko van Treeck},
  \bibinfo{person}{Roman~D Buelow}, \bibinfo{person}{Rupert Langer},
  \bibinfo{person}{Bastian Dislich}, \bibinfo{person}{Peter Boor},
  \bibinfo{person}{Volkmar Schulz}, {et~al\mbox{.}}}
  \bibinfo{year}{2022}\natexlab{}.
\newblock \showarticletitle{Adversarial attacks and adversarial robustness in
  computational pathology}.
\newblock \bibinfo{journal}{\emph{Nature Communications}}  \bibinfo{volume}{13}
  (\bibinfo{year}{2022}).
\newblock


\bibitem[Goldblum et~al\mbox{.}(2020)]%
        {goldblum2020adversarially}
\bibfield{author}{\bibinfo{person}{Micah Goldblum}, \bibinfo{person}{Liam
  Fowl}, \bibinfo{person}{Soheil Feizi}, {and} \bibinfo{person}{Tom
  Goldstein}.} \bibinfo{year}{2020}\natexlab{}.
\newblock \showarticletitle{Adversarially robust distillation}. In
  \bibinfo{booktitle}{\emph{AAAI}}.
\newblock


\bibitem[Gondal et~al\mbox{.}(2017)]%
        {gondal2017weakly}
\bibfield{author}{\bibinfo{person}{Waleed~M Gondal}, \bibinfo{person}{Jan~M
  K{\"o}hler}, \bibinfo{person}{Ren{\'e} Grzeszick}, \bibinfo{person}{Gernot~A
  Fink}, {and} \bibinfo{person}{Michael Hirsch}.}
  \bibinfo{year}{2017}\natexlab{}.
\newblock \showarticletitle{Weakly-supervised localization of diabetic
  retinopathy lesions in retinal fundus images}. In
  \bibinfo{booktitle}{\emph{IEEE ICIP}}. \bibinfo{pages}{2069--2073}.
\newblock


\bibitem[Gongye et~al\mbox{.}(2020)]%
        {gongye2020new_A24}
\bibfield{author}{\bibinfo{person}{Cheng Gongye}, \bibinfo{person}{Hongjia Li},
  \bibinfo{person}{Xiang Zhang}, \bibinfo{person}{Majid Sabbagh},
  \bibinfo{person}{Geng Yuan}, \bibinfo{person}{Xue Lin},
  \bibinfo{person}{Thomas Wahl}, {and} \bibinfo{person}{Yunsi Fei}.}
  \bibinfo{year}{2020}\natexlab{}.
\newblock \showarticletitle{New passive and active attacks on deep neural
  networks in medical applications}. In \bibinfo{booktitle}{\emph{IEEE
  International Conference on Computer-Aided Design}}. \bibinfo{pages}{1--9}.
\newblock


\bibitem[Goodfellow et~al\mbox{.}(2015)]%
        {GoodfellowSS14}
\bibfield{author}{\bibinfo{person}{Ian~J. Goodfellow},
  \bibinfo{person}{Jonathon Shlens}, {and} \bibinfo{person}{Christian
  Szegedy}.} \bibinfo{year}{2015}\natexlab{}.
\newblock \showarticletitle{Explaining and Harnessing Adversarial Examples}. In
  \bibinfo{booktitle}{\emph{ICLR}}.
\newblock


\bibitem[Gougeh(2021)]%
        {gougeh2021adversarial_A23}
\bibfield{author}{\bibinfo{person}{Reza~Amini Gougeh}.}
  \bibinfo{year}{2021}\natexlab{}.
\newblock \showarticletitle{How Adversarial attacks affect Deep Neural Networks
  Detecting COVID-19?}
\newblock  (\bibinfo{year}{2021}).
\newblock


\bibitem[Gupta and Pal(2022)]%
        {gupta2022vulnerability_D56}
\bibfield{author}{\bibinfo{person}{Debashis Gupta} {and}
  \bibinfo{person}{Biprodip Pal}.} \bibinfo{year}{2022}\natexlab{}.
\newblock \showarticletitle{Vulnerability Analysis and robust training with
  additive noise for FGSM attack on transfer learning-based brain tumor
  detection from MRI}. In \bibinfo{booktitle}{\emph{International Conference on
  Big Data, IoT, and Machine Learning}}. \bibinfo{pages}{103--114}.
\newblock


\bibitem[Han et~al\mbox{.}(2021)]%
        {han2021advancing_D35}
\bibfield{author}{\bibinfo{person}{Tianyu Han}, \bibinfo{person}{Sven
  Nebelung}, \bibinfo{person}{Federico Pedersoli}, \bibinfo{person}{Markus
  Zimmermann}, \bibinfo{person}{Maximilian Schulze-Hagen},
  \bibinfo{person}{Michael Ho}, \bibinfo{person}{Christoph Haarburger},
  \bibinfo{person}{Fabian Kiessling}, \bibinfo{person}{Christiane Kuhl},
  \bibinfo{person}{Volkmar Schulz}, {et~al\mbox{.}}}
  \bibinfo{year}{2021}\natexlab{}.
\newblock \showarticletitle{Advancing diagnostic performance and clinical
  usability of neural networks via adversarial training and dual batch
  normalization}.
\newblock \bibinfo{journal}{\emph{Nature communications}} \bibinfo{volume}{12},
  \bibinfo{number}{1} (\bibinfo{year}{2021}), \bibinfo{pages}{4315}.
\newblock


\bibitem[He et~al\mbox{.}(2016)]%
        {he2016deep}
\bibfield{author}{\bibinfo{person}{Kaiming He}, \bibinfo{person}{Xiangyu
  Zhang}, \bibinfo{person}{Shaoqing Ren}, {and} \bibinfo{person}{Jian Sun}.}
  \bibinfo{year}{2016}\natexlab{}.
\newblock \showarticletitle{Deep residual learning for image recognition}. In
  \bibinfo{booktitle}{\emph{IEEE CVPR}}. \bibinfo{pages}{770--778}.
\newblock


\bibitem[He et~al\mbox{.}(2019)]%
        {he2019non_D16}
\bibfield{author}{\bibinfo{person}{Xiang He}, \bibinfo{person}{Sibei Yang},
  \bibinfo{person}{Guanbin Li}, \bibinfo{person}{Haofeng Li},
  \bibinfo{person}{Huiyou Chang}, {and} \bibinfo{person}{Yizhou Yu}.}
  \bibinfo{year}{2019}\natexlab{}.
\newblock \showarticletitle{Non-local context encoder: Robust biomedical image
  segmentation against adversarial attacks}. In
  \bibinfo{booktitle}{\emph{AAAI}}, Vol.~\bibinfo{volume}{33}.
  \bibinfo{pages}{8417--8424}.
\newblock


\bibitem[Hirano et~al\mbox{.}(2021)]%
        {hirano2021universal_A7_D55}
\bibfield{author}{\bibinfo{person}{Hokuto Hirano}, \bibinfo{person}{Akinori
  Minagi}, {and} \bibinfo{person}{Kazuhiro Takemoto}.}
  \bibinfo{year}{2021}\natexlab{}.
\newblock \showarticletitle{Universal adversarial attacks on deep neural
  networks for medical image classification}.
\newblock \bibinfo{journal}{\emph{BMC medical imaging}} \bibinfo{volume}{21},
  \bibinfo{number}{1} (\bibinfo{year}{2021}), \bibinfo{pages}{1--13}.
\newblock


\bibitem[Hu et~al\mbox{.}(2024)]%
        {hu2024protecting}
\bibfield{author}{\bibinfo{person}{Lei Hu}, \bibinfo{person}{Dawei Zhou},
  \bibinfo{person}{Jiahua Xu}, \bibinfo{person}{Cheng Lu}, \bibinfo{person}{Chu
  Han}, \bibinfo{person}{Zhenwei Shi}, \bibinfo{person}{Qikui Zhu},
  \bibinfo{person}{Xinbo Gao}, \bibinfo{person}{Nannan Wang}, {and}
  \bibinfo{person}{Zaiyi Liu}.} \bibinfo{year}{2024}\natexlab{}.
\newblock \showarticletitle{Protecting Prostate Cancer Classification from
  Rectal Artifacts via Targeted Adversarial Training}.
\newblock \bibinfo{journal}{\emph{IEEE Journal of Biomedical and Health
  Informatics}} (\bibinfo{year}{2024}).
\newblock


\bibitem[Hu et~al\mbox{.}(2022)]%
        {hu2022adversarial_D29}
\bibfield{author}{\bibinfo{person}{Lei Hu}, \bibinfo{person}{Da-Wei Zhou},
  \bibinfo{person}{Xiang-Yu Guo}, \bibinfo{person}{Wen-Hao Xu},
  \bibinfo{person}{Li-Ming Wei}, {and} \bibinfo{person}{Jun-Gong Zhao}.}
  \bibinfo{year}{2022}\natexlab{}.
\newblock \showarticletitle{Adversarial training for prostate cancer
  classification using magnetic resonance imaging}.
\newblock \bibinfo{journal}{\emph{Quantitative Imaging in Medicine and
  Surgery}} \bibinfo{volume}{12}, \bibinfo{number}{6} (\bibinfo{year}{2022}),
  \bibinfo{pages}{3276--3287}.
\newblock


\bibitem[Huang et~al\mbox{.}(2018)]%
        {huang2018some_D21}
\bibfield{author}{\bibinfo{person}{Yixing Huang}, \bibinfo{person}{Tobias
  W{\"u}rfl}, \bibinfo{person}{Katharina Breininger}, \bibinfo{person}{Ling
  Liu}, \bibinfo{person}{G{\"u}nter Lauritsch}, {and} \bibinfo{person}{Andreas
  Maier}.} \bibinfo{year}{2018}\natexlab{}.
\newblock \showarticletitle{Some investigations on robustness of deep learning
  in limited angle tomography}. In \bibinfo{booktitle}{\emph{MICCAI}}.
  \bibinfo{pages}{145--153}.
\newblock


\bibitem[Huq and Pervin(2020)]%
        {huq2020analysis_D47}
\bibfield{author}{\bibinfo{person}{Aminul Huq} {and}
  \bibinfo{person}{Mst~Tasnim Pervin}.} \bibinfo{year}{2020}\natexlab{}.
\newblock \showarticletitle{Analysis of adversarial attacks on skin cancer
  recognition}. In \bibinfo{booktitle}{\emph{ICoDSA}}.
\newblock


\bibitem[Ilyas et~al\mbox{.}(2019a)]%
        {ilyas2019prior}
\bibfield{author}{\bibinfo{person}{Andrew Ilyas}, \bibinfo{person}{Logan
  Engstrom}, {and} \bibinfo{person}{Aleksander Madry}.}
  \bibinfo{year}{2019}\natexlab{a}.
\newblock \showarticletitle{Prior Convictions: Black-box Adversarial Attacks
  with Bandits and Priors}. In \bibinfo{booktitle}{\emph{ICLR}}.
\newblock


\bibitem[Ilyas et~al\mbox{.}(2019b)]%
        {ilyas2019adversarial}
\bibfield{author}{\bibinfo{person}{Andrew Ilyas}, \bibinfo{person}{Shibani
  Santurkar}, \bibinfo{person}{Dimitris Tsipras}, \bibinfo{person}{Logan
  Engstrom}, \bibinfo{person}{Brandon Tran}, {and} \bibinfo{person}{Aleksander
  Madry}.} \bibinfo{year}{2019}\natexlab{b}.
\newblock \showarticletitle{Adversarial examples are not bugs, they are
  features}.
\newblock \bibinfo{journal}{\emph{Advances in Neural Information Processing
  Systems}}  \bibinfo{volume}{32} (\bibinfo{year}{2019}).
\newblock


\bibitem[Irvin et~al\mbox{.}(2019)]%
        {irvin2019chexpert}
\bibfield{author}{\bibinfo{person}{Jeremy Irvin}, \bibinfo{person}{Pranav
  Rajpurkar}, \bibinfo{person}{Michael Ko}, \bibinfo{person}{Yifan Yu},
  \bibinfo{person}{Silviana Ciurea-Ilcus}, \bibinfo{person}{Chris Chute},
  \bibinfo{person}{Henrik Marklund}, \bibinfo{person}{Behzad Haghgoo},
  \bibinfo{person}{Robyn Ball}, \bibinfo{person}{Katie Shpanskaya},
  {et~al\mbox{.}}} \bibinfo{year}{2019}\natexlab{}.
\newblock \showarticletitle{Chexpert: A large chest radiograph dataset with
  uncertainty labels and expert comparison}. In
  \bibinfo{booktitle}{\emph{Proceedings of the AAAI conference on artificial
  intelligence}}, Vol.~\bibinfo{volume}{33}. \bibinfo{pages}{590--597}.
\newblock


\bibitem[Jaiswal et~al\mbox{.}(2022)]%
        {jaiswal2022ros_D66}
\bibfield{author}{\bibinfo{person}{Ajay Jaiswal}, \bibinfo{person}{Kumar
  Ashutosh}, \bibinfo{person}{Justin~F Rousseau}, \bibinfo{person}{Yifan Peng},
  \bibinfo{person}{Zhangyang Wang}, {and} \bibinfo{person}{Ying Ding}.}
  \bibinfo{year}{2022}\natexlab{}.
\newblock \showarticletitle{RoS-KD: A Robust Stochastic Knowledge Distillation
  Approach for Noisy Medical Imaging}.
\newblock \bibinfo{journal}{\emph{arXiv preprint arXiv:2210.08388}}
  (\bibinfo{year}{2022}).
\newblock


\bibitem[Jandial et~al\mbox{.}(2019)]%
        {jandial2019advgan++}
\bibfield{author}{\bibinfo{person}{Surgan Jandial}, \bibinfo{person}{Puneet
  Mangla}, \bibinfo{person}{Sakshi Varshney}, {and} \bibinfo{person}{Vineeth
  Balasubramanian}.} \bibinfo{year}{2019}\natexlab{}.
\newblock \showarticletitle{Advgan++: Harnessing latent layers for adversary
  generation}. In \bibinfo{booktitle}{\emph{IEEE International Conference on
  Computer Vision Workshops}}.
\newblock


\bibitem[Jiang et~al\mbox{.}(2019)]%
        {jiang2019black}
\bibfield{author}{\bibinfo{person}{Linxi Jiang}, \bibinfo{person}{Xingjun Ma},
  \bibinfo{person}{Shaoxiang Chen}, \bibinfo{person}{James Bailey}, {and}
  \bibinfo{person}{Yu-Gang Jiang}.} \bibinfo{year}{2019}\natexlab{}.
\newblock \showarticletitle{Black-box adversarial attacks on video recognition
  models}. In \bibinfo{booktitle}{\emph{ACM International Conference on
  Multimedia}}. \bibinfo{pages}{864--872}.
\newblock


\bibitem[Joel et~al\mbox{.}(2021)]%
        {joel2021adversarial_A33}
\bibfield{author}{\bibinfo{person}{Marina~Z Joel}, \bibinfo{person}{Sachin
  Umrao}, \bibinfo{person}{Enoch Chang}, \bibinfo{person}{Rachel Choi},
  \bibinfo{person}{Daniel Yang}, \bibinfo{person}{James Duncan},
  \bibinfo{person}{Antonio Omuro}, \bibinfo{person}{Roy Herbst},
  \bibinfo{person}{Harlan Krumholz}, \bibinfo{person}{Sanjay Aneja},
  {et~al\mbox{.}}} \bibinfo{year}{2021}\natexlab{}.
\newblock \showarticletitle{Adversarial attack vulnerability of deep learning
  models for oncologic images}.
\newblock \bibinfo{journal}{\emph{medRxiv}} (\bibinfo{year}{2021}).
\newblock


\bibitem[Joel et~al\mbox{.}(2022)]%
        {joel2022using_D32}
\bibfield{author}{\bibinfo{person}{Marina~Z Joel}, \bibinfo{person}{Sachin
  Umrao}, \bibinfo{person}{Enoch Chang}, \bibinfo{person}{Rachel Choi},
  \bibinfo{person}{Daniel~X Yang}, \bibinfo{person}{James~S Duncan},
  \bibinfo{person}{Antonio Omuro}, \bibinfo{person}{Roy Herbst},
  \bibinfo{person}{Harlan~M Krumholz}, {and} \bibinfo{person}{Sanjay Aneja}.}
  \bibinfo{year}{2022}\natexlab{}.
\newblock \showarticletitle{Using Adversarial Images to Assess the Robustness
  of Deep Learning Models Trained on Diagnostic Images in Oncology}.
\newblock \bibinfo{journal}{\emph{JCO Clinical Cancer Informatics}}
  \bibinfo{volume}{6} (\bibinfo{year}{2022}), \bibinfo{pages}{e2100170}.
\newblock


\bibitem[Kansal et~al\mbox{.}(2022)]%
        {kansal2022defending_D51}
\bibfield{author}{\bibinfo{person}{Keshav Kansal}, \bibinfo{person}{P~Sai
  Krishna}, \bibinfo{person}{Parshva~B Jain}, \bibinfo{person}{R Surya},
  \bibinfo{person}{Prasad Honnavalli}, {and} \bibinfo{person}{Sivaraman
  Eswaran}.} \bibinfo{year}{2022}\natexlab{}.
\newblock \showarticletitle{Defending against adversarial attacks on Covid-19
  classifier: A denoiser-based approach}.
\newblock \bibinfo{journal}{\emph{Heliyon}}  \bibinfo{volume}{8}
  (\bibinfo{year}{2022}).
\newblock


\bibitem[Kaviani et~al\mbox{.}(2022)]%
        {kaviani2022adversarial}
\bibfield{author}{\bibinfo{person}{Sara Kaviani}, \bibinfo{person}{Ki~Jin Han},
  {and} \bibinfo{person}{Insoo Sohn}.} \bibinfo{year}{2022}\natexlab{}.
\newblock \showarticletitle{Adversarial attacks and defenses on AI in medical
  imaging informatics: A survey}.
\newblock \bibinfo{journal}{\emph{Expert Systems with Applications}}
  (\bibinfo{year}{2022}), \bibinfo{pages}{116815}.
\newblock


\bibitem[Koga and Takemoto(2021)]%
        {koga2021simple_A29}
\bibfield{author}{\bibinfo{person}{Kazuki Koga} {and} \bibinfo{person}{Kazuhiro
  Takemoto}.} \bibinfo{year}{2021}\natexlab{}.
\newblock \showarticletitle{Simple black-box universal adversarial attacks on
  medical image classification based on deep neural networks}.
\newblock \bibinfo{journal}{\emph{arXiv preprint arXiv:2108.04979}}
  (\bibinfo{year}{2021}).
\newblock


\bibitem[Kotia et~al\mbox{.}(2020)]%
        {kotia2020risk_D49}
\bibfield{author}{\bibinfo{person}{Jai Kotia}, \bibinfo{person}{Adit Kotwal},
  {and} \bibinfo{person}{Rishika Bharti}.} \bibinfo{year}{2020}\natexlab{}.
\newblock \showarticletitle{Risk susceptibility of brain tumor classification
  to adversarial attacks}. In \bibinfo{booktitle}{\emph{International
  Conference on Man-Machine Interactions}}. Springer,
  \bibinfo{pages}{181--187}.
\newblock


\bibitem[Kovalev et~al\mbox{.}(2021)]%
        {kovalev2021biomedical_A51}
\bibfield{author}{\bibinfo{person}{VA Kovalev}, \bibinfo{person}{VA Liauchuk},
  \bibinfo{person}{DM Voynov}, {and} \bibinfo{person}{AV Tuzikov}.}
  \bibinfo{year}{2021}\natexlab{}.
\newblock \showarticletitle{Biomedical Image Recognition in Pulmonology and
  Oncology with the Use of Deep Learning}.
\newblock \bibinfo{journal}{\emph{Pattern Recognition and Image Analysis}}
  (\bibinfo{year}{2021}).
\newblock


\bibitem[Kovalev and Voynov(2019)]%
        {kovalev2019influence_A5}
\bibfield{author}{\bibinfo{person}{Vassili Kovalev} {and}
  \bibinfo{person}{Dmitry Voynov}.} \bibinfo{year}{2019}\natexlab{}.
\newblock \showarticletitle{Influence of Control Parameters and the Size of
  Biomedical Image Datasets on the Success of Adversarial Attacks}.
\newblock \bibinfo{journal}{\emph{arXiv preprint arXiv:1904.06964}}
  (\bibinfo{year}{2019}).
\newblock


\bibitem[Kulkarni and Bhambani(2021)]%
        {kulkarni2021kryptonite_A28}
\bibfield{author}{\bibinfo{person}{Yogesh Kulkarni} {and}
  \bibinfo{person}{Krisha Bhambani}.} \bibinfo{year}{2021}\natexlab{}.
\newblock \showarticletitle{Kryptonite: An Adversarial Attack Using Regional
  Focus}. In \bibinfo{booktitle}{\emph{International Conference on Applied
  Cryptography and Network Security}}. Springer, \bibinfo{pages}{463--481}.
\newblock


\bibitem[Kwon and Jeong(2022)]%
        {kwon2022advu_A35}
\bibfield{author}{\bibinfo{person}{Hyun Kwon} {and} \bibinfo{person}{Jongwook
  Jeong}.} \bibinfo{year}{2022}\natexlab{}.
\newblock \showarticletitle{AdvU-Net: Generating Adversarial Example Based on
  Medical Image and Targeting U-Net Model}.
\newblock \bibinfo{journal}{\emph{Journal of Sensors}} (\bibinfo{year}{2022}).
\newblock


\bibitem[Lal et~al\mbox{.}(2021)]%
        {lal2021adversarial_D3}
\bibfield{author}{\bibinfo{person}{Sheeba Lal}, \bibinfo{person}{Saeed~Ur
  Rehman}, \bibinfo{person}{Jamal~Hussain Shah}, \bibinfo{person}{Talha Meraj},
  \bibinfo{person}{Hafiz~Tayyab Rauf}, \bibinfo{person}{Robertas
  Dama{\v{s}}evi{\v{c}}ius}, \bibinfo{person}{Mazin~Abed Mohammed}, {and}
  \bibinfo{person}{Karrar~Hameed Abdulkareem}.}
  \bibinfo{year}{2021}\natexlab{}.
\newblock \showarticletitle{Adversarial attack and defence through adversarial
  training and feature fusion for diabetic retinopathy recognition}.
\newblock \bibinfo{journal}{\emph{Sensors}} \bibinfo{volume}{21},
  \bibinfo{number}{11} (\bibinfo{year}{2021}), \bibinfo{pages}{3922}.
\newblock


\bibitem[Le et~al\mbox{.}(2022)]%
        {le2022efficient_D65}
\bibfield{author}{\bibinfo{person}{Linh~D Le}, \bibinfo{person}{Huazhu Fu},
  \bibinfo{person}{Xinxing Xu}, \bibinfo{person}{Yong Liu},
  \bibinfo{person}{Yanyu Xu}, \bibinfo{person}{Jiawei Du},
  \bibinfo{person}{Joey~T Zhou}, {and} \bibinfo{person}{Rick Goh}.}
  \bibinfo{year}{2022}\natexlab{}.
\newblock \showarticletitle{An Efficient Defending Mechanism Against Image
  Attacking on Medical Image Segmentation Models}. In
  \bibinfo{booktitle}{\emph{MICCAI Workshop}}.
\newblock


\bibitem[Lee and Kim(2023)]%
        {lee2023robust}
\bibfield{author}{\bibinfo{person}{Minjong Lee} {and} \bibinfo{person}{Dongwoo
  Kim}.} \bibinfo{year}{2023}\natexlab{}.
\newblock \showarticletitle{Robust evaluation of diffusion-based adversarial
  purification}. In \bibinfo{booktitle}{\emph{Proceedings of the IEEE/CVF
  International Conference on Computer Vision}}. \bibinfo{pages}{134--144}.
\newblock


\bibitem[Lee et~al\mbox{.}(2024)]%
        {lee2024adversarial_A2024_1}
\bibfield{author}{\bibinfo{person}{Woonghee Lee}, \bibinfo{person}{Mingeon Ju},
  \bibinfo{person}{Yura Sim}, \bibinfo{person}{Young~Kul Jung},
  \bibinfo{person}{Tae~Hyung Kim}, {and} \bibinfo{person}{Younghoon Kim}.}
  \bibinfo{year}{2024}\natexlab{}.
\newblock \showarticletitle{Adversarial Attacks on Medical Segmentation Model
  via Transformation of Feature Statistics}.
\newblock \bibinfo{journal}{\emph{Applied Sciences}} \bibinfo{volume}{14},
  \bibinfo{number}{6} (\bibinfo{year}{2024}), \bibinfo{pages}{2576}.
\newblock


\bibitem[Levy et~al\mbox{.}(2022)]%
        {levy2022security_A27}
\bibfield{author}{\bibinfo{person}{Moshe Levy}, \bibinfo{person}{Guy Amit},
  \bibinfo{person}{Yuval Elovici}, {and} \bibinfo{person}{Yisroel Mirsky}.}
  \bibinfo{year}{2022}\natexlab{}.
\newblock \showarticletitle{The Security of Deep Learning Defences for Medical
  Imaging}.
\newblock \bibinfo{journal}{\emph{arXiv preprint arXiv:2201.08661}}
  (\bibinfo{year}{2022}).
\newblock


\bibitem[Li et~al\mbox{.}(2022)]%
        {li2022query_A50}
\bibfield{author}{\bibinfo{person}{Siyuan Li}, \bibinfo{person}{Guangji Huang},
  \bibinfo{person}{Xing Xu}, {and} \bibinfo{person}{Huimin Lu}.}
  \bibinfo{year}{2022}\natexlab{}.
\newblock \showarticletitle{Query-based black-box attack against medical image
  segmentation model}.
\newblock \bibinfo{journal}{\emph{Future Generation Computer Systems}}
  \bibinfo{volume}{133} (\bibinfo{year}{2022}), \bibinfo{pages}{331--337}.
\newblock


\bibitem[Li et~al\mbox{.}(2024)]%
        {li2024dynamic_D2024_4}
\bibfield{author}{\bibinfo{person}{Shuai Li}, \bibinfo{person}{Xiaoguang Ma},
  \bibinfo{person}{Shancheng Jiang}, {and} \bibinfo{person}{Lu Meng}.}
  \bibinfo{year}{2024}\natexlab{}.
\newblock \showarticletitle{Dynamic Perturbation-Adaptive Adversarial Training
  on Medical Image Classification}.
\newblock \bibinfo{journal}{\emph{arXiv preprint arXiv:2403.06798}}
  (\bibinfo{year}{2024}).
\newblock


\bibitem[Li et~al\mbox{.}(2021)]%
        {li2021defending_D1}
\bibfield{author}{\bibinfo{person}{Xin Li}, \bibinfo{person}{Deng Pan}, {and}
  \bibinfo{person}{Dongxiao Zhu}.} \bibinfo{year}{2021}\natexlab{}.
\newblock \showarticletitle{Defending against adversarial attacks on medical
  imaging AI system, classification or detection?}. In
  \bibinfo{booktitle}{\emph{IEEE International Symposium on Biomedical Imaging
  (ISBI)}}.
\newblock


\bibitem[Li and Zhu(2020)]%
        {li2020robust_D24}
\bibfield{author}{\bibinfo{person}{Xin Li} {and} \bibinfo{person}{Dongxiao
  Zhu}.} \bibinfo{year}{2020}\natexlab{}.
\newblock \showarticletitle{Robust detection of adversarial attacks on medical
  images}. In \bibinfo{booktitle}{\emph{International Symposium on Biomedical
  Imaging (ISBI)}}. \bibinfo{pages}{1154--1158}.
\newblock


\bibitem[Li and Liu(2023)]%
        {li2023threat_A2023_5}
\bibfield{author}{\bibinfo{person}{Yang Li} {and} \bibinfo{person}{Shaoying
  Liu}.} \bibinfo{year}{2023}\natexlab{}.
\newblock \showarticletitle{The threat of adversarial attack on a COVID-19 CT
  image-based deep learning system}.
\newblock \bibinfo{journal}{\emph{Bioengineering}} \bibinfo{volume}{10},
  \bibinfo{number}{2} (\bibinfo{year}{2023}), \bibinfo{pages}{194}.
\newblock


\bibitem[Li et~al\mbox{.}(2020)]%
        {li2020anatomical_D40}
\bibfield{author}{\bibinfo{person}{Yi Li}, \bibinfo{person}{Huahong Zhang},
  \bibinfo{person}{Camilo Bermudez}, \bibinfo{person}{Yifan Chen},
  \bibinfo{person}{Bennett~A Landman}, {and} \bibinfo{person}{Yevgeniy
  Vorobeychik}.} \bibinfo{year}{2020}\natexlab{}.
\newblock \showarticletitle{Anatomical context protects deep learning from
  adversarial perturbations in medical imaging}.
\newblock \bibinfo{journal}{\emph{Neurocomputing}} (\bibinfo{year}{2020}).
\newblock


\bibitem[Li et~al\mbox{.}(2019)]%
        {li2019volumetric_D46}
\bibfield{author}{\bibinfo{person}{Yingwei Li}, \bibinfo{person}{Zhuotun Zhu},
  \bibinfo{person}{Yuyin Zhou}, \bibinfo{person}{Yingda Xia},
  \bibinfo{person}{Wei Shen}, \bibinfo{person}{Elliot~K Fishman}, {and}
  \bibinfo{person}{Alan~L Yuille}.} \bibinfo{year}{2019}\natexlab{}.
\newblock \showarticletitle{Volumetric medical image segmentation: a 3D deep
  coarse-to-fine framework and its adversarial examples}.
\newblock In \bibinfo{booktitle}{\emph{Deep Learning and Convolutional Neural
  Networks for Medical Imaging and Clinical Informatics}}.
  \bibinfo{pages}{69--91}.
\newblock


\bibitem[Liu et~al\mbox{.}(2020a)]%
        {liu2020defending_D23}
\bibfield{author}{\bibinfo{person}{Qi Liu}, \bibinfo{person}{Han Jiang},
  \bibinfo{person}{Tao Liu}, \bibinfo{person}{Zihao Liu},
  \bibinfo{person}{Sicheng Li}, \bibinfo{person}{Wujie Wen}, {and}
  \bibinfo{person}{Yiyu Shi}.} \bibinfo{year}{2020}\natexlab{a}.
\newblock \showarticletitle{Defending deep learning-based biomedical image
  segmentation from adversarial attacks: a low-cost frequency refinement
  approach}. In \bibinfo{booktitle}{\emph{MICCAI}}.
\newblock


\bibitem[Liu et~al\mbox{.}(2020b)]%
        {liu2020no_D12}
\bibfield{author}{\bibinfo{person}{Siqi Liu}, \bibinfo{person}{Arnaud
  Arindra~Adiyoso Setio}, \bibinfo{person}{Florin~C Ghesu},
  \bibinfo{person}{Eli Gibson}, \bibinfo{person}{Sasa Grbic},
  \bibinfo{person}{Bogdan Georgescu}, {and} \bibinfo{person}{Dorin Comaniciu}.}
  \bibinfo{year}{2020}\natexlab{b}.
\newblock \showarticletitle{No surprises: Training robust lung nodule detection
  for low-dose CT scans by augmenting with adversarial attacks}.
\newblock \bibinfo{journal}{\emph{IEEE Transactions on Medical Imaging}}
  (\bibinfo{year}{2020}).
\newblock


\bibitem[Liu et~al\mbox{.}(2017)]%
        {LiuCLS17}
\bibfield{author}{\bibinfo{person}{Yanpei Liu}, \bibinfo{person}{Xinyun Chen},
  \bibinfo{person}{Chang Liu}, {and} \bibinfo{person}{Dawn Song}.}
  \bibinfo{year}{2017}\natexlab{}.
\newblock \showarticletitle{Delving into Transferable Adversarial Examples and
  Black-box Attacks}. In \bibinfo{booktitle}{\emph{ICLR}}.
\newblock


\bibitem[Liu et~al\mbox{.}(2019)]%
        {liu2019robustifying_A31}
\bibfield{author}{\bibinfo{person}{Zheng Liu}, \bibinfo{person}{Jinnian Zhang},
  \bibinfo{person}{Varun Jog}, \bibinfo{person}{Po-Ling Loh}, {and}
  \bibinfo{person}{Alan~B McMillan}.} \bibinfo{year}{2019}\natexlab{}.
\newblock \showarticletitle{Robustifying deep networks for image segmentation}.
\newblock \bibinfo{journal}{\emph{arXiv preprint arXiv:1908.00656}}
  (\bibinfo{year}{2019}).
\newblock


\bibitem[Liu et~al\mbox{.}(2021)]%
        {liu2021robustifying_D6}
\bibfield{author}{\bibinfo{person}{Zheng Liu}, \bibinfo{person}{Jinnian Zhang},
  \bibinfo{person}{Varun Jog}, \bibinfo{person}{Po-Ling Loh}, {and}
  \bibinfo{person}{Alan~B McMillan}.} \bibinfo{year}{2021}\natexlab{}.
\newblock \showarticletitle{Robustifying deep networks for medical image
  segmentation}.
\newblock \bibinfo{journal}{\emph{Journal of Digital Imaging}}
  \bibinfo{volume}{34} (\bibinfo{year}{2021}), \bibinfo{pages}{1279--1293}.
\newblock


\bibitem[Lu et~al\mbox{.}(2020)]%
        {lu2020enhancing}
\bibfield{author}{\bibinfo{person}{Yantao Lu}, \bibinfo{person}{Yunhan Jia},
  \bibinfo{person}{Jianyu Wang}, \bibinfo{person}{Bai Li},
  \bibinfo{person}{Weiheng Chai}, \bibinfo{person}{Lawrence Carin}, {and}
  \bibinfo{person}{Senem Velipasalar}.} \bibinfo{year}{2020}\natexlab{}.
\newblock \showarticletitle{Enhancing cross-task black-box transferability of
  adversarial examples with dispersion reduction}. In
  \bibinfo{booktitle}{\emph{IEEE CVPR}}.
\newblock


\bibitem[Ma et~al\mbox{.}(2021a)]%
        {ma2021finding}
\bibfield{author}{\bibinfo{person}{Chen Ma}, \bibinfo{person}{Xiangyu Guo},
  \bibinfo{person}{Li Chen}, \bibinfo{person}{Jun-Hai Yong}, {and}
  \bibinfo{person}{Yisen Wang}.} \bibinfo{year}{2021}\natexlab{a}.
\newblock \showarticletitle{Finding optimal tangent points for reducing
  distortions of hard-label attacks}.
\newblock \bibinfo{journal}{\emph{Advances in Neural Information Processing
  Systems}}  \bibinfo{volume}{34} (\bibinfo{year}{2021}),
  \bibinfo{pages}{19288--19300}.
\newblock


\bibitem[Ma and Liang(2020)]%
        {ma2020increasing_D5}
\bibfield{author}{\bibinfo{person}{Linhai Ma} {and} \bibinfo{person}{Liang
  Liang}.} \bibinfo{year}{2020}\natexlab{}.
\newblock \showarticletitle{Increasing-margin adversarial (IMA) training to
  improve adversarial robustness of neural networks}.
\newblock \bibinfo{journal}{\emph{arXiv preprint arXiv:2005.09147}}
  (\bibinfo{year}{2020}).
\newblock


\bibitem[Ma and Liang(2022)]%
        {ma2022adaptive_D53}
\bibfield{author}{\bibinfo{person}{Linhai Ma} {and} \bibinfo{person}{Liang
  Liang}.} \bibinfo{year}{2022}\natexlab{}.
\newblock \showarticletitle{Adaptive Adversarial Training to Improve
  Adversarial Robustness of DNNs for Medical Image Segmentation and Detection}.
\newblock \bibinfo{journal}{\emph{arXiv preprint arXiv:2206.01736}}
  (\bibinfo{year}{2022}).
\newblock


\bibitem[Ma et~al\mbox{.}(2021b)]%
        {ma2021understanding_D10}
\bibfield{author}{\bibinfo{person}{Xingjun Ma}, \bibinfo{person}{Yuhao Niu},
  \bibinfo{person}{Lin Gu}, \bibinfo{person}{Yisen Wang},
  \bibinfo{person}{Yitian Zhao}, \bibinfo{person}{James Bailey}, {and}
  \bibinfo{person}{Feng Lu}.} \bibinfo{year}{2021}\natexlab{b}.
\newblock \showarticletitle{Understanding adversarial attacks on deep learning
  based medical image analysis systems}.
\newblock \bibinfo{journal}{\emph{Pattern Recognition}} (\bibinfo{year}{2021}).
\newblock


\bibitem[Madry et~al\mbox{.}(2018)]%
        {MadryMSTV18}
\bibfield{author}{\bibinfo{person}{Aleksander Madry},
  \bibinfo{person}{Aleksandar Makelov}, \bibinfo{person}{Ludwig Schmidt},
  \bibinfo{person}{Dimitris Tsipras}, {and} \bibinfo{person}{Adrian Vladu}.}
  \bibinfo{year}{2018}\natexlab{}.
\newblock \showarticletitle{Towards Deep Learning Models Resistant to
  Adversarial Attacks}. In \bibinfo{booktitle}{\emph{International Conference
  on Learning Representations (ICLR)}}.
\newblock


\bibitem[Maliamanis et~al\mbox{.}(2022)]%
        {maliamanis2022resilient_D39}
\bibfield{author}{\bibinfo{person}{Theodore~V Maliamanis},
  \bibinfo{person}{Kyriakos~D Apostolidis}, {and} \bibinfo{person}{George~A
  Papakostas}.} \bibinfo{year}{2022}\natexlab{}.
\newblock \showarticletitle{How Resilient Are Deep Learning Models in Medical
  Image Analysis? The Case of the Moment-Based Adversarial Attack (Mb-AdA)}.
\newblock \bibinfo{journal}{\emph{Biomedicines}} (\bibinfo{year}{2022}).
\newblock


\bibitem[Mao et~al\mbox{.}(2023)]%
        {MaoGYWV23}
\bibfield{author}{\bibinfo{person}{Chengzhi Mao}, \bibinfo{person}{Scott Geng},
  \bibinfo{person}{Junfeng Yang}, \bibinfo{person}{Xin Wang}, {and}
  \bibinfo{person}{Carl Vondrick}.} \bibinfo{year}{2023}\natexlab{}.
\newblock \showarticletitle{Understanding Zero-shot Adversarial Robustness for
  Large-Scale Models}. In \bibinfo{booktitle}{\emph{The Eleventh International
  Conference on Learning Representations,{ICLR}}}.
\newblock


\bibitem[Minagi et~al\mbox{.}(2022)]%
        {minagi2022natural_A19}
\bibfield{author}{\bibinfo{person}{Akinori Minagi}, \bibinfo{person}{Hokuto
  Hirano}, {and} \bibinfo{person}{Kauzhiro Takemoto}.}
  \bibinfo{year}{2022}\natexlab{}.
\newblock \showarticletitle{Natural Images Allow Universal Adversarial Attacks
  on Medical Image Classification Using Deep Neural Networks with Transfer
  Learning}.
\newblock \bibinfo{journal}{\emph{Journal of Imaging}} (\bibinfo{year}{2022}).
\newblock


\bibitem[Moosavi-Dezfooli et~al\mbox{.}(2017)]%
        {moosavi2017universal}
\bibfield{author}{\bibinfo{person}{Seyed-Mohsen Moosavi-Dezfooli},
  \bibinfo{person}{Alhussein Fawzi}, \bibinfo{person}{Omar Fawzi}, {and}
  \bibinfo{person}{Pascal Frossard}.} \bibinfo{year}{2017}\natexlab{}.
\newblock \showarticletitle{Universal adversarial perturbations}. In
  \bibinfo{booktitle}{\emph{IEEE CVPR}}.
\newblock


\bibitem[Moosavi-Dezfooli et~al\mbox{.}(2016)]%
        {moosavi2016deepfool}
\bibfield{author}{\bibinfo{person}{Seyed-Mohsen Moosavi-Dezfooli},
  \bibinfo{person}{Alhussein Fawzi}, {and} \bibinfo{person}{Pascal Frossard}.}
  \bibinfo{year}{2016}\natexlab{}.
\newblock \showarticletitle{Deepfool: a simple and accurate method to fool deep
  neural networks}. In \bibinfo{booktitle}{\emph{CVPR}}.
  \bibinfo{pages}{2574--2582}.
\newblock


\bibitem[Morshuis et~al\mbox{.}(2022)]%
        {morshuis2022adversarial_A26}
\bibfield{author}{\bibinfo{person}{Jan~Nikolas Morshuis},
  \bibinfo{person}{Sergios Gatidis}, \bibinfo{person}{Matthias Hein}, {and}
  \bibinfo{person}{Christian~F Baumgartner}.} \bibinfo{year}{2022}\natexlab{}.
\newblock \showarticletitle{Adversarial Robustness of MR Image Reconstruction
  Under Realistic Perturbations}. In \bibinfo{booktitle}{\emph{International
  Workshop on Machine Learning for Medical Image Reconstruction}}.
  \bibinfo{pages}{24--33}.
\newblock


\bibitem[Nesterov(1983)]%
        {Nesterov1983AMF}
\bibfield{author}{\bibinfo{person}{Yurii Nesterov}.}
  \bibinfo{year}{1983}\natexlab{}.
\newblock \showarticletitle{A method for solving the convex programming problem
  with convergence rate $O(1/k^2)$}.
\newblock \bibinfo{journal}{\emph{Proceedings of the USSR Academy of Sciences}}
   \bibinfo{volume}{269} (\bibinfo{year}{1983}), \bibinfo{pages}{543--547}.
\newblock


\bibitem[Nguyen et~al\mbox{.}(2012)]%
        {nguyen2012evolutionary}
\bibfield{author}{\bibinfo{person}{Trung~Thanh Nguyen},
  \bibinfo{person}{Shengxiang Yang}, {and} \bibinfo{person}{Juergen Branke}.}
  \bibinfo{year}{2012}\natexlab{}.
\newblock \showarticletitle{Evolutionary dynamic optimization: A survey of the
  state of the art}.
\newblock \bibinfo{journal}{\emph{Swarm and Evolutionary Computation}}
  \bibinfo{volume}{6} (\bibinfo{year}{2012}), \bibinfo{pages}{1--24}.
\newblock


\bibitem[Ozbulak et~al\mbox{.}(2019)]%
        {ozbulak2019impact_A12}
\bibfield{author}{\bibinfo{person}{Utku Ozbulak}, \bibinfo{person}{Arnout
  Van~Messem}, {and} \bibinfo{person}{Wesley~De Neve}.}
  \bibinfo{year}{2019}\natexlab{}.
\newblock \showarticletitle{Impact of adversarial examples on deep learning
  models for biomedical image segmentation}. In
  \bibinfo{booktitle}{\emph{MICCAI}}. \bibinfo{pages}{300--308}.
\newblock


\bibitem[Pal et~al\mbox{.}(2021)]%
        {pal2021vulnerability_A8}
\bibfield{author}{\bibinfo{person}{Biprodip Pal}, \bibinfo{person}{Debashis
  Gupta}, \bibinfo{person}{Md Rashed-Al-Mahfuz}, \bibinfo{person}{Salem~A
  Alyami}, {and} \bibinfo{person}{Mohammad~Ali Moni}.}
  \bibinfo{year}{2021}\natexlab{}.
\newblock \showarticletitle{Vulnerability in deep transfer learning models to
  adversarial fast gradient sign attack for covid-19 prediction from chest
  radiography images}.
\newblock \bibinfo{journal}{\emph{Applied Sciences}}  \bibinfo{volume}{11}
  (\bibinfo{year}{2021}).
\newblock


\bibitem[Pandey et~al\mbox{.}(2022)]%
        {pandey2022adversarially_D52}
\bibfield{author}{\bibinfo{person}{Prashant Pandey}, \bibinfo{person}{Aleti
  Vardhan}, \bibinfo{person}{Mustafa Chasmai}, \bibinfo{person}{Tanuj Sur},
  {and} \bibinfo{person}{Brejesh Lall}.} \bibinfo{year}{2022}\natexlab{}.
\newblock \showarticletitle{Adversarially Robust Prototypical Few-Shot
  Segmentation with Neural-ODEs}. In \bibinfo{booktitle}{\emph{MICCAI}}.
  \bibinfo{pages}{77--87}.
\newblock


\bibitem[Papernot et~al\mbox{.}(2016)]%
        {papernot2016limitations}
\bibfield{author}{\bibinfo{person}{Nicolas Papernot}, \bibinfo{person}{Patrick
  McDaniel}, \bibinfo{person}{Somesh Jha}, \bibinfo{person}{Matt Fredrikson},
  \bibinfo{person}{Z~Berkay Celik}, {and} \bibinfo{person}{Ananthram Swami}.}
  \bibinfo{year}{2016}\natexlab{}.
\newblock \showarticletitle{The limitations of deep learning in adversarial
  settings}. In \bibinfo{booktitle}{\emph{European Symposium on Security and
  Privacy}}.
\newblock


\bibitem[Park et~al\mbox{.}(2020)]%
        {park2020robustification_D19}
\bibfield{author}{\bibinfo{person}{Hanwool Park}, \bibinfo{person}{Amirhossein
  Bayat}, \bibinfo{person}{Mohammad Sabokrou}, \bibinfo{person}{Jan~S
  Kirschke}, {and} \bibinfo{person}{Bjoern~H Menze}.}
  \bibinfo{year}{2020}\natexlab{}.
\newblock \showarticletitle{Robustification of Segmentation Models Against
  Adversarial Perturbations in Medical Imaging}. In
  \bibinfo{booktitle}{\emph{International Workshop on PRedictive Intelligence
  In MEdicine}}. Springer, \bibinfo{pages}{46--57}.
\newblock


\bibitem[Paschali et~al\mbox{.}(2018)]%
        {paschali2018generalizability_A1}
\bibfield{author}{\bibinfo{person}{Magdalini Paschali},
  \bibinfo{person}{Sailesh Conjeti}, \bibinfo{person}{Fernando Navarro}, {and}
  \bibinfo{person}{Nassir Navab}.} \bibinfo{year}{2018}\natexlab{}.
\newblock \showarticletitle{Generalizability vs. robustness: investigating
  medical imaging networks using adversarial examples}. In
  \bibinfo{booktitle}{\emph{MICCAI}}.
\newblock


\bibitem[Patel et~al\mbox{.}(2022)]%
        {patel2022predictive_A49}
\bibfield{author}{\bibinfo{person}{Pavan Patel}, \bibinfo{person}{Mohit
  Bhadla}, \bibinfo{person}{Jinal Upadhyay}, \bibinfo{person}{Dhruvi Suthar},
  {and} \bibinfo{person}{Drashti Darji}.} \bibinfo{year}{2022}\natexlab{}.
\newblock \showarticletitle{Predictive COVID-19 Risk and Virus Mutation
  isolation with CNN based Machine learning Technique}. In
  \bibinfo{booktitle}{\emph{International Conference on Innovative Practices in
  Technology and Management}}, Vol.~\bibinfo{volume}{2}.
\newblock


\bibitem[Paul et~al\mbox{.}(2020)]%
        {paul2020mitigating_D8}
\bibfield{author}{\bibinfo{person}{Rahul Paul}, \bibinfo{person}{Matthew
  Schabath}, \bibinfo{person}{Robert Gillies}, \bibinfo{person}{Lawrence Hall},
  {and} \bibinfo{person}{Dmitry Goldgof}.} \bibinfo{year}{2020}\natexlab{}.
\newblock \showarticletitle{Mitigating adversarial attacks on medical image
  understanding systems}. In \bibinfo{booktitle}{\emph{International Symposium
  on Biomedical Imaging}}.
\newblock


\bibitem[Pereira et~al\mbox{.}(2016)]%
        {pereira2016brain}
\bibfield{author}{\bibinfo{person}{S{\'e}rgio Pereira},
  \bibinfo{person}{Adriano Pinto}, \bibinfo{person}{Victor Alves}, {and}
  \bibinfo{person}{Carlos~A Silva}.} \bibinfo{year}{2016}\natexlab{}.
\newblock \showarticletitle{Brain tumor segmentation using convolutional neural
  networks in MRI images}.
\newblock \bibinfo{journal}{\emph{IEEE Transactions on Medical Imaging}}
  \bibinfo{volume}{35} (\bibinfo{year}{2016}), \bibinfo{pages}{1240--1251}.
\newblock


\bibitem[Pervin et~al\mbox{.}(2021)]%
        {pervin2021adversarial_D25}
\bibfield{author}{\bibinfo{person}{Mst Pervin}, \bibinfo{person}{Linmi Tao},
  \bibinfo{person}{Aminul Huq}, \bibinfo{person}{Zuoxiang He},
  \bibinfo{person}{Li Huo}, {et~al\mbox{.}}} \bibinfo{year}{2021}\natexlab{}.
\newblock \showarticletitle{Adversarial Attack Driven Data Augmentation for
  Accurate And Robust Medical Image Segmentation}.
\newblock \bibinfo{journal}{\emph{arXiv preprint arXiv:2105.12106}}
  (\bibinfo{year}{2021}).
\newblock


\bibitem[Pitropakis et~al\mbox{.}(2019)]%
        {pitropakis2019taxonomy}
\bibfield{author}{\bibinfo{person}{Nikolaos Pitropakis},
  \bibinfo{person}{Emmanouil Panaousis}, \bibinfo{person}{Thanassis
  Giannetsos}, \bibinfo{person}{Eleftherios Anastasiadis}, {and}
  \bibinfo{person}{George Loukas}.} \bibinfo{year}{2019}\natexlab{}.
\newblock \showarticletitle{A taxonomy and survey of attacks against machine
  learning}.
\newblock \bibinfo{journal}{\emph{Computer Science Review}}
  \bibinfo{volume}{34} (\bibinfo{year}{2019}), \bibinfo{pages}{100199}.
\newblock


\bibitem[Poursaeed et~al\mbox{.}(2018)]%
        {poursaeed2018generative}
\bibfield{author}{\bibinfo{person}{Omid Poursaeed}, \bibinfo{person}{Isay
  Katsman}, \bibinfo{person}{Bicheng Gao}, {and} \bibinfo{person}{Serge
  Belongie}.} \bibinfo{year}{2018}\natexlab{}.
\newblock \showarticletitle{Generative adversarial perturbations}. In
  \bibinfo{booktitle}{\emph{CVPR}}.
\newblock


\bibitem[Qi et~al\mbox{.}(2021)]%
        {QiGS0Z21_A17}
\bibfield{author}{\bibinfo{person}{Gege Qi}, \bibinfo{person}{Lijun Gong},
  \bibinfo{person}{Yibing Song}, \bibinfo{person}{Kai Ma}, {and}
  \bibinfo{person}{Yefeng Zheng}.} \bibinfo{year}{2021}\natexlab{}.
\newblock \showarticletitle{Stabilized Medical Image Attacks}. In
  \bibinfo{booktitle}{\emph{ICLR}}.
\newblock


\bibitem[Rade and Moosavi{-}Dezfooli(2022)]%
        {RadeM22}
\bibfield{author}{\bibinfo{person}{Rahul Rade} {and}
  \bibinfo{person}{Seyed{-}Mohsen Moosavi{-}Dezfooli}.}
  \bibinfo{year}{2022}\natexlab{}.
\newblock \showarticletitle{Reducing Excessive Margin to Achieve a Better
  Accuracy vs. Robustness Trade-off}. In \bibinfo{booktitle}{\emph{ICLR}}.
\newblock


\bibitem[Rahman et~al\mbox{.}(2020)]%
        {rahman2020adversarial_A22}
\bibfield{author}{\bibinfo{person}{Abdur Rahman}, \bibinfo{person}{M~Shamim
  Hossain}, \bibinfo{person}{Nabil~A Alrajeh}, {and} \bibinfo{person}{Fawaz
  Alsolami}.} \bibinfo{year}{2020}\natexlab{}.
\newblock \showarticletitle{Adversarial examples—Security threats to COVID-19
  deep learning systems in medical IoT devices}.
\newblock \bibinfo{journal}{\emph{IEEE Internet of Things Journal}}
  (\bibinfo{year}{2020}).
\newblock


\bibitem[Raj et~al\mbox{.}(2020)]%
        {raj2020improving_D61}
\bibfield{author}{\bibinfo{person}{Ankit Raj}, \bibinfo{person}{Yoram Bresler},
  {and} \bibinfo{person}{Bo Li}.} \bibinfo{year}{2020}\natexlab{}.
\newblock \showarticletitle{Improving robustness of deep-learning-based image
  reconstruction}. In \bibinfo{booktitle}{\emph{ICML}}.
\newblock


\bibitem[Rao et~al\mbox{.}(2020)]%
        {rao2020thorough_A45_D60}
\bibfield{author}{\bibinfo{person}{Chendi Rao}, \bibinfo{person}{Jiezhang Cao},
  \bibinfo{person}{Runhao Zeng}, \bibinfo{person}{Qi Chen},
  \bibinfo{person}{Huazhu Fu}, \bibinfo{person}{Yanwu Xu}, {and}
  \bibinfo{person}{Mingkui Tan}.} \bibinfo{year}{2020}\natexlab{}.
\newblock \showarticletitle{A thorough comparison study on adversarial attacks
  and defenses for common thorax disease classification in chest x-rays}.
\newblock \bibinfo{journal}{\emph{arXiv preprint arXiv:2003.13969}}
  (\bibinfo{year}{2020}).
\newblock


\bibitem[Rasaee and Rivaz(2021)]%
        {rasaee2021explainable_A44}
\bibfield{author}{\bibinfo{person}{Hamza Rasaee} {and} \bibinfo{person}{Hassan
  Rivaz}.} \bibinfo{year}{2021}\natexlab{}.
\newblock \showarticletitle{Explainable AI and susceptibility to adversarial
  attacks: a case study in classification of breast ultrasound images}. In
  \bibinfo{booktitle}{\emph{IEEE International Ultrasonics Symposium}}.
\newblock


\bibitem[Ren et~al\mbox{.}(2019)]%
        {ren2019brain_D11}
\bibfield{author}{\bibinfo{person}{Xuhua Ren}, \bibinfo{person}{Lichi Zhang},
  \bibinfo{person}{Dongming Wei}, \bibinfo{person}{Dinggang Shen}, {and}
  \bibinfo{person}{Qian Wang}.} \bibinfo{year}{2019}\natexlab{}.
\newblock \showarticletitle{Brain MR image segmentation in small dataset with
  adversarial defense and task reorganization}. In
  \bibinfo{booktitle}{\emph{Machine Learning in Medical Imaging}}.
\newblock


\bibitem[Rodriguez et~al\mbox{.}(2022)]%
        {rodriguez2022role_D30}
\bibfield{author}{\bibinfo{person}{David Rodriguez}, \bibinfo{person}{Tapsya
  Nayak}, \bibinfo{person}{Yidong Chen}, \bibinfo{person}{Ram Krishnan}, {and}
  \bibinfo{person}{Yufei Huang}.} \bibinfo{year}{2022}\natexlab{}.
\newblock \showarticletitle{On the role of deep learning model complexity in
  adversarial robustness for medical images}.
\newblock \bibinfo{journal}{\emph{BMC Medical Informatics and Decision Making}}
  (\bibinfo{year}{2022}).
\newblock


\bibitem[Roh(2022)]%
        {roh2022impact_D59}
\bibfield{author}{\bibinfo{person}{Jaechul Roh}.}
  \bibinfo{year}{2022}\natexlab{}.
\newblock \showarticletitle{Impact of Adversarial Training on the Robustness of
  Deep Neural Networks}. In \bibinfo{booktitle}{\emph{IEEE International
  Conference on Information Systems and Computer Aided Education}}.
  \bibinfo{pages}{560--566}.
\newblock


\bibitem[Ronneberger et~al\mbox{.}(2015)]%
        {ronneberger2015u}
\bibfield{author}{\bibinfo{person}{Olaf Ronneberger}, \bibinfo{person}{Philipp
  Fischer}, {and} \bibinfo{person}{Thomas Brox}.}
  \bibinfo{year}{2015}\natexlab{}.
\newblock \showarticletitle{U-net: Convolutional networks for biomedical image
  segmentation}. In \bibinfo{booktitle}{\emph{MICCAI}}.
\newblock


\bibitem[Sandler et~al\mbox{.}(2018)]%
        {sandler2018mobilenetv2}
\bibfield{author}{\bibinfo{person}{Mark Sandler}, \bibinfo{person}{Andrew
  Howard}, \bibinfo{person}{Menglong Zhu}, \bibinfo{person}{Andrey Zhmoginov},
  {and} \bibinfo{person}{Liang-Chieh Chen}.} \bibinfo{year}{2018}\natexlab{}.
\newblock \showarticletitle{Mobilenetv2: Inverted residuals and linear
  bottlenecks}. In \bibinfo{booktitle}{\emph{IEEE CVPR}}.
  \bibinfo{pages}{4510--4520}.
\newblock


\bibitem[Schlarmann et~al\mbox{.}(2024)]%
        {schlarmann2024robust}
\bibfield{author}{\bibinfo{person}{Christian Schlarmann},
  \bibinfo{person}{Naman~Deep Singh}, \bibinfo{person}{Francesco Croce}, {and}
  \bibinfo{person}{Matthias Hein}.} \bibinfo{year}{2024}\natexlab{}.
\newblock \showarticletitle{Robust clip: Unsupervised adversarial fine-tuning
  of vision embeddings for robust large vision-language models}.
\newblock \bibinfo{journal}{\emph{arXiv preprint arXiv:2402.12336}}
  (\bibinfo{year}{2024}).
\newblock


\bibitem[Selvakkumar et~al\mbox{.}(2022)]%
        {selvakkumar2022addressing_A25}
\bibfield{author}{\bibinfo{person}{Arawinkumaar Selvakkumar},
  \bibinfo{person}{Shantanu Pal}, {and} \bibinfo{person}{Zahra Jadidi}.}
  \bibinfo{year}{2022}\natexlab{}.
\newblock \showarticletitle{Addressing Adversarial Machine Learning Attacks in
  Smart Healthcare Perspectives}.
\newblock In \bibinfo{booktitle}{\emph{Sensing Technology}}.
  \bibinfo{pages}{269--282}.
\newblock


\bibitem[Selvaraju et~al\mbox{.}(2017)]%
        {selvaraju2017grad}
\bibfield{author}{\bibinfo{person}{Ramprasaath~R Selvaraju},
  \bibinfo{person}{Michael Cogswell}, \bibinfo{person}{Abhishek Das},
  \bibinfo{person}{Ramakrishna Vedantam}, \bibinfo{person}{Devi Parikh}, {and}
  \bibinfo{person}{Dhruv Batra}.} \bibinfo{year}{2017}\natexlab{}.
\newblock \showarticletitle{Grad-cam: Visual explanations from deep networks
  via gradient-based localization}. In \bibinfo{booktitle}{\emph{IEEE ICCV}}.
  \bibinfo{pages}{618--626}.
\newblock


\bibitem[Shafahi et~al\mbox{.}(2019)]%
        {shafahi2019adversarial}
\bibfield{author}{\bibinfo{person}{Ali Shafahi}, \bibinfo{person}{Mahyar
  Najibi}, \bibinfo{person}{Mohammad~Amin Ghiasi}, \bibinfo{person}{Zheng Xu},
  \bibinfo{person}{John Dickerson}, \bibinfo{person}{Christoph Studer},
  \bibinfo{person}{Larry~S Davis}, \bibinfo{person}{Gavin Taylor}, {and}
  \bibinfo{person}{Tom Goldstein}.} \bibinfo{year}{2019}\natexlab{}.
\newblock \showarticletitle{Adversarial training for free!}
\newblock \bibinfo{journal}{\emph{NeurIPS}} (\bibinfo{year}{2019}).
\newblock


\bibitem[Shah et~al\mbox{.}(2018)]%
        {shah2018susceptibility_A4}
\bibfield{author}{\bibinfo{person}{Abhay Shah}, \bibinfo{person}{Stephanie
  Lynch}, \bibinfo{person}{Meindert Niemeijer}, \bibinfo{person}{Ryan Amelon},
  \bibinfo{person}{Warren Clarida}, \bibinfo{person}{James Folk},
  \bibinfo{person}{Stephen Russell}, \bibinfo{person}{Xiaodong Wu}, {and}
  \bibinfo{person}{Michael~D Abr{\`a}moff}.} \bibinfo{year}{2018}\natexlab{}.
\newblock \showarticletitle{Susceptibility to misdiagnosis of adversarial
  images by deep learning based retinal image analysis algorithms}. In
  \bibinfo{booktitle}{\emph{IEEE International Symposium on Biomedical Imaging
  (ISBI)}}. \bibinfo{pages}{1454--1457}.
\newblock


\bibitem[Shamshiri and Sohn(2022)]%
        {shamshiri2022security}
\bibfield{author}{\bibinfo{person}{Samaneh Shamshiri} {and}
  \bibinfo{person}{Insoo Sohn}.} \bibinfo{year}{2022}\natexlab{}.
\newblock \showarticletitle{Security methods for AI based COVID-19 analysis
  system: A survey}.
\newblock \bibinfo{journal}{\emph{ICT Express}} (\bibinfo{year}{2022}).
\newblock


\bibitem[Shao et~al\mbox{.}(2021)]%
        {shao2021target_A16}
\bibfield{author}{\bibinfo{person}{Mingwen Shao}, \bibinfo{person}{Gaozhi
  Zhang}, \bibinfo{person}{Wangmeng Zuo}, {and} \bibinfo{person}{Deyu Meng}.}
  \bibinfo{year}{2021}\natexlab{}.
\newblock \showarticletitle{Target attack on biomedical image segmentation
  model based on multi-scale gradients}.
\newblock \bibinfo{journal}{\emph{Information Sciences}}  \bibinfo{volume}{554}
  (\bibinfo{year}{2021}), \bibinfo{pages}{33--46}.
\newblock


\bibitem[Shen et~al\mbox{.}(2020)]%
        {shen2020robustness_D36}
\bibfield{author}{\bibinfo{person}{Chenyang Shen}, \bibinfo{person}{Min-Yu
  Tsai}, \bibinfo{person}{Liyuan Chen}, \bibinfo{person}{Shulong Li},
  \bibinfo{person}{Dan Nguyen}, \bibinfo{person}{Jing Wang},
  \bibinfo{person}{Steve~B Jiang}, {and} \bibinfo{person}{Xun Jia}.}
  \bibinfo{year}{2020}\natexlab{}.
\newblock \showarticletitle{On the robustness of deep learning-based
  lung-nodule classification for CT images with respect to image noise}.
\newblock \bibinfo{journal}{\emph{Physics in Medicine \& Biology}}
  \bibinfo{volume}{65}, \bibinfo{number}{24} (\bibinfo{year}{2020}),
  \bibinfo{pages}{245037}.
\newblock


\bibitem[Shi et~al\mbox{.}(2022)]%
        {shi2022robust_D57}
\bibfield{author}{\bibinfo{person}{Xiaoshuang Shi}, \bibinfo{person}{Yifan
  Peng}, \bibinfo{person}{Qingyu Chen}, \bibinfo{person}{Tiarnan Keenan},
  \bibinfo{person}{Alisa~T Thavikulwat}, \bibinfo{person}{Sungwon Lee},
  \bibinfo{person}{Yuxing Tang}, \bibinfo{person}{Emily~Y Chew},
  \bibinfo{person}{Ronald~M Summers}, {and} \bibinfo{person}{Zhiyong Lu}.}
  \bibinfo{year}{2022}\natexlab{}.
\newblock \showarticletitle{Robust convolutional neural networks against
  adversarial attacks on medical images}.
\newblock \bibinfo{journal}{\emph{Pattern Recognition}}  \bibinfo{volume}{132}
  (\bibinfo{year}{2022}), \bibinfo{pages}{108923}.
\newblock


\bibitem[Sipola et~al\mbox{.}(2020)]%
        {sipola2020model}
\bibfield{author}{\bibinfo{person}{Tuomo Sipola}, \bibinfo{person}{Samir
  Puuska}, {and} \bibinfo{person}{Tero Kokkonen}.}
  \bibinfo{year}{2020}\natexlab{}.
\newblock \showarticletitle{Model fooling attacks against medical imaging: a
  short survey}.
\newblock \bibinfo{journal}{\emph{Information \& Security: An International
  Journal (ISIJ)}}  \bibinfo{volume}{46} (\bibinfo{year}{2020}),
  \bibinfo{pages}{215--224}.
\newblock


\bibitem[Smith and Topin(2019)]%
        {smith2019super}
\bibfield{author}{\bibinfo{person}{Leslie~N Smith} {and}
  \bibinfo{person}{Nicholay Topin}.} \bibinfo{year}{2019}\natexlab{}.
\newblock \showarticletitle{Super-convergence: Very fast training of neural
  networks using large learning rates}. In \bibinfo{booktitle}{\emph{Artificial
  Intelligence and Machine Learning for Multi-domain Operations Applications}}.
\newblock


\bibitem[Stimpel et~al\mbox{.}(2019)]%
        {stimpel2019multi_D62}
\bibfield{author}{\bibinfo{person}{Bernhard Stimpel},
  \bibinfo{person}{Christopher Syben}, \bibinfo{person}{Franziska
  Schirrmacher}, \bibinfo{person}{Philip Hoelter}, \bibinfo{person}{Arnd
  D{\"o}rfler}, {and} \bibinfo{person}{Andreas Maier}.}
  \bibinfo{year}{2019}\natexlab{}.
\newblock \showarticletitle{Multi-modal deep guided filtering for
  comprehensible medical image processing}.
\newblock \bibinfo{journal}{\emph{IEEE TMI}} (\bibinfo{year}{2019}).
\newblock


\bibitem[Su et~al\mbox{.}(2019)]%
        {su2019one}
\bibfield{author}{\bibinfo{person}{Jiawei Su},
  \bibinfo{person}{Danilo~Vasconcellos Vargas}, {and} \bibinfo{person}{Kouichi
  Sakurai}.} \bibinfo{year}{2019}\natexlab{}.
\newblock \showarticletitle{One pixel attack for fooling deep neural networks}.
\newblock \bibinfo{journal}{\emph{IEEE Transactions on Evolutionary
  Computation}} \bibinfo{volume}{23}, \bibinfo{number}{5}
  (\bibinfo{year}{2019}), \bibinfo{pages}{828--841}.
\newblock


\bibitem[Sun et~al\mbox{.}(2023)]%
        {sun2023critical}
\bibfield{author}{\bibinfo{person}{Jiachen Sun}, \bibinfo{person}{Jiongxiao
  Wang}, \bibinfo{person}{Weili Nie}, \bibinfo{person}{Zhiding Yu},
  \bibinfo{person}{Zhuoqing Mao}, {and} \bibinfo{person}{Chaowei Xiao}.}
  \bibinfo{year}{2023}\natexlab{}.
\newblock \showarticletitle{A critical revisit of adversarial robustness in 3D
  point cloud recognition with diffusion-driven purification}. In
  \bibinfo{booktitle}{\emph{ICML}}. \bibinfo{pages}{33100--33114}.
\newblock


\bibitem[Sun et~al\mbox{.}(2022)]%
        {SunXVG22_D44}
\bibfield{author}{\bibinfo{person}{Shoukun Sun}, \bibinfo{person}{Min Xian},
  \bibinfo{person}{Aleksandar Vakanski}, {and} \bibinfo{person}{Hossny
  Ghanem}.} \bibinfo{year}{2022}\natexlab{}.
\newblock \showarticletitle{{MIRST-DM:} Multi-instance {RST} with Drop-Max
  Layer for Robust Classification of Breast Cancer}. In
  \bibinfo{booktitle}{\emph{MICCAI}}. \bibinfo{pages}{401--410}.
\newblock


\bibitem[Szegedy et~al\mbox{.}(2014)]%
        {SzegedyZSBEGF13}
\bibfield{author}{\bibinfo{person}{Christian Szegedy},
  \bibinfo{person}{Wojciech Zaremba}, \bibinfo{person}{Ilya Sutskever},
  \bibinfo{person}{Joan Bruna}, \bibinfo{person}{Dumitru Erhan},
  \bibinfo{person}{Ian~J. Goodfellow}, {and} \bibinfo{person}{Rob Fergus}.}
  \bibinfo{year}{2014}\natexlab{}.
\newblock \showarticletitle{Intriguing properties of neural networks}. In
  \bibinfo{booktitle}{\emph{International Conference on Learning
  Representations (ICLR)}}.
\newblock


\bibitem[Taghanaki et~al\mbox{.}(2019)]%
        {taghanaki2019kernelized_D20}
\bibfield{author}{\bibinfo{person}{Saeid~Asgari Taghanaki},
  \bibinfo{person}{Kumar Abhishek}, \bibinfo{person}{Shekoofeh Azizi}, {and}
  \bibinfo{person}{Ghassan Hamarneh}.} \bibinfo{year}{2019}\natexlab{}.
\newblock \showarticletitle{A kernelized manifold mapping to diminish the
  effect of adversarial perturbations}. In \bibinfo{booktitle}{\emph{IEEE/CVF
  CVPR}}. \bibinfo{pages}{11340--11349}.
\newblock


\bibitem[Terzi et~al\mbox{.}(2021)]%
        {terzi2021adversarial}
\bibfield{author}{\bibinfo{person}{Matteo Terzi}, \bibinfo{person}{Alessandro
  Achille}, \bibinfo{person}{Marco Maggipinto}, {and}
  \bibinfo{person}{Gian~Antonio Susto}.} \bibinfo{year}{2021}\natexlab{}.
\newblock \showarticletitle{Adversarial training reduces information and
  improves transferability}. In \bibinfo{booktitle}{\emph{Proceedings of the
  AAAI Conference on Artificial Intelligence}}.
\newblock


\bibitem[Tian et~al\mbox{.}(2021)]%
        {tian2021bias_A14}
\bibfield{author}{\bibinfo{person}{Binyu Tian}, \bibinfo{person}{Qing Guo},
  \bibinfo{person}{Felix Juefei-Xu}, \bibinfo{person}{Wen Le~Chan},
  \bibinfo{person}{Yupeng Cheng}, \bibinfo{person}{Xiaohong Li},
  \bibinfo{person}{Xiaofei Xie}, {and} \bibinfo{person}{Shengchao Qin}.}
  \bibinfo{year}{2021}\natexlab{}.
\newblock \showarticletitle{Bias field poses a threat to dnn-based x-ray
  recognition}. In \bibinfo{booktitle}{\emph{IEEE ICME}}.
\newblock


\bibitem[Tiu et~al\mbox{.}(2022)]%
        {tiu2022expert}
\bibfield{author}{\bibinfo{person}{Ekin Tiu}, \bibinfo{person}{Ellie Talius},
  \bibinfo{person}{Pujan Patel}, \bibinfo{person}{Curtis~P Langlotz},
  \bibinfo{person}{Andrew~Y Ng}, {and} \bibinfo{person}{Pranav Rajpurkar}.}
  \bibinfo{year}{2022}\natexlab{}.
\newblock \showarticletitle{Expert-level detection of pathologies from
  unannotated chest X-ray images via self-supervised learning}.
\newblock \bibinfo{journal}{\emph{Nature Biomedical Engineering}}
  (\bibinfo{year}{2022}).
\newblock


\bibitem[Tramer et~al\mbox{.}(2020)]%
        {tramer2020adaptive}
\bibfield{author}{\bibinfo{person}{Florian Tramer}, \bibinfo{person}{Nicholas
  Carlini}, \bibinfo{person}{Wieland Brendel}, {and}
  \bibinfo{person}{Aleksander Madry}.} \bibinfo{year}{2020}\natexlab{}.
\newblock \showarticletitle{On adaptive attacks to adversarial example
  defenses}.
\newblock \bibinfo{journal}{\emph{Advances in Neural Information Processing
  Systems}}  \bibinfo{volume}{33} (\bibinfo{year}{2020}),
  \bibinfo{pages}{1633--1645}.
\newblock


\bibitem[Tripathi and Mishra(2020)]%
        {tripathi2020fuzzy_D4}
\bibfield{author}{\bibinfo{person}{Achyut~Mani Tripathi} {and}
  \bibinfo{person}{Ashish Mishra}.} \bibinfo{year}{2020}\natexlab{}.
\newblock \showarticletitle{Fuzzy unique image transformation: Defense against
  adversarial attacks on deep covid-19 models}.
\newblock \bibinfo{journal}{\emph{arXiv preprint arXiv:2009.04004}}
  (\bibinfo{year}{2020}).
\newblock


\bibitem[Tsai et~al\mbox{.}(2023)]%
        {tsai2023adversarial_A2023_3}
\bibfield{author}{\bibinfo{person}{Min-Jen Tsai}, \bibinfo{person}{Ping-Yi
  Lin}, {and} \bibinfo{person}{Ming-En Lee}.} \bibinfo{year}{2023}\natexlab{}.
\newblock \showarticletitle{Adversarial Attacks on Medical Image
  Classification}.
\newblock \bibinfo{journal}{\emph{Cancers}} (\bibinfo{year}{2023}).
\newblock


\bibitem[Uesato et~al\mbox{.}(2018)]%
        {uesato2018adversarial}
\bibfield{author}{\bibinfo{person}{Jonathan Uesato}, \bibinfo{person}{Brendan
  O’donoghue}, \bibinfo{person}{Pushmeet Kohli}, {and} \bibinfo{person}{Aaron
  Oord}.} \bibinfo{year}{2018}\natexlab{}.
\newblock \showarticletitle{Adversarial risk and the dangers of evaluating
  against weak attacks}. In \bibinfo{booktitle}{\emph{International Conference
  on Machine Learning}}. PMLR, \bibinfo{pages}{5025--5034}.
\newblock


\bibitem[Uwimana and Senanayake(2021)]%
        {uwimana2021out_D26}
\bibfield{author}{\bibinfo{person}{Anisie Uwimana} {and}
  \bibinfo{person}{Ransalu Senanayake}.} \bibinfo{year}{2021}\natexlab{}.
\newblock \showarticletitle{Out of Distribution Detection and Adversarial
  Attacks on Deep Neural Networks for Robust Medical Image Analysis}. In
  \bibinfo{booktitle}{\emph{ICML Workshop}}.
\newblock


\bibitem[Vatian et~al\mbox{.}(2019)]%
        {vatian2019impact_D13}
\bibfield{author}{\bibinfo{person}{Aleksandra Vatian}, \bibinfo{person}{Natalia
  Gusarova}, \bibinfo{person}{Natalia Dobrenko}, \bibinfo{person}{Sergey
  Dudorov}, \bibinfo{person}{Niyaz Nigmatullin}, \bibinfo{person}{Anatoly
  Shalyto}, {and} \bibinfo{person}{Artem Lobantsev}.}
  \bibinfo{year}{2019}\natexlab{}.
\newblock \showarticletitle{Impact of adversarial examples on the efficiency of
  interpretation and use of information from high-tech medical images}. In
  \bibinfo{booktitle}{\emph{Conference of Open Innovations Association
  (FRUCT)}}. \bibinfo{pages}{472--478}.
\newblock


\bibitem[Villegas-Ortega et~al\mbox{.}(2021)]%
        {villegas2021fourteen}
\bibfield{author}{\bibinfo{person}{Jos{\'e} Villegas-Ortega},
  \bibinfo{person}{Luciana Bellido-Boza}, {and} \bibinfo{person}{David
  Mauricio}.} \bibinfo{year}{2021}\natexlab{}.
\newblock \showarticletitle{Fourteen years of manifestations and factors of
  health insurance fraud, 2006--2020: a scoping review}.
\newblock \bibinfo{journal}{\emph{Health \& justice}}  \bibinfo{volume}{9}
  (\bibinfo{year}{2021}).
\newblock


\bibitem[Wang et~al\mbox{.}(2023b)]%
        {wang2023adversarial_D2023_5}
\bibfield{author}{\bibinfo{person}{Jian Wang}, \bibinfo{person}{Sainan Zhang},
  \bibinfo{person}{Yanting Xie}, \bibinfo{person}{Hongen Liao}, {and}
  \bibinfo{person}{Fang Chen}.} \bibinfo{year}{2023}\natexlab{b}.
\newblock \showarticletitle{Adversarial Detection and Defense for Medical
  Ultrasound Images: From a Frequency Perspective}. In
  \bibinfo{booktitle}{\emph{Asian-Pacific Conference on Medical and Biological
  Engineering}}. Springer, \bibinfo{pages}{73--82}.
\newblock


\bibitem[Wang et~al\mbox{.}(2024)]%
        {wang2024pre}
\bibfield{author}{\bibinfo{person}{Sibo Wang}, \bibinfo{person}{Jie Zhang},
  \bibinfo{person}{Zheng Yuan}, {and} \bibinfo{person}{Shiguang Shan}.}
  \bibinfo{year}{2024}\natexlab{}.
\newblock \showarticletitle{Pre-trained Model Guided Fine-Tuning for Zero-Shot
  Adversarial Robustness}. In \bibinfo{booktitle}{\emph{Proceedings of the
  IEEE/CVF conference on computer vision and pattern recognition}}.
\newblock


\bibitem[Wang et~al\mbox{.}(2019a)]%
        {wang2019gan}
\bibfield{author}{\bibinfo{person}{Xiaosen Wang}, \bibinfo{person}{Kun He},
  {and} \bibinfo{person}{John~E Hopcroft}.} \bibinfo{year}{2019}\natexlab{a}.
\newblock \showarticletitle{AT-GAN: A generative attack model for adversarial
  transferring on generative adversarial nets}.
\newblock \bibinfo{journal}{\emph{arXiv preprint arXiv:1904.07793}}
  \bibinfo{volume}{3}, \bibinfo{number}{4} (\bibinfo{year}{2019}).
\newblock


\bibitem[Wang et~al\mbox{.}(2021)]%
        {wang2021adversarial_A37}
\bibfield{author}{\bibinfo{person}{Xiaoyin Wang}, \bibinfo{person}{Shuo Lv},
  \bibinfo{person}{Jiaze Sun}, {and} \bibinfo{person}{Shuyan Wang}.}
  \bibinfo{year}{2021}\natexlab{}.
\newblock \showarticletitle{Adversarial Attacks Medical Diagnosis Model with
  Generative Adversarial Networks}. In \bibinfo{booktitle}{\emph{The
  International Conference on Natural Computation, Fuzzy Systems and Knowledge
  Discovery}}. \bibinfo{pages}{678--685}.
\newblock


\bibitem[Wang et~al\mbox{.}(2017)]%
        {wang2017chestx}
\bibfield{author}{\bibinfo{person}{Xiaosong Wang}, \bibinfo{person}{Yifan
  Peng}, \bibinfo{person}{Le Lu}, \bibinfo{person}{Zhiyong Lu},
  \bibinfo{person}{Mohammadhadi Bagheri}, {and} \bibinfo{person}{Ronald~M
  Summers}.} \bibinfo{year}{2017}\natexlab{}.
\newblock \showarticletitle{Chestx-ray8: Hospital-scale chest x-ray database
  and benchmarks on weakly-supervised classification and localization of common
  thorax diseases}. In \bibinfo{booktitle}{\emph{IEEE CVPR}}.
  \bibinfo{pages}{2097--2106}.
\newblock


\bibitem[Wang et~al\mbox{.}(2022a)]%
        {wang2022fight_D43}
\bibfield{author}{\bibinfo{person}{Yongwei Wang}, \bibinfo{person}{Yuan Li},
  {and} \bibinfo{person}{Zhiqi Shen}.} \bibinfo{year}{2022}\natexlab{a}.
\newblock \showarticletitle{Fight Fire With Fire: Reversing Skin Adversarial
  Examples by Multiscale Diffusive and Denoising Aggregation Mechanism}.
\newblock \bibinfo{journal}{\emph{arXiv preprint arXiv:2208.10373}}
  (\bibinfo{year}{2022}).
\newblock


\bibitem[Wang et~al\mbox{.}(2023a)]%
        {wang2023reversing_D2023_2}
\bibfield{author}{\bibinfo{person}{Yongwei Wang}, \bibinfo{person}{Yuan Li},
  \bibinfo{person}{Zhiqi Shen}, {and} \bibinfo{person}{Yuhui Qiao}.}
  \bibinfo{year}{2023}\natexlab{a}.
\newblock \showarticletitle{Reversing skin cancer adversarial examples by
  multiscale diffusive and denoising aggregation mechanism}.
\newblock \bibinfo{journal}{\emph{Computers in Biology and Medicine}}
  \bibinfo{volume}{164} (\bibinfo{year}{2023}), \bibinfo{pages}{107310}.
\newblock


\bibitem[Wang et~al\mbox{.}(2019b)]%
        {wang2019improving}
\bibfield{author}{\bibinfo{person}{Yisen Wang}, \bibinfo{person}{Difan Zou},
  \bibinfo{person}{Jinfeng Yi}, \bibinfo{person}{James Bailey},
  \bibinfo{person}{Xingjun Ma}, {and} \bibinfo{person}{Quanquan Gu}.}
  \bibinfo{year}{2019}\natexlab{b}.
\newblock \showarticletitle{Improving adversarial robustness requires
  revisiting misclassified examples}. In
  \bibinfo{booktitle}{\emph{International Conference on Learning
  Representations (ICLR)}}.
\newblock


\bibitem[Wang et~al\mbox{.}(2020)]%
        {wang2020deep}
\bibfield{author}{\bibinfo{person}{Zhihao Wang}, \bibinfo{person}{Jian Chen},
  {and} \bibinfo{person}{Steven~CH Hoi}.} \bibinfo{year}{2020}\natexlab{}.
\newblock \showarticletitle{Deep learning for image super-resolution: A
  survey}.
\newblock \bibinfo{journal}{\emph{IEEE Transactions on Pattern Analysis and
  Machine Intelligence}} \bibinfo{volume}{43}, \bibinfo{number}{10}
  (\bibinfo{year}{2020}), \bibinfo{pages}{3365--3387}.
\newblock


\bibitem[Wang et~al\mbox{.}(2022b)]%
        {wang2022feature_A20}
\bibfield{author}{\bibinfo{person}{Zizhou Wang}, \bibinfo{person}{Xin Shu},
  \bibinfo{person}{Yan Wang}, \bibinfo{person}{Yangqin Feng},
  \bibinfo{person}{Lei Zhang}, {and} \bibinfo{person}{Zhang Yi}.}
  \bibinfo{year}{2022}\natexlab{b}.
\newblock \showarticletitle{A Feature Space-Restricted Attention Attack on
  Medical Deep Learning Systems}.
\newblock \bibinfo{journal}{\emph{IEEE Transactions on Cybernetics}}
  (\bibinfo{year}{2022}).
\newblock


\bibitem[Watson and Al~Moubayed(2021)]%
        {watson2021attack_D7}
\bibfield{author}{\bibinfo{person}{Matthew Watson} {and} \bibinfo{person}{Noura
  Al~Moubayed}.} \bibinfo{year}{2021}\natexlab{}.
\newblock \showarticletitle{Attack-agnostic adversarial detection on medical
  data using explainable machine learning}. In
  \bibinfo{booktitle}{\emph{International Conference on Pattern Recognition
  (ICPR)}}.
\newblock


\bibitem[Wei et~al\mbox{.}(2022)]%
        {wei2022analysis_A42}
\bibfield{author}{\bibinfo{person}{Chuyang Wei}, \bibinfo{person}{Rui Sun},
  \bibinfo{person}{Peilin Li}, {and} \bibinfo{person}{Jiaxuan Wei}.}
  \bibinfo{year}{2022}\natexlab{}.
\newblock \showarticletitle{Analysis of the No-sign Adversarial Attack on the
  COVID Chest X-ray Classification}. In \bibinfo{booktitle}{\emph{Image
  Processing and Media Computing}}. \bibinfo{pages}{73--79}.
\newblock


\bibitem[Wei et~al\mbox{.}(2019)]%
        {WeiLCC19}
\bibfield{author}{\bibinfo{person}{Xingxing Wei}, \bibinfo{person}{Siyuan
  Liang}, \bibinfo{person}{Ning Chen}, {and} \bibinfo{person}{Xiaochun Cao}.}
  \bibinfo{year}{2019}\natexlab{}.
\newblock \showarticletitle{Transferable Adversarial Attacks for Image and
  Video Object Detection}. In \bibinfo{booktitle}{\emph{IJCAI}}.
\newblock


\bibitem[Wu et~al\mbox{.}(2020)]%
        {wu2020classification_D15}
\bibfield{author}{\bibinfo{person}{Dawen Wu}, \bibinfo{person}{Shishi Liu},
  {and} \bibinfo{person}{Jian Ban}.} \bibinfo{year}{2020}\natexlab{}.
\newblock \showarticletitle{Classification of Diabetic Retinopathy Using
  Adversarial Training}. In \bibinfo{booktitle}{\emph{IOP Conference Series:
  Materials Science and Engineering}}, Vol.~\bibinfo{volume}{806}.
\newblock


\bibitem[Xiang et~al\mbox{.}(2023)]%
        {xiang2023toward_D2023_3}
\bibfield{author}{\bibinfo{person}{Kun Xiang}, \bibinfo{person}{Xing Zhang},
  \bibinfo{person}{Jinwen She}, \bibinfo{person}{Jinpeng Liu},
  \bibinfo{person}{Haohan Wang}, \bibinfo{person}{Shiqi Deng}, {and}
  \bibinfo{person}{Shancheng Jiang}.} \bibinfo{year}{2023}\natexlab{}.
\newblock \showarticletitle{Toward robust diagnosis: A contour attention
  preserving adversarial defense for covid-19 detection}. In
  \bibinfo{booktitle}{\emph{AAAI}}.
\newblock


\bibitem[Xiao et~al\mbox{.}(2018)]%
        {xiao2018generating}
\bibfield{author}{\bibinfo{person}{Chaowei Xiao}, \bibinfo{person}{Bo Li},
  \bibinfo{person}{Jun~Yan Zhu}, \bibinfo{person}{Warren He},
  \bibinfo{person}{Mingyan Liu}, {and} \bibinfo{person}{Dawn Song}.}
  \bibinfo{year}{2018}\natexlab{}.
\newblock \showarticletitle{Generating adversarial examples with adversarial
  networks}. In \bibinfo{booktitle}{\emph{International Joint Conference on
  Artificial Intelligence (IJCAI)}}.
\newblock


\bibitem[Xie et~al\mbox{.}(2018)]%
        {XieWZRY18}
\bibfield{author}{\bibinfo{person}{Cihang Xie}, \bibinfo{person}{Jianyu Wang},
  \bibinfo{person}{Zhishuai Zhang}, \bibinfo{person}{Zhou Ren}, {and}
  \bibinfo{person}{Alan~L. Yuille}.} \bibinfo{year}{2018}\natexlab{}.
\newblock \showarticletitle{Mitigating Adversarial Effects Through
  Randomization}. In \bibinfo{booktitle}{\emph{International Conference on
  Learning Representations (ICLR)}}.
\newblock


\bibitem[Xie et~al\mbox{.}(2017)]%
        {xie2017adversarial}
\bibfield{author}{\bibinfo{person}{Cihang Xie}, \bibinfo{person}{Jianyu Wang},
  \bibinfo{person}{Zhishuai Zhang}, \bibinfo{person}{Yuyin Zhou},
  \bibinfo{person}{Lingxi Xie}, {and} \bibinfo{person}{Alan Yuille}.}
  \bibinfo{year}{2017}\natexlab{}.
\newblock \showarticletitle{Adversarial examples for semantic segmentation and
  object detection}. In \bibinfo{booktitle}{\emph{IEEE International Conference
  on Computer Vision (ICCV)}}.
\newblock


\bibitem[Xie and Yuille(2020)]%
        {XieY20}
\bibfield{author}{\bibinfo{person}{Cihang Xie} {and} \bibinfo{person}{Alan~L.
  Yuille}.} \bibinfo{year}{2020}\natexlab{}.
\newblock \showarticletitle{Intriguing Properties of Adversarial Training at
  Scale}. In \bibinfo{booktitle}{\emph{ICLR}}.
\newblock


\bibitem[Xie et~al\mbox{.}(2019)]%
        {xie2019improving}
\bibfield{author}{\bibinfo{person}{Cihang Xie}, \bibinfo{person}{Zhishuai
  Zhang}, \bibinfo{person}{Yuyin Zhou}, \bibinfo{person}{Song Bai},
  \bibinfo{person}{Jianyu Wang}, \bibinfo{person}{Zhou Ren}, {and}
  \bibinfo{person}{Alan~L Yuille}.} \bibinfo{year}{2019}\natexlab{}.
\newblock \showarticletitle{Improving transferability of adversarial examples
  with input diversity}. In \bibinfo{booktitle}{\emph{IEEE CVPR}}.
  \bibinfo{pages}{2730--2739}.
\newblock


\bibitem[Xie and Fetit(2022)]%
        {xie2022effective_D38}
\bibfield{author}{\bibinfo{person}{Yiming Xie} {and} \bibinfo{person}{Ahmed~E
  Fetit}.} \bibinfo{year}{2022}\natexlab{}.
\newblock \showarticletitle{How Effective is Adversarial Training of CNNs in
  Medical Image Analysis?}. In \bibinfo{booktitle}{\emph{Medical Image
  Understanding and Analysis}}. Springer, \bibinfo{pages}{443--457}.
\newblock


\bibitem[Xu et~al\mbox{.}(2021)]%
        {xu2021towards_D9}
\bibfield{author}{\bibinfo{person}{Mengting Xu}, \bibinfo{person}{Tao Zhang},
  \bibinfo{person}{Zhongnian Li}, \bibinfo{person}{Mingxia Liu}, {and}
  \bibinfo{person}{Daoqiang Zhang}.} \bibinfo{year}{2021}\natexlab{}.
\newblock \showarticletitle{Towards evaluating the robustness of deep
  diagnostic models by adversarial attack}.
\newblock \bibinfo{journal}{\emph{Medical Image Analysis}}
  \bibinfo{volume}{69} (\bibinfo{year}{2021}), \bibinfo{pages}{101977}.
\newblock


\bibitem[Xu et~al\mbox{.}(2022b)]%
        {xu2022infoat}
\bibfield{author}{\bibinfo{person}{Mengting Xu}, \bibinfo{person}{Tao Zhang},
  \bibinfo{person}{Zhongnian Li}, {and} \bibinfo{person}{Daoqiang Zhang}.}
  \bibinfo{year}{2022}\natexlab{b}.
\newblock \showarticletitle{Infoat: Improving adversarial training using the
  information bottleneck principle}.
\newblock \bibinfo{journal}{\emph{IEEE Transactions on Neural Networks and
  Learning Systems}} \bibinfo{volume}{35}, \bibinfo{number}{1}
  (\bibinfo{year}{2022}), \bibinfo{pages}{1255--1264}.
\newblock


\bibitem[Xu et~al\mbox{.}(2022a)]%
        {xu2022medrdf_D2}
\bibfield{author}{\bibinfo{person}{Mengting Xu}, \bibinfo{person}{Tao Zhang},
  {and} \bibinfo{person}{Daoqiang Zhang}.} \bibinfo{year}{2022}\natexlab{a}.
\newblock \showarticletitle{Medrdf: a robust and retrain-less diagnostic
  framework for medical pretrained models against adversarial attack}.
\newblock \bibinfo{journal}{\emph{IEEE Transactions on Medical Imaging}}
  (\bibinfo{year}{2022}).
\newblock


\bibitem[Xue et~al\mbox{.}(2019)]%
        {xue2019improving_D22}
\bibfield{author}{\bibinfo{person}{Fei-Fei Xue}, \bibinfo{person}{Jin Peng},
  \bibinfo{person}{Ruixuan Wang}, \bibinfo{person}{Qiong Zhang}, {and}
  \bibinfo{person}{Wei-Shi Zheng}.} \bibinfo{year}{2019}\natexlab{}.
\newblock \showarticletitle{Improving robustness of medical image diagnosis
  with denoising convolutional neural networks}. In
  \bibinfo{booktitle}{\emph{MICCAI}}. \bibinfo{pages}{846--854}.
\newblock


\bibitem[Yang et~al\mbox{.}(2022)]%
        {yang2022defense_D67}
\bibfield{author}{\bibinfo{person}{Yanan Yang}, \bibinfo{person}{Frank~Y Shih},
  {and} \bibinfo{person}{Usman Roshan}.} \bibinfo{year}{2022}\natexlab{}.
\newblock \showarticletitle{Defense Against Adversarial Attacks Based on
  Stochastic Descent Sign Activation Networks on Medical Images}.
\newblock \bibinfo{journal}{\emph{International Journal of Pattern Recognition
  and Artificial Intelligence}} \bibinfo{volume}{36}, \bibinfo{number}{03}
  (\bibinfo{year}{2022}), \bibinfo{pages}{2254005}.
\newblock


\bibitem[Yao et~al\mbox{.}(2020)]%
        {yao2020miss_A34}
\bibfield{author}{\bibinfo{person}{Qingsong Yao}, \bibinfo{person}{Zecheng He},
  \bibinfo{person}{Hu Han}, {and} \bibinfo{person}{S~Kevin Zhou}.}
  \bibinfo{year}{2020}\natexlab{}.
\newblock \showarticletitle{Miss the point: Targeted adversarial attack on
  multiple landmark detection}. In \bibinfo{booktitle}{\emph{Medical Image
  Computing and Computer-Assisted Intervention}}.
\newblock


\bibitem[Yao et~al\mbox{.}(2023)]%
        {yao2023adversarial_A2023_1}
\bibfield{author}{\bibinfo{person}{Qingsong Yao}, \bibinfo{person}{Zecheng He},
  \bibinfo{person}{Yuexiang Li}, \bibinfo{person}{Yi Lin}, \bibinfo{person}{Kai
  Ma}, \bibinfo{person}{Yefeng Zheng}, {and} \bibinfo{person}{S~Kevin Zhou}.}
  \bibinfo{year}{2023}\natexlab{}.
\newblock \showarticletitle{Adversarial Medical Image with Hierarchical Feature
  Hiding}.
\newblock \bibinfo{journal}{\emph{IEEE Transactions on Medical Imaging}}
  (\bibinfo{year}{2023}).
\newblock


\bibitem[Yao et~al\mbox{.}(2021b)]%
        {yao2021hierarchical_A15}
\bibfield{author}{\bibinfo{person}{Qingsong Yao}, \bibinfo{person}{Zecheng He},
  \bibinfo{person}{Yi Lin}, \bibinfo{person}{Kai Ma}, \bibinfo{person}{Yefeng
  Zheng}, {and} \bibinfo{person}{S~Kevin Zhou}.}
  \bibinfo{year}{2021}\natexlab{b}.
\newblock \showarticletitle{A hierarchical feature constraint to camouflage
  medical adversarial attacks}. In \bibinfo{booktitle}{\emph{MICCAI}}.
  \bibinfo{pages}{36--47}.
\newblock


\bibitem[Yao et~al\mbox{.}(2021a)]%
        {yao2021medical_D45}
\bibfield{author}{\bibinfo{person}{Qingsong Yao}, \bibinfo{person}{Zecheng He},
  {and} \bibinfo{person}{S~Kevin Zhou}.} \bibinfo{year}{2021}\natexlab{a}.
\newblock \showarticletitle{Medical Aegis: Robust adversarial protectors for
  medical images}.
\newblock \bibinfo{journal}{\emph{arXiv preprint:2111.10969}}
  (\bibinfo{year}{2021}).
\newblock


\bibitem[Yap et~al\mbox{.}(2003)]%
        {yap2003image}
\bibfield{author}{\bibinfo{person}{P-T Yap}, \bibinfo{person}{Raveendran
  Paramesran}, {and} \bibinfo{person}{Seng-Huat Ong}.}
  \bibinfo{year}{2003}\natexlab{}.
\newblock \showarticletitle{Image analysis by Krawtchouk moments}.
\newblock \bibinfo{journal}{\emph{IEEE TIP}} (\bibinfo{year}{2003}).
\newblock


\bibitem[Yatsura et~al\mbox{.}(2021)]%
        {yatsura2021meta}
\bibfield{author}{\bibinfo{person}{Maksym Yatsura}, \bibinfo{person}{Jan
  Metzen}, {and} \bibinfo{person}{Matthias Hein}.}
  \bibinfo{year}{2021}\natexlab{}.
\newblock \showarticletitle{Meta-Learning the Search Distribution of Black-Box
  Random Search Based Adversarial Attacks}.
\newblock \bibinfo{journal}{\emph{Advances in Neural Information Processing
  Systems}} (\bibinfo{year}{2021}).
\newblock


\bibitem[Yilmaz et~al\mbox{.}(2021)]%
        {yilmaz2021assessment_A10}
\bibfield{author}{\bibinfo{person}{Ibrahim Yilmaz}, \bibinfo{person}{Mohamed
  Baza}, \bibinfo{person}{Ramy Amer}, \bibinfo{person}{Amar Rasheed},
  \bibinfo{person}{Fathi Amsaad}, {and} \bibinfo{person}{Rasha Morsi}.}
  \bibinfo{year}{2021}\natexlab{}.
\newblock \showarticletitle{On the assessment of robustness of telemedicine
  applications against adversarial machine learning attacks}. In
  \bibinfo{booktitle}{\emph{International Conference on Industrial, Engineering
  and Other Applications of Applied Intelligent Systems}}.
  \bibinfo{pages}{519--529}.
\newblock


\bibitem[Yin et~al\mbox{.}(2021)]%
        {YinWYGKDLL21}
\bibfield{author}{\bibinfo{person}{Bangjie Yin}, \bibinfo{person}{Wenxuan
  Wang}, \bibinfo{person}{Taiping Yao}, \bibinfo{person}{Junfeng Guo},
  \bibinfo{person}{Zelun Kong}, \bibinfo{person}{Shouhong Ding},
  \bibinfo{person}{Jilin Li}, {and} \bibinfo{person}{Cong Liu}.}
  \bibinfo{year}{2021}\natexlab{}.
\newblock \showarticletitle{Adv-Makeup: {A} New Imperceptible and Transferable
  Attack on Face Recognition}. In \bibinfo{booktitle}{\emph{IJCAI}}.
  \bibinfo{pages}{1252--1258}.
\newblock


\bibitem[Yoo and Choi(2020)]%
        {yoo2020outcomes_A32}
\bibfield{author}{\bibinfo{person}{Tae~Keun Yoo} {and}
  \bibinfo{person}{Joon~Yul Choi}.} \bibinfo{year}{2020}\natexlab{}.
\newblock \showarticletitle{Outcomes of adversarial attacks on deep learning
  models for ophthalmology imaging domains}.
\newblock \bibinfo{journal}{\emph{JAMA ophthalmology}} \bibinfo{volume}{138},
  \bibinfo{number}{11} (\bibinfo{year}{2020}), \bibinfo{pages}{1213--1215}.
\newblock


\bibitem[Yoon et~al\mbox{.}(2021)]%
        {yoon2021adversarial}
\bibfield{author}{\bibinfo{person}{Jongmin Yoon}, \bibinfo{person}{Sung~Ju
  Hwang}, {and} \bibinfo{person}{Juho Lee}.} \bibinfo{year}{2021}\natexlab{}.
\newblock \showarticletitle{Adversarial purification with score-based
  generative models}. In \bibinfo{booktitle}{\emph{International Conference on
  Machine Learning}}. PMLR, \bibinfo{pages}{12062--12072}.
\newblock


\bibitem[Yu et~al\mbox{.}(2022)]%
        {yu2022understanding}
\bibfield{author}{\bibinfo{person}{Chaojian Yu}, \bibinfo{person}{Bo Han},
  \bibinfo{person}{Li Shen}, \bibinfo{person}{Jun Yu}, \bibinfo{person}{Chen
  Gong}, \bibinfo{person}{Mingming Gong}, {and} \bibinfo{person}{Tongliang
  Liu}.} \bibinfo{year}{2022}\natexlab{}.
\newblock \showarticletitle{Understanding robust overfitting of adversarial
  training and beyond}. In \bibinfo{booktitle}{\emph{International Conference
  on Machine Learning}}. PMLR, \bibinfo{pages}{25595--25610}.
\newblock


\bibitem[Zafar et~al\mbox{.}(2024)]%
        {zafar2024robust_D2024_6}
\bibfield{author}{\bibinfo{person}{Aasim Zafar} {et~al\mbox{.}}}
  \bibinfo{year}{2024}\natexlab{}.
\newblock \showarticletitle{Robust Medical Diagnosis: A Novel Two-Phase Deep
  Learning Framework for Adversarial Proof Disease Detection in Radiology
  Images}.
\newblock \bibinfo{journal}{\emph{Journal of Imaging Informatics in Medicine}}
  (\bibinfo{year}{2024}), \bibinfo{pages}{1--31}.
\newblock


\bibitem[Zagoruyko and Komodakis(2016)]%
        {ZagoruykoK16}
\bibfield{author}{\bibinfo{person}{Sergey Zagoruyko} {and}
  \bibinfo{person}{Nikos Komodakis}.} \bibinfo{year}{2016}\natexlab{}.
\newblock \showarticletitle{Wide Residual Networks}. In
  \bibinfo{booktitle}{\emph{Proceedings of the British Machine Vision
  Conference 2016, {BMVC}}}, \bibfield{editor}{\bibinfo{person}{Richard~C.
  Wilson}, \bibinfo{person}{Edwin~R. Hancock}, {and} \bibinfo{person}{William
  A.~P. Smith}} (Eds.). \bibinfo{publisher}{{BMVA} Press}.
\newblock


\bibitem[Zemskova et~al\mbox{.}(2022)]%
        {zemskova2022deep}
\bibfield{author}{\bibinfo{person}{Varvara~E Zemskova},
  \bibinfo{person}{Tai-Long He}, \bibinfo{person}{Zirui Wan}, {and}
  \bibinfo{person}{Nicolas Grisouard}.} \bibinfo{year}{2022}\natexlab{}.
\newblock \showarticletitle{A deep-learning estimate of the decadal trends in
  the Southern Ocean carbon storage}.
\newblock \bibinfo{journal}{\emph{Nature communications}}  \bibinfo{volume}{13}
  (\bibinfo{year}{2022}), \bibinfo{pages}{1--11}.
\newblock


\bibitem[Zhang et~al\mbox{.}(2019)]%
        {zhang2019theoretically}
\bibfield{author}{\bibinfo{person}{Hongyang Zhang}, \bibinfo{person}{Yaodong
  Yu}, \bibinfo{person}{Jiantao Jiao}, \bibinfo{person}{Eric Xing},
  \bibinfo{person}{Laurent El~Ghaoui}, {and} \bibinfo{person}{Michael Jordan}.}
  \bibinfo{year}{2019}\natexlab{}.
\newblock \showarticletitle{Theoretically principled trade-off between
  robustness and accuracy}. In \bibinfo{booktitle}{\emph{International
  Conference on Machine Learning}}.
\newblock


\bibitem[Zhang et~al\mbox{.}(2020)]%
        {zhang2020attacks}
\bibfield{author}{\bibinfo{person}{Jingfeng Zhang}, \bibinfo{person}{Xilie Xu},
  \bibinfo{person}{Bo Han}, \bibinfo{person}{Gang Niu}, \bibinfo{person}{Lizhen
  Cui}, \bibinfo{person}{Masashi Sugiyama}, {and} \bibinfo{person}{Mohan
  Kankanhalli}.} \bibinfo{year}{2020}\natexlab{}.
\newblock \showarticletitle{Attacks which do not kill training make adversarial
  learning stronger}. In \bibinfo{booktitle}{\emph{ICML}}.
  \bibinfo{pages}{11278--11287}.
\newblock


\bibitem[Zhang et~al\mbox{.}(2022)]%
        {ZhangGL000S022}
\bibfield{author}{\bibinfo{person}{Yonggang Zhang}, \bibinfo{person}{Mingming
  Gong}, \bibinfo{person}{Tongliang Liu}, \bibinfo{person}{Gang Niu},
  \bibinfo{person}{Xinmei Tian}, \bibinfo{person}{Bo Han},
  \bibinfo{person}{Bernhard Sch{\"{o}}lkopf}, {and} \bibinfo{person}{Kun
  Zhang}.} \bibinfo{year}{2022}\natexlab{}.
\newblock \showarticletitle{Adversarial Robustness Through the Lens of
  Causality}. In \bibinfo{booktitle}{\emph{ICLR}}.
\newblock


\bibitem[Zheng et~al\mbox{.}(2019)]%
        {zheng2019distributionally}
\bibfield{author}{\bibinfo{person}{Tianhang Zheng}, \bibinfo{person}{Changyou
  Chen}, {and} \bibinfo{person}{Kui Ren}.} \bibinfo{year}{2019}\natexlab{}.
\newblock \showarticletitle{Distributionally adversarial attack}. In
  \bibinfo{booktitle}{\emph{AAAI}}.
\newblock


\bibitem[Zhou et~al\mbox{.}(2023)]%
        {zhou2023eliminating}
\bibfield{author}{\bibinfo{person}{Dawei Zhou}, \bibinfo{person}{Yukun Chen},
  \bibinfo{person}{Nannan Wang}, \bibinfo{person}{Decheng Liu},
  \bibinfo{person}{Xinbo Gao}, {and} \bibinfo{person}{Tongliang Liu}.}
  \bibinfo{year}{2023}\natexlab{}.
\newblock \showarticletitle{Eliminating adversarial noise via information
  discard and robust representation restoration}. In
  \bibinfo{booktitle}{\emph{ICML}}. PMLR, \bibinfo{pages}{42517--42530}.
\newblock


\bibitem[Zhou et~al\mbox{.}(2021a)]%
        {zhou2021towards}
\bibfield{author}{\bibinfo{person}{Dawei Zhou}, \bibinfo{person}{Tongliang
  Liu}, \bibinfo{person}{Bo Han}, \bibinfo{person}{Nannan Wang},
  \bibinfo{person}{Chunlei Peng}, {and} \bibinfo{person}{Xinbo Gao}.}
  \bibinfo{year}{2021}\natexlab{a}.
\newblock \showarticletitle{Towards defending against adversarial examples via
  attack-invariant features}. In \bibinfo{booktitle}{\emph{ICML}}. PMLR,
  \bibinfo{pages}{12835--12845}.
\newblock


\bibitem[Zhou et~al\mbox{.}(2022a)]%
        {zhou2022improving}
\bibfield{author}{\bibinfo{person}{Dawei Zhou}, \bibinfo{person}{Nannan Wang},
  \bibinfo{person}{Xinbo Gao}, \bibinfo{person}{Bo Han},
  \bibinfo{person}{Xiaoyu Wang}, \bibinfo{person}{Yibing Zhan}, {and}
  \bibinfo{person}{Tongliang Liu}.} \bibinfo{year}{2022}\natexlab{a}.
\newblock \showarticletitle{Improving adversarial robustness via mutual
  information estimation}. In \bibinfo{booktitle}{\emph{ICML}}. PMLR,
  \bibinfo{pages}{27338--27352}.
\newblock


\bibitem[Zhou et~al\mbox{.}(2022b)]%
        {zhou2022modeling}
\bibfield{author}{\bibinfo{person}{Dawei Zhou}, \bibinfo{person}{Nannan Wang},
  \bibinfo{person}{Bo Han}, {and} \bibinfo{person}{Tongliang Liu}.}
  \bibinfo{year}{2022}\natexlab{b}.
\newblock \showarticletitle{Modeling adversarial noise for adversarial
  training}. In \bibinfo{booktitle}{\emph{International Conference on Machine
  Learning}}. PMLR, \bibinfo{pages}{27353--27366}.
\newblock


\bibitem[Zhou et~al\mbox{.}(2021c)]%
        {zhou2021removing}
\bibfield{author}{\bibinfo{person}{Dawei Zhou}, \bibinfo{person}{Nannan Wang},
  \bibinfo{person}{Chunlei Peng}, \bibinfo{person}{Xinbo Gao},
  \bibinfo{person}{Xiaoyu Wang}, \bibinfo{person}{Jun Yu}, {and}
  \bibinfo{person}{Tongliang Liu}.} \bibinfo{year}{2021}\natexlab{c}.
\newblock \showarticletitle{Removing adversarial noise in class activation
  feature space}. In \bibinfo{booktitle}{\emph{IEEE/CVF ICCV}}.
  \bibinfo{pages}{7878--7887}.
\newblock


\bibitem[Zhou et~al\mbox{.}(2021b)]%
        {zhou2021ssmd_D41}
\bibfield{author}{\bibinfo{person}{Hong-Yu Zhou}, \bibinfo{person}{Chengdi
  Wang}, \bibinfo{person}{Haofeng Li}, \bibinfo{person}{Gang Wang},
  \bibinfo{person}{Shu Zhang}, \bibinfo{person}{Weimin Li}, {and}
  \bibinfo{person}{Yizhou Yu}.} \bibinfo{year}{2021}\natexlab{b}.
\newblock \showarticletitle{SSMD: semi-supervised medical image detection with
  adaptive consistency and heterogeneous perturbation}.
\newblock \bibinfo{journal}{\emph{MIA}} (\bibinfo{year}{2021}).
\newblock


\bibitem[Zhou et~al\mbox{.}(2021d)]%
        {zhou2021machine_A46}
\bibfield{author}{\bibinfo{person}{Qianwei Zhou}, \bibinfo{person}{Margarita
  Zuley}, \bibinfo{person}{Yuan Guo}, \bibinfo{person}{Lu Yang},
  \bibinfo{person}{Bronwyn Nair}, \bibinfo{person}{Adrienne Vargo},
  \bibinfo{person}{Suzanne Ghannam}, \bibinfo{person}{Dooman Arefan}, {and}
  \bibinfo{person}{Shandong Wu}.} \bibinfo{year}{2021}\natexlab{d}.
\newblock \showarticletitle{A machine and human reader study on AI diagnosis
  model safety under attacks of adversarial images}.
\newblock \bibinfo{journal}{\emph{Nature communications}} \bibinfo{volume}{12},
  \bibinfo{number}{1} (\bibinfo{year}{2021}), \bibinfo{pages}{1--11}.
\newblock


\bibitem[Zhu et~al\mbox{.}(2022)]%
        {ZhuY0ZL0ZXY22}
\bibfield{author}{\bibinfo{person}{Jianing Zhu}, \bibinfo{person}{Jiangchao
  Yao}, \bibinfo{person}{Bo Han}, \bibinfo{person}{Jingfeng Zhang},
  \bibinfo{person}{Tongliang Liu}, \bibinfo{person}{Gang Niu},
  \bibinfo{person}{Jingren Zhou}, \bibinfo{person}{Jianliang Xu}, {and}
  \bibinfo{person}{Hongxia Yang}.} \bibinfo{year}{2022}\natexlab{}.
\newblock \showarticletitle{Reliable Adversarial Distillation with Unreliable
  Teachers}. In \bibinfo{booktitle}{\emph{ICLR}}.
\newblock


\bibitem[Zi et~al\mbox{.}(2021)]%
        {zi2021revisiting}
\bibfield{author}{\bibinfo{person}{Bojia Zi}, \bibinfo{person}{Shihao Zhao},
  \bibinfo{person}{Xingjun Ma}, {and} \bibinfo{person}{Yu-Gang Jiang}.}
  \bibinfo{year}{2021}\natexlab{}.
\newblock \showarticletitle{Revisiting adversarial robustness distillation:
  Robust soft labels make student better}. In
  \bibinfo{booktitle}{\emph{IEEE/CVF International Conference on Computer
  Vision}}.
\newblock


\end{thebibliography}
}

\end{document}